\documentclass[prb,aps,floatfix,amsmath,eqsecnum,twocolumn]{revtex4-1}

\usepackage{amsmath}
\usepackage{epsfig}
\usepackage{array}
\usepackage{verbatim}

\usepackage{amssymb}
\usepackage{multirow}
\usepackage{graphicx}
\usepackage[export]{adjustbox}
\usepackage[caption=false]{subfig}

\usepackage{ulem}

\usepackage[cspex,bbgreekl]{mathbbol}

\usepackage{color}
	\definecolor{redish}{rgb}{0.5, 0.0, 0.13}
	 \definecolor{greenish}{rgb}{0.0, 0.5, 0.0}
\usepackage[linktocpage=true]{hyperref}
\hypersetup{
    colorlinks,
    citecolor=blue,
    filecolor=blue,
    linkcolor=blue,
    urlcolor=blue
 }

\newcommand{\nocontentsline}[3]{}
\newcommand{\tocless}[2]{\bgroup\let\addcontentsline=\nocontentsline#1{#2}\egroup}

\newcommand{\re}[1]{(\ref{#1})}

\newcommand{\beg}{\begin{equation}}

\newcommand{\en}{\end{equation}}

\newcommand{\eps}{\varepsilon}

\newcommand{\rr}{\boldsymbol{r}}

\newcommand{\eref}[1]{Eq.~(\ref{#1})}
\newcommand{\esref}[1]{Eqs.~(\ref{#1})}

\newcommand{\dinf}{\Delta_{\infty}}

\newcommand{\up}{\uparrow}

\renewcommand{\emph}{\textit}

\newcommand{\beq}{\begin{equation}}

\newcommand{\eeq}{\end{equation}}
\newcommand{\barray}{\begin{eqnarray}}
\newcommand{\earray}{\end{eqnarray}}

\usepackage{braket}

\renewcommand{\rr}{\mathfrak{r}}

\newcommand{\msf}[1]{\mathsf{#1}}
\newcommand{\Emin}{E_{\msf{min}}}

\begin{document}

\title{Consequences of integrability breaking in quench dynamics of pairing Hamiltonians}

\author{Jasen A. Scaramazza, Pietro Smacchia and Emil A. Yuzbashyan}
\date{\today}							

\affiliation{Department of Physics and Astronomy, Rutgers University, Piscataway, New Jersey 08854, USA}

\begin{abstract}
We study the collisionless  dynamics of two classes of nonintegrable pairing models. One  is a  BCS model   with  separable energy-dependent interactions,  the other -- a 2D topological superconductor with spin-orbit coupling and a band-splitting external field. 
The long-time quantum quench   dynamics   at integrable points of these models are well understood. Namely, the squared magnitude of the time-dependent order parameter $\Delta(t)$ can either vanish (Phase I), reach a nonzero constant (Phase II), or periodically oscillate as an elliptic function (Phase III). We demonstrate that nonintegrable models too exhibit some or all of these nonequilibrium phases. Remarkably, elliptic periodic oscillations persist, even though both their amplitude and functional form change drastically  with integrability breaking. 
Striking  new phenomena   accompany  loss of integrability.  First, an extremely long time scale  emerges in the relaxation to Phase III, such that short-time numerical simulations risk erroneously classifying the asymptotic state. This  time scale  diverges near integrable points.  
Second, an entirely new Phase IV of quasiperiodic oscillations of $|\Delta|$ emerges in the quantum quench phase diagrams of nonintegrable pairing models. As integrability techniques do not apply for the models we study, we develop the concept of asymptotic self-consistency  and a linear stability analysis of the asymptotic phases.  With the help of these new tools, we determine the phase boundaries, characterize
the asymptotic state, and clarify the physical meaning of the quantum quench phase diagrams of BCS superconductors. We also propose an explanation of these diagrams in terms of bifurcation theory. 
 \end{abstract}

\date{\today}
\maketitle

 \tableofcontents

\newpage

\section{Introduction}
\label{intro}

 \parskip=6pt plus 1 pt

The past fifteen years have borne witness to impressive advances in the ability to experimentally control many-body systems where dissipative and decoherence effects are strongly suppressed. Studies of cold atomic gases\cite{kinoshita2006,ligner,Hofferberth:2007,weiler,widera,Gring:2012,Langen:2015,tang,norcia,smale}, solid state pump-probe experiments\cite{ftdk,ktn,gcff,mhmutws,mtfsmutwas} and quantum information processing\cite{riberio,gorshkov0,zhou,barends,nichol,song,lukin,gorshkov} can now explore coherent many-body dynamics for long time scales, paving the way for the characterization of new phenomena. In particular, cold atomic gases with tunable interactions\cite{jin,ketterle,bdz,zass,gps,bdn} are an instrumental experimental tool in the quest to understand previously inaccessible aspects of far from equilibrium many-body dynamics.

A major focus of recent theory and experiment has been the unitary time evolution of a system, initially in the ground state, subject to a sudden perturbation\cite{pssv,gogolin,moore}. This experimental protocol, known as a quantum quench, can induce long-lived states with properties strikingly different from those of equilibrium states at similar energy scales. In this work, we focus on the quench dynamics of various superconducting models, which is a modern reformulation of the longstanding problem of nonequilibrium superconductivity in the collisionless regime\cite{anderson,galaiko,volkov,galperin}. A canonical result is that the infinitesimal perturbation of a Bardeen-Cooper-Schrieffer (BCS) $s$-wave superconductor leads to power law oscillatory relaxation of the order parameter amplitude $|\Delta|$ to a constant value\cite{volkov}.

Decades later, it was discovered that larger deviations could give rise to different dynamical phases identified by the asymptotic behavior of the amplitude of the order parameter\cite{barlev0,amin,yuzbashyan-jpa,Szymanska,yuzbashyan2005,yta,barlev,yuzdze}. Consider the dynamics of $\Delta$ after quenches of the coupling $g$ in various superconducting models. When the final coupling $g_f$ is small enough, $\Delta$ vanishes rapidly in time; this behavior characterizes what we call Phase~I. For intermediate $g_f$, $|\Delta|$ exhibits oscillatory power law decay to a nonzero constant (Phase~II). For larger $g_f$, $|\Delta|$ exhibits persistent periodic oscillations (Phase~III) -- a nonlinear manifestation of what is known in the literature as the Higgs or amplitude mode\cite{barlev3,pekvar,pashleit,barvar,nmfkd,hkcvak,kbums,mve}.

The exact quantum quench phase diagrams of the $s$-wave superconductor were eventually constructed using a sophisticated analytical method that relies on the model's integrability\cite{ydgf}. It turns out that the integrable $p+ip$ topological superconductor exhibits the same three phases, and similar analytical tools lead to the construction of its phase diagrams\cite{fdgy}. Thus, there may appear to be some profound connection between integrability and these three dynamical phases, but nonintegrable models  also have Phases~I and II\cite{barlev3,Szymanska,ddgp,dky,psc} and Phase~III-like behavior is thought to persist in some nonintegrable models as well. On the other hand, the existence of Phase~III in such models has not been convincingly established beyond the linear regime and aspects of quench dynamics unique to the nonintegrable case have not been explored.


Overall, the description of these nonequilibrium dynamical phases lacks a unifying mechanism applicable to finite quenches of nonintegrable pairing models. Here we present an in-depth study of the nonequilibrium phases of various nonintegrable superconducting models with and without spin-orbit coupling.  A common feature of models we consider is that the order parameter takes the form of a single complex number. We establish that Phase~III persists when integrability is broken
\cite{note} and give strong numerical evidence that the persistent oscillations are always elliptic, which generalizes the known behavior of integrable models\cite{barlev0,ydgf,fdgy}.

\begin{figure*}
\subfloat[]{
	\includegraphics[width=.48\linewidth]{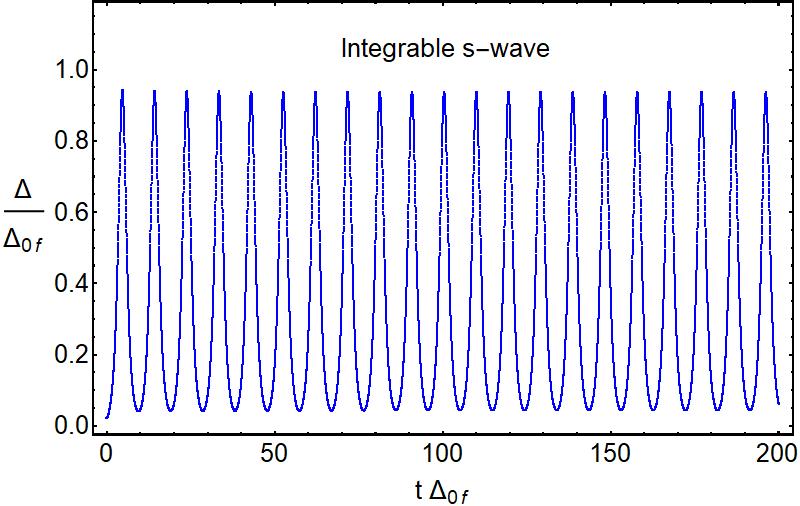}
}
\hfill
\subfloat[]{
\includegraphics[width=.48\linewidth]{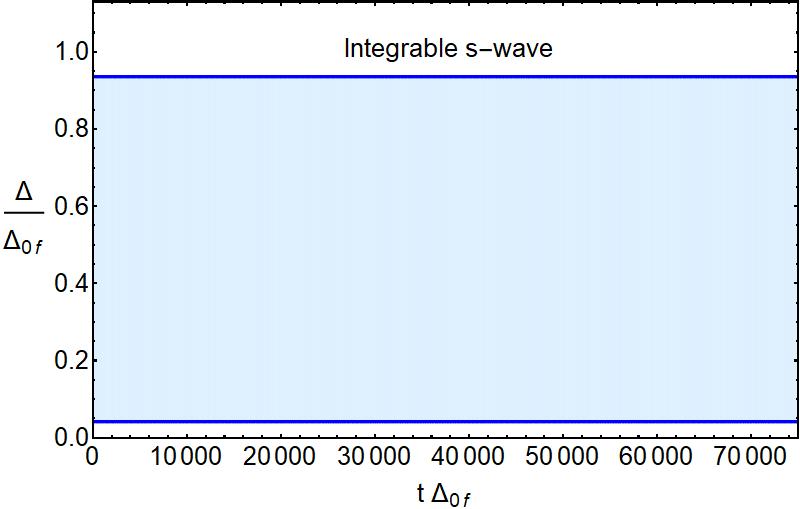}
}

\subfloat[]{
	\includegraphics[width=.48\linewidth]{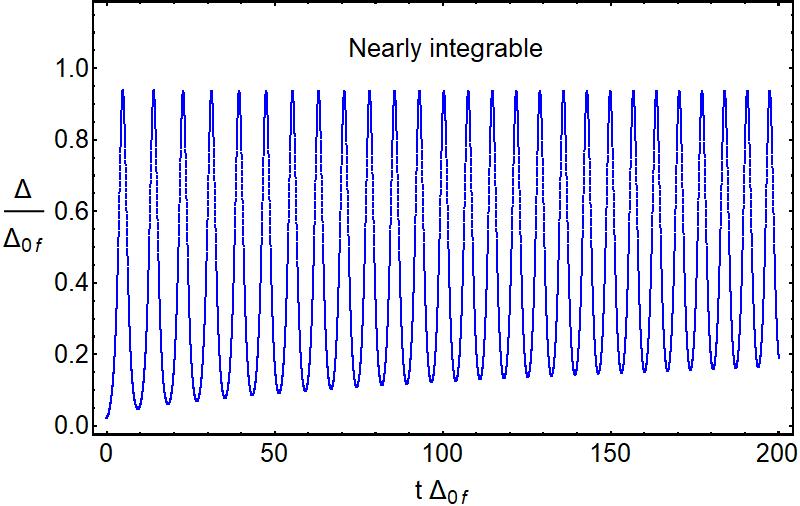}
}
\hfill
\subfloat[]{
\includegraphics[width=.48\linewidth]{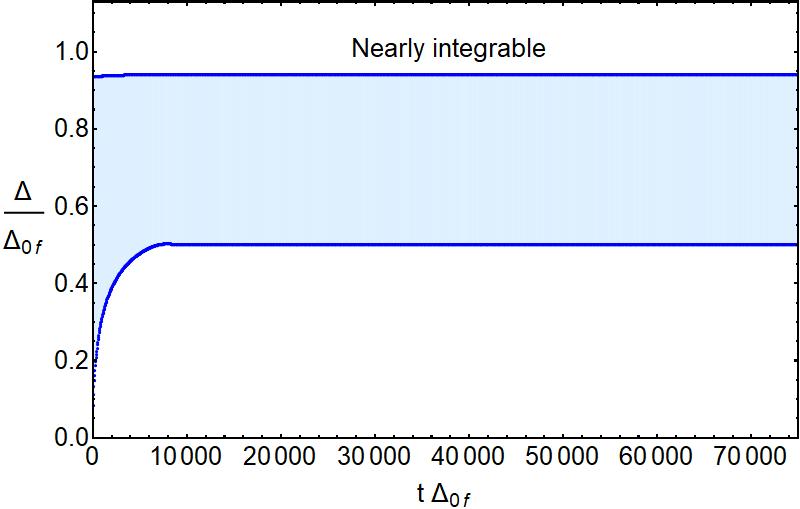}
}

\subfloat[]{
	\includegraphics[width=.48\linewidth]{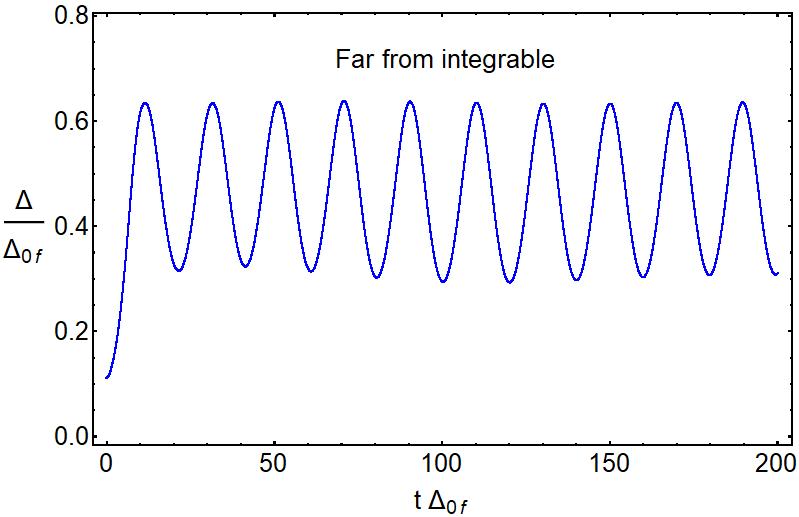}
}
\hfill
\subfloat[]{
\includegraphics[width=.48\linewidth]{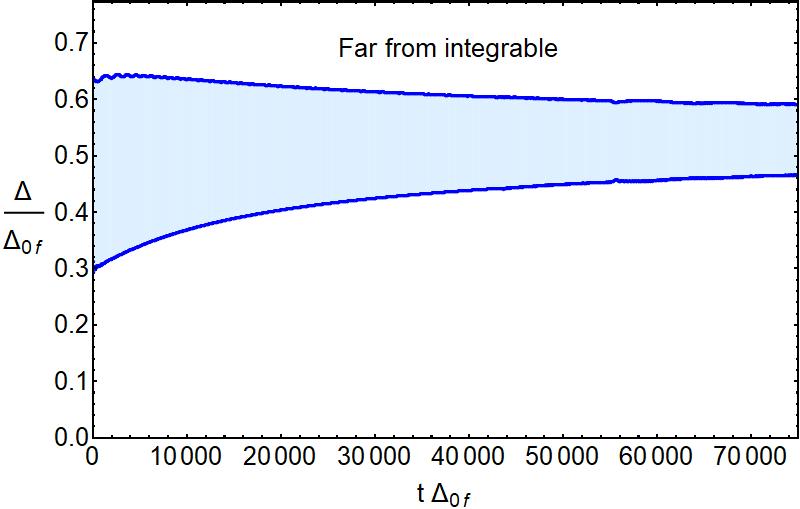}
}
\caption{Illustration of the large time scale $\tau$ that emerges in Phase~III quenches $g_i\to g_f$ of nonintegrable pairing models. In all plots, the equilibrium gap corresponding to the initial coupling $g_i$ is $\Delta_{0i} = 1.33\times10^{-3}W,$ while that  for the final coupling $g_f$ is $\Delta_{0f} = 0.4W,$ and we took $N=2\times 10^5$ equally spaced single-particle energy levels   on
the interval $[-W/2,W/2]$. The lines in the plots on the right are the local minima and maxima of the oscillations. In terms of
the single-particle level spacing $\delta$, the evolution in the right column goes
out to $t_{\max}=0.94\delta^{-1}$.
 In (a) and (b), we see that the persistent elliptic oscillations in the integrable $s$-wave case stabilize after a small number of oscillations. In (c) and (d), the amplitude of the oscillations takes roughly a thousand times longer to stop changing. In (e) and (f), integrability is strongly broken and it is not even clear whether the oscillations stabilize to a constant amplitude. The nonintegrable model used was the separable BCS model \re{HfComplex} with  $f(\eps)$ from \eref{fLorIntro}. The nearly integrable version uses $\gamma = W,$ while the far from integrable one has  $\gamma = 1.33\times10^{-2}W.$}
\label{time_scales}
\end{figure*}

Although the integrable and nonintegrable phenomenology are similar, we find that integrability breaking has profound consequences. Unique to nonintegrable models is an \textit{extremely long relaxation time scale} $\tau$ which diverges as one approaches integrable points and is most prominent in quenches to Phase~III. One must analyze dynamics beyond $\tau$ to truly observe Phase~III, which has not been done in other studies. As illustrated in Fig.~\ref{time_scales}, for $t<\tau$, $|\Delta|$ may oscillate with several frequencies and a slowly evolving amplitude, both of which undermine naive analyses restricted to $t < \tau$. One may incorrectly conclude from the transient dynamics that the asymptotic nonequilibrium phase has several undamped frequencies, or that $|\Delta|$ is oscillating periodically while in fact the amplitude is still changing. Nonintegrable Phase~III oscillations further require comparatively more elaborate elliptic functions to describe the oscillations.

To complicate the picture even further, certain quantum quenches of nonintegrable pairing models genuinely do not fit into any of the Phases I, II and III. Here the asymptotic $|\Delta|$ is truly
   \textit{quasiperiodic}, leading us to  conclude that there are regions of quasiperiodicity -- \textit{a new Phase~IV} -- in the quantum quench phase diagrams of these models.

Another consequence of integrability breaking arises in the analytical description of the three nonequilibrium phases. In the integrable case, there is a dynamical reduction in the number of degrees of freedom of the system\cite{ydgf,fdgy} such that Phases I, II and III correspond to an effective classical spin Hamiltonian with 0, 1 and 2 spins, respectively. Phase~III in the general case, however, does not admit such a 2-spin representation. As a surrogate to this analytical method, we propose a stability analysis of Phases I and II that applies generally to finite quenches. The stability analysis is based on linearizing around the asymptotic solutions to the equations of motion in each of the phases. We can then nonperturbatively  determine the phase I-II boundary as well as the phase II-III boundary in nonintegrable pairing models. Finally, we return to Phase~III and argue that the self-consistency condition (gap equation) is responsible not only for the existence of persistent periodic oscillations of $|\Delta|$, but also for selecting elliptic functions amongst all possible periodic functions.

 \parskip=0pt plus 1pt

\section{Models and pseudospin representation}
\label{models}

In this paper, we consider quantum quenches in two types of nonintegrable pairing models
\beg
\begin{split}
\hat{H}_f &= \sum_{j\lambda}\eps_j \hat{c}^{\dag}_{j\lambda}\hat{c}_{j\lambda}-\frac{1}{g}\hat{\Delta}^{\dag}\hat{\Delta},\quad
\hat{\Delta} \equiv g\sum_{j}f_j\hat{c}_{j\downarrow}\hat{c}_{j\up},\\
\hat{H}_{so} &= \sum_{\mathbf{k}ab}\bigg[(\eps_k\delta_{ab}-h \sigma^z_{ab}) + \alpha(k_y\sigma^x_{ab}-k_x\sigma^y_{ab})\bigg]\hat{c}^{\dag}_{\mathbf{k}a}\hat{c}_{\mathbf{k}b}-\\
&-\frac{1}{g}\hat{\Delta}^{\dag}\hat{\Delta}, \quad \hat{\Delta} \equiv g\sum_{\mathbf{k}}\hat{c}_{-\mathbf{k}\downarrow}\hat{c}_{\mathbf{k}\up}.
\end{split}
\label{ham1}
\en
The Hamiltonian $\hat{H}_f$ is a separable BCS Hamiltonian where the $\eps_j$ are $N$ single-particle energy levels, $\hat{c}^{\dag}_{j\lambda}$ ($\hat{c}_{j\lambda}$) is a fermion creation (annihilation) operator for an electron with energy $\eps_j$ and spin index $\lambda$, $g>0$ is the pairing interaction strength and $f_j \equiv f(\eps_j)$ is a generic function of $\eps_j$. The Hamiltonian $\hat{H}_{so}$ describes a 2D topological spin-orbit coupled superconductor with $s$-wave interactions\cite{so1,so2}. Here $\mathbf{k} = (k_x,k_y)$ is a two-dimensional momentum vector, $\sigma^{j}$ are Pauli matrices, $h$ is a Zeeman field and $\alpha$ is the Rashba spin-orbit coupling. We will take the density of states to be constant for both models, which is the case in 2D or at weak coupling, so that the single-particle energy levels are distributed uniformly on an interval of length $W$, called the bandwidth.

Apart from certain choices of $f(x)$, the separable BCS Hamiltonian $\hat{H}_f$ is a toy model for breaking integrability. The choice of $f^2(x)=C_1+C_2 x$ produces a quantum integrable Hamiltonian\cite{richardson,ortiz2005}; for example, $f(x) = 1$ and $f(x) = \sqrt{x}$ correspond to the $s$-wave\cite{yuzbashyan-jpa} and $p+ip$\cite{dunning,dukelsky} BCS models, respectively. A notable nonintegrable case is the $d+id$ model\cite{Marquette}, where $f(x) = x$. The spin-orbit Hamiltonian $\hat{H}_{so}$, on the other hand, can be realized with cold Fermi gases\cite{mamatu,ffvkzk,dkwhmhlhv,wyfmhczz,cshybz,wbljs,fhmwzzzzz,mamatu2,hmwpzclzz,huizhai}.

As both Hamiltonians in \eref{ham1} have infinite range interactions, the mean-field approximation is expected to be exact in the thermodynamic ($N\to\infty$) limit. We therefore replace 2-body operators as follows $\hat{c}^{\dag}\hat{c}^{\dag}\hat{c}\hat{c} \approx \langle\hat{c}^{\dag}\hat{c}^{\dag}\rangle\hat{c}\hat{c} + \hat{c}^{\dag}\hat{c}^{\dag}\langle\hat{c}\hat{c}\rangle - \langle\hat{c}^{\dag}\hat{c}^{\dag}\rangle\langle\hat{c}\hat{c}\rangle$ in the equations of motion. We also diagonalize the noninteracting part of $\hat{H}_{so}$ through a unitary transformation $U_{\mathbf{k}}$ which is detailed in Appendix~\ref{appA}. Up to additive constants, the effective mean-field Hamiltonians of \eref{ham1} are
\beg
\begin{split}
\hat{H}_f &= \sum_{j\lambda=\up\downarrow}\eps_j \hat{c}^{\dag}_{j\lambda}\hat{c}_{j\lambda}-\sum_{j}f_j\bigg[\Delta^*\hat{c}_{j\downarrow}\hat{c}_{j\up}+h.c.\bigg],\\
\hat{H}_{so} &= \sum_{\mathbf{k}\lambda=\pm}\eps_{k\lambda}\hat{a}^{\dag}_{\mathbf{k}\lambda}\hat{a}_{\mathbf{k}\lambda}-\bigg(\frac{\Delta}{2}\sum_{\mathbf{k}\lambda}e^{-i \theta_{\mathbf{k}}}\bigg[\lambda\sin\phi_{k}\hat{a}^{\dag}_{\mathbf{k}\lambda}\hat{a}^{\dag}_{-\mathbf{k}\lambda}+\\
&+\cos\phi_{k}\hat{a}^{\dag}_{-\mathbf{k}\lambda}\hat{a}^{\dag}_{\mathbf{k}\bar{\lambda}}\bigg] + h.c.\bigg)\\
\end{split}
\label{hamMF}
\en
The new parameters in $\hat{H}_{so}$ are
\beg
\begin{split}
&\cos\phi_{k}=\frac{h}{R_k},\quad \sin\phi_{k}=\frac{\alpha k}{R_k},\\
&R_k = \sqrt{h^2+\alpha^2k^2},\\
&\eps_{k\lambda} = \eps_k-\lambda R_k,\quad \lambda=\pm,\quad \bar{\lambda}=-\lambda,\\
&\mathbf{k} = k_x + i\,k_y = k e^{i\,\theta_{\mathbf{k}}}.
\end{split}
\label{SOparams}
\en
Note that both $\alpha=0$ and $h=0$ correspond to integrable points of the spin-orbit model; in both cases, $\hat{H}_{so}$ becomes a Hamiltonian for two bands of independent $s$-wave BCS models. Most importantly, the mean-field order parameters $\Delta \equiv \Delta(t)$ are defined in terms of expectation values
\beg
\begin{split}
\Delta &= g\sum_jf_j\langle\hat{c}_{j\downarrow}\hat{c}_{j\up}\rangle,\\
\Delta &= \frac{g}{2}\sum_{\mathbf{k}\lambda=\pm}e^{i\,\theta_{\mathbf{k}}}\bigg[\lambda\sin\phi_{k}\langle\hat{a}_{-\mathbf{k}\lambda}\hat{a}_{\mathbf{k}\lambda}\rangle+\cos\phi_{k}\langle\hat{a}_{\mathbf{k}\lambda}\hat{a}_{-\mathbf{k}\bar{\lambda}}\rangle\bigg],
\end{split}
\label{OPs1}
\en
for their respective models.

We will discuss the mean-field dynamics generated by Hamiltonians \re{hamMF} in terms of Anderson pseudospins $\mathbf{\hat{s}}_j=(\hat{s}^x_j,\hat{s}^y_j,\hat{s}^z_j)$ which will allow for intuitive visualizations of the dynamics of different nonequilibrium phases. The transformation from fermions to pseudospins is given by
\beg
\hat{s}^-_j = \hat{s}^x_j - i\,\hat{s}^y_j = \hat{c}_{j\downarrow}\hat{c}_{j\up},\quad \hat{s}^z_j = \frac{1}{2}(\hat{c}^{\dag}_{j\up}\hat{c}_{j\up} + \hat{c}^{\dag}_{j\downarrow}\hat{c}_{j\downarrow} - 1).
\label{FtoPs}
\en
In the spin-orbit case the pseudospin representation requires an additional set of auxiliary  variables. For the sake of brevity, we relegate the derivations of the pseudospin equations of motion to Appendix~\ref{appA} and simply state them here.

In the mean-field equations of motion that follow, $\mathbf{s} = \langle \mathbf{\hat{s}} \rangle$ are to be understood as classical variables satisfying the angular momentum Poisson brackets~$\{s^a_j,s^b_k\} = -\delta_{jk}\epsilon_{abc}s_j^c$. In the separable BCS model, we have
\beg
\begin{split}
\dot{\mathbf{s}}_j &= \mathbf{b}_j\times\mathbf{s}_j,\quad 
\mathbf{b}_j =(-2f_j\Delta_x,-2f_j\Delta_y,2\eps_j),\\
\end{split}
\label{eomHF}
\en
where self-consistency requires 
\beg
\Delta = g\sum_jf_js_j^-=\Delta_x-i\Delta_y.
\label{self13}
\en
The spin-length $s_j = 1/2$ is conserved by \esref{eomHF}, which together with \eref{self13} are the equations of motion of the following classical spin Hamiltonian:
\beg
\begin{split}
H_f &= \sum_j2\eps_js_j^z-g\sum_{j,k}f_jf_ks_j^+s_k^-\\
&=\sum_j2\eps_js^z_j-|\Delta|^2/g.
\end{split}
\label{HFSpin}
\en
Note that without loss of generality, we can choose  $f_j$ to be real and nonnegative as we have done above. Indeed, let $f_j=|f_j| e^{-i\theta_j}$ be general complex numbers and
\beg
\begin{split}
H_f &=\sum_j  2 \eps_j s^z_j-  g \sum_{j,k} f_j  f_k^* s^+_j s^-_k.
\end{split}
\label{HfComplex}
\en
We redefine the spins by making local rotations around the z-axis, $s^-_j \to s^-_j e^{-i\theta_j}$.  In terms of the new spins the Hamiltonian becomes
\beg
\begin{split}
H_f &=\sum_j  2 \eps_j s^z_j-  g \sum_{j,k} |f_j|  |f_k| s^+_j s^-_k, 
\end{split}
\label{HfComplex2}
\en
and the order parameter is $\Delta=\sum_j |f_j| s^-_j$. This transformation does not affect spin (angular momentum) Poisson brackets and therefore the equations of motion retain their form. We thus arrive at the same problem only with $f_j\to |f_j|$.

We use capital letters $\mathbf{S}_{\mathbf{k}\lambda}$ to denote the classical pseudospins in the spin-orbit model and must introduce (see Appendix~\ref{appA}) a set of auxiliary variables: the scalars $T_{\mathbf{k}}$ and vectors $\mathbf{L}_{\mathbf{k}\pm}$, where $\mathbf{L}_{\mathbf{k}+}$ and $\mathbf{L}_{\mathbf{k}-}$ differ only in sign of the z-component. The equations of motion are
\beg
\begin{split}
\dot{\mathbf{S}}_{\mathbf{k}\lambda} &= \mathbf{B}_{k\lambda}\times\mathbf{S}_{\mathbf{k}\lambda}+\mathbf{m}_k\times \mathbf{L}_{\mathbf{k}\lambda}-\mathbf{m}_kT_{\mathbf{k}},\\
\dot{L}^x_{\mathbf{k}\lambda} &= -2\eps_kL^y_{\mathbf{k}\lambda}+\frac{m^y_k}{2}\big[S^z_{\mathbf{k}+}+S^z_{\mathbf{k}-}\big]+B^x_{k\lambda}T_{\mathbf{k}},\\
\dot{L}^y_{\mathbf{k}\lambda} &= 2\eps_kL^x_{\mathbf{k}\lambda}-\frac{m^x_k}{2}\big[S^z_{\mathbf{k}+}+S^z_{\mathbf{k}-}\big]+B^y_{k\lambda}T_{\mathbf{k}},\\
\dot{L}^z_{\mathbf{k}\lambda} &= -2 R_k \lambda T_{\mathbf{k}}+\frac{m^x_k}{2}\big[ S^y_{\mathbf{k}\lambda} - S^y_{\mathbf{k}\bar{\lambda}} \big]-\frac{m^y_k}{2}\big[ S^x_{\mathbf{k}\lambda} - S^x_{\mathbf{k}\bar{\lambda}} \big],\\
\dot{T}_{\mathbf{k}} &= 2R_k L^z_{\mathbf{k}+} - B^x_{k+}L^x_{\mathbf{k}+} - B^y_{k+}L^y_{\mathbf{k}+}+\\
 &+ \frac{1}{2}\mathbf{m}_k\cdot\big[ \mathbf{S}_{\mathbf{k}+} + \mathbf{S}_{\mathbf{k}-} \big],
\end{split}
\label{eomHSO}
\en
where the momentum dependent fields $\mathbf{B}_{k\lambda}$ and $\mathbf{m}_k$ are defined in terms of the order parameter $\Delta$
\beg
\begin{split}
\Delta &= \frac{g}{2}\sum_{\mathbf{k}\lambda}\big[ \sin\phi_kS^-_{\mathbf{k}\lambda} + \cos\phi_k L^-_{\mathbf{k}\lambda}\big]\\
&= \Delta_x - i \Delta_y,\\
\mathbf{B}_{k\lambda}&=(-2\sin\phi_k\Delta_x,-2\sin\phi_k\Delta_y,2\eps_{k\lambda}),\\
\mathbf{m}_k&=(-2\cos\phi_k\Delta_x,-2\cos\phi_k\Delta_y,0).
\end{split}
\label{fieldsHSO}
\en
The first of these equations is the self-consistency relationship for the spin-orbit model.
The equation for $\dot{\mathbf{S}}_{\mathbf{k}\lambda}$ in \eref{eomHSO} corrects an error in a previous paper\cite{dky}, which is missing the last term. For each $\mathbf{k}$, there is a conserved quantity analogous to pseudospin length
\beg
N^2_{\mathbf{k}} = 2T^2_{\mathbf{k}} +  \sum_{\lambda}\big[\mathbf{S}^2_{\mathbf{k}\lambda} + \mathbf{L}^2_{\mathbf{k}\lambda}\big]= \frac{1}{4}.
\label{NSO}
\en
Similar to \eref{HFSpin}, the classical spin-orbit Hamiltonian in pseudospin notation has a simple and compact expression
\beg
\begin{split}
H_{so} 
&=\sum_{\mathbf{k}\lambda}2\eps_{k\lambda}S^z_{\mathbf{k}\lambda}-2|\Delta|^2/g.
\end{split}
\label{HSOSpin}
\en
Because of the simple relationship connecting $\mathbf{L}_{\mathbf{k}+}$ to $\mathbf{L}_{\mathbf{k}-}$, each momentum vector $\mathbf{k}$ corresponds to ten dynamical variables $(\mathbf{S}_{\mathbf{k}+},\mathbf{S}_{\mathbf{k}-},\mathbf{L}_{\mathbf{k}+},T_{\mathbf{k}})$ constrained by \eref{NSO}. Note that $T_{\mathbf{k}}$ and $L^z_{\mathbf{k}\lambda}$ do not appear in \re{HSOSpin}, but as discussed in Appendix~\ref{appA}, they are necessary for the closure of the equations of motion. From now on we simplify notation to $\mathbf{L}_{\mathbf{k}}\equiv\mathbf{L}_{\mathbf{k}+}$ and define the 10-dimensional vector $\mathbf{\Gamma}_{\mathbf{k}} \equiv (\mathbf{S}_{\mathbf{k}+},\mathbf{S}_{\mathbf{k}-},\mathbf{L}_{\mathbf{k}},T_{\mathbf{k}})$.

Finally, the conservation of the total number of fermions $N_f$ in each model corresponds to the conservation of total z-component in the pseudospin language
\beg
\begin{split}
N_f = \sum_j (2s^z_j+1),
\end{split}
\label{HFzComp}
\en
for the separable BCS model and
\beg
\begin{split}
N_f = \sum_{\mathbf{k}\lambda} \bigg(S^z_{\mathbf{k}\lambda}+\frac{1}{2}\bigg),
\end{split}
\label{HSOzComp}
\en
for the spin-orbit model.

\section{Main results}
\label{mainresults}
The main purpose of this work is to compare the nonequilibrium phases of quenches from the ground state of nonintegrable pairing Hamiltonians, such as those in \eref{ham1}, to those of the integrable $s$-wave\cite{ydgf} and $p$-wave\cite{fdgy} models. Some qualitative aspects of the primary phases are independent of integrability insofar as the squared modulus of the order parameter $\Delta$ may exhibit any of three distinct asymptotic behaviors in the continuum limit: it can relax to zero (Phase~I), relax to a nonzero constant value (Phase~II), or display persistent periodic elliptic oscillations (Phase~III).

We first show the existence of these three phases in Sects.~\ref{subphase1}-\ref{subphase3} through direct numerical simulation of the dynamics. In Sect.~\ref{subStAn} we present a stability analysis of the phases of the separable BCS models which leads to conditions for nonequilibrium phase transitions. The stability analysis applied to integrable cases reduces to the known results that relied on exact solvability\cite{fdgy,ydgf}.   Our analysis provides a physical explanation for the transitions in terms of the frequencies of linearized perturbations $\delta\Delta(t)$ of the asymptotic $\Delta$. The transition from Phase~I to Phase~II occurs through an exponential instability characterized by a pair of conjugate imaginary frequencies in the linearization spectrum, while that of Phase~II to III occurs either when small harmonic oscillations fail to dephase or when an exponential instability occurs.

The appearance of some or all of Phases~I-III in nonintegrable models suggests an underlying universality to quench dynamics, but we show that the story is less straightforward. One the one hand, these phases are understood in the integrable cases\cite{ydgf,fdgy}. There is a dynamical reduction of the number of effective degrees of freedom, so that at large times the dynamics are governed by a Hamiltonian of the same form, but which has just a few collective degrees of freedom. The three phases correspond to 0, 1 or 2 effective spins for each phase, respectively. On the other hand, the nonintegrable dynamics admit no known analogous reduction because the 2-spin solutions to the equations of motion do not reproduce the observed asymptotic behavior of $\Delta$ in Phase~III. If such a reducing ``flow'' in time of the Hamiltonian occurs in the nonintegrable case, then the form of the Hamiltonian itself must change. For specifics on this latter point, see Appendix~\ref{AppReductionMech}.

Importantly, nonintegrable pairing models also display dynamics markedly different from those in the main three phases. We illustrate this behavior with two  examples in Sect.~\ref{quasisect} -- one for the spin-orbit Hamiltonian and one for a particle-hole symmetric separable BCS Hamiltonian -- where the magnitude of the order
parameter   oscillates  quasiperiodically. We interpret this observation as an indication of a new quasiperiodic phase (Phase IV) unique to  quantum quench phase diagrams of these models.

More subtle details of the dynamics in the main three phases change drastically   once integrability is broken. We show in Sect.~\ref{subsubRel} that nonintegrable models take an extremely long time to relax to Phase~III. This time scale is absent in the integrable case, yet it diverges when one approaches the integrable limit. One must take this time scale into account when studying Phase~III on the basis of numerical simulation alone. For example, in the weak coupling regime, the nonintegrable $d+id$ model may appear to quickly enter Phase~III\cite{didShortTime} while in fact the minima of $|\Delta|$ oscillations have not converged to a fixed value. The further into the weak coupling regime one explores, the longer the relaxation time. Quenches outside of weak coupling have  faster dynamics, but exhibit behavior that markedly contrasts with Phase~III, and above a certain energy threshold the asymptotic state collapses rapidly to Phase~II. This long relaxation time is typical in the nonintegrable case.  

Despite these consequences of breaking integrability, our mixed strategy of simulation and stability analysis applies to the two rather different classes of nonintegrable pairing models found in \eref{ham1}. The separable BCS permits a standard Anderson pseudospin representation and is a single band model, while the spin-orbit model requires an expanded pseudospin representation, has multiple bands and a topological quantum phase transition. Yet both models have a single complex order parameter, which we believe is the essential characteristic that leads to the three phases.

The self-consistency relationship \re{self13} for the order parameter is central to both our stability analysis of Phases~I and II in Sect.~\ref{subStAn} and our investigations of Phase~III in Sect.~\ref{asympPerSol}. In the former case, the frequencies of harmonic perturbations of a given nonequilibrium phase are constrained by the self-consistency requirement. As for Phase~III, we show in Sect.~\ref{asympPerSol} that there is always a periodic solution to the spin equations of motion when $\Delta(t)$ is periodic, and that the general spin solution precesses around the periodic one. We then argue through numerical examples that further imposing the self-consistency requirement on $\Delta(t)$ selects elliptic functions amongst all possible periodic $\Delta(t)$.

\section{Ground state and quench protocol}
\label{GSandQP}
In a quantum quench, we prepare the system in the ground state with an initial order parameter $\Delta = \Delta_0e^{-2i\mu t}$, which corresponds to system parameters such as the interaction strength $g$, the equilibrium chemical potential $\mu$, the magnetic field $h$ and the spin-orbit strength $\alpha$. The amplitude $\Delta_0$ is constant in the ground state. At time $t=0$, we suddenly change one of these parameters, which throws the system out of equilibrium. In the separable BCS model we will consider quenches $g_i\to g_f$, but we will label the initial and final states by the coordinates $\Delta_{0i}\equiv\Delta_0(g_i)$ and $\Delta_{0f}\equiv\Delta_0(g_f)$. In the spin-orbit model, we will consider quenches of the magnetic field $h_i \to h_f$. The fermion number $N_f$ is fixed across the quench in both cases, which implies that the equilibrium chemical potential $\mu$ changes with $h$.

For a given $\Delta_0$ and $\mu$, we express the ground state configuration of the separable BCS model in a frame that rotates around the z-axis with frequency $2\mu$.  We then orient each $\mathbf{s}_j$ against the magnetic $\mathbf{b}_j$, the z-component of which is shifted by $2\mu$,
\beg
\begin{split}
s^-_{j0} &= \frac{f_j\Delta_0}{2E_j},\quad s^z_{j0} = -\frac{\eps_j-\mu}{2E_j},\\
E_j(\Delta) &\equiv \sqrt{(\eps_j-\mu)^2+f^2(\eps_j)|\Delta|^2}.
\end{split}
\label{GSHF}
\en
The relationship between $\Delta_0$, $g$, $N_f$ and $\mu$ obtains from the application of the definition of $\Delta$ in \re{self13} to \re{HFzComp} and the configuration in \re{GSHF},
\beg
\begin{split}
\frac{1}{g} = \sum_j\frac{f_j^2}{2E_j},\quad N_f = \sum_j\bigg(1-\frac{\eps_j-\mu}{E_j}\bigg)
\end{split}
\label{SelfConHF}
\en
We will assert without loss of generality that $\Delta_{0i}$ is real in both models, which can always be achieved by a time-independent rotation in the $xy$-plane in pseudospin space.

Unless otherwise stated, we will simplify the analysis of the separable BCS model by restricting ourselves to cases where the order parameter $\Delta$ remains real for all time, i.e., $\Delta_y(t) = 0$. To achieve this, we will consider the particle-hole symmetric case where the energies $\eps_j$ are symmetrically distributed around the chemical potential $\mu$, which is set to zero without loss of generality. We will also only consider even functions $f(x)=f(-x)$. Under these conditions, any initial spin configuration that satisfies the symmetry conditions $s^z(\eps_j) = -s^z(-\eps_j)$, $s^+(\eps_j) = s^-(-\eps_j)$, as does the ground state \re{GSHF}, will do so for all time. This fact can be verified with the equations of motion \re{eomHF} by considering time derivatives of quantities such as $s^z(\eps_j)+ s^z(-\eps_j)$, which vanish under the aforementioned assumptions. We will not use particle-hole symmetry in the $d+id$ model, where $f(x) = x$ and $\eps_j$ will be distributed on a positive interval.
Further, \esref{eomHF} and \re{self13} are invariant under the time-reversal transformation
\beg
\begin{split}
s_j^z(t)\to s_j^z(-t),\quad s_j^\pm(t)\to s_j^\mp(-t),\\  
\Delta(t)\to\Delta^*(-t). \\
\end{split}
\label{treversal}
\en
Since the initial conditions \re{GSHF} at $t=0$ also have this property, it holds at all times.

The ground state of the spin-orbit model is less obvious\cite{dky}
\beg
\begin{split}
S^x_{\mathbf{k}\lambda0} &= \frac{\Delta_0\sin\phi_k}{D_k}\bigg[\Delta_0^2 + \xi^2_{k\bar{\lambda}} + E_{k+}E_{k-}\bigg],\\
S^z_{\mathbf{k}\lambda0} &= -\frac{1}{D_k}\bigg[ \xi_{k\lambda}(E_{k+}E_{k-} +\xi^2_{k\bar{\lambda}}+ \Delta_0^2\sin^2\phi_k) +\\ &+\Delta_0^2\cos^2\phi_k\xi_{k\bar{\lambda}} \bigg],\\
L^x_{\mathbf{k}0} &= \frac{\Delta_0 \cos\phi_k}{D_k}\bigg[ \Delta_0^2 + \xi_{k+}\xi_{k-} + E_{k+}E_{k-} \bigg],\\
L^z_{\mathbf{k}0} &= \frac{1}{D_k}\bigg[ 2R_k\Delta_0^2\cos\phi_k\sin\phi_k \bigg],\\
\xi_{k(\lambda)} &\equiv \eps_{k(\lambda)}  - \mu,\\
E_{k\lambda}(\Delta) &\equiv \bigg[ \xi_k^2 + \Delta^2 + R_k^2 -2R_k\lambda\sqrt{\xi_k^2 +\cos^2\phi_k\Delta^2} \bigg]^{1/2},\\
D_k &\equiv 2E_{k+}E_{k-}(E_{k+}+E_{k-}),
\end{split}
\label{GSHSO}
\en
while $S^y_{\mathbf{k}\lambda0}=L^y_{\mathbf{k}0}=T_{\mathbf{k}0} = 0$. The corresponding self-consistent equation relating $\Delta_0$ to $g$ is
\beg
\begin{split}
\frac{2}{g} = \sum_{\mathbf{k}\lambda}\frac{ E_{k+}E_{k-} + \Delta_0^2 + \sin^2\phi_k\xi^2_{\mathbf{k}\lambda} + \cos^2\phi_k\xi_{\mathbf{k}\lambda}\xi_{\mathbf{k}\bar{\lambda}} }{ 2E_{k+}E_{k-}(E_{k+}+E_{k-}) }.
\end{split}
\label{SelfConHSO}
\en
The quantities $2E_j(\Delta)$ and $2E_{\mathbf{k}\lambda}(\Delta)$ in \re{GSHF} and \re{GSHSO} are the excitation energies obtained by diagonalization of the quadratic mean-field Hamiltonians in \esref{hamMF} at a given $\Delta$.

For given values of $g$, $N_f$, $\alpha$ and $h$, one can simultaneously solve \eref{HSOzComp} and \eref{SelfConHSO} using the ground state configurations to obtain the corresponding chemical potential $\mu$ and ground state gap $\Delta_0$. As the ground state is rotationally symmetric in $\mathbf{k}$, and the equations of motion preserve this symmetry, in our numerics we always replace sums over momenta with sums over energies with a flat density of states $\sum_{\mathbf{k}}\to \sum_{\eps}$. The level spacing $\delta$ is related to the number of spins $N$ and the bandwidth $W$ through
\beg
\delta = \frac{W}{N-1}.
\en

\begin{figure}
\includegraphics[width=\linewidth]{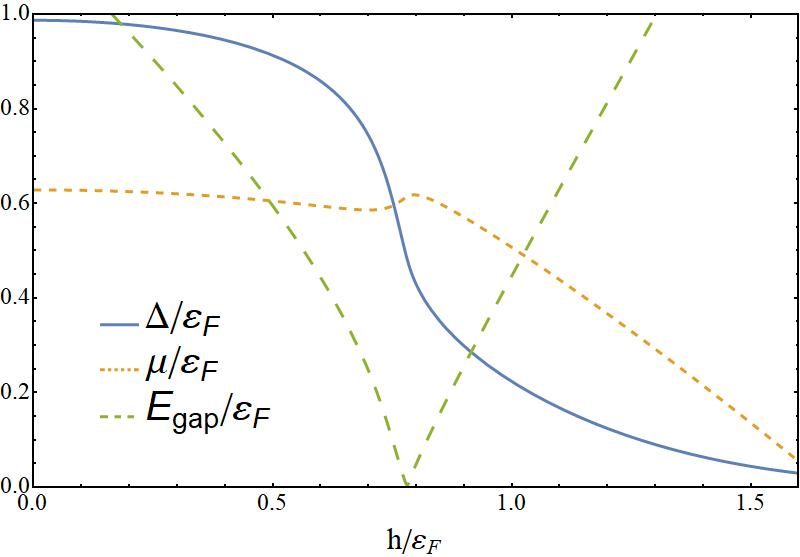}
\caption{Ground state order parameter $\Delta_0$, chemical potential $\mu$ and $E_{\textrm{gap}} = E_{k=0,+} = (\sqrt{\Delta_0^2 + \mu^2} - h)^2$ as functions of the external field $h$ in the spin-orbit model. One simultaneously solves the fermion number equation \re{HSOzComp} and the self-consistency relationship \eref{SelfConHSO} with the ground state configuration \re{GSHSO}. The vanishing of $E_{\textrm{gap}}$ corresponds to a topological quantum phase transition. The number of fermions is $N_f = 0.65N$, where $N$ is the number of spins. We express energies in units of the bandwidth $W$, including the spin-orbit coupling $\alpha^2 = 0.1W$, the level spacing $\delta = W/(N-1)$, and the BCS coupling $g=0.9 \delta$. The Fermi energy in these units is $\eps_F = \frac{W}{2N}N_f = 0.325W$. These spin-orbit model parameters remain the same for the remainder of this work, up to adjusting the value of $N$. We do not consider a similar plot for the separable BCS model because in the particle-hole symmetric case considered, the fermion number $N_f=N$  and thus $\mu = 0$.}
\label{HSOparams}
\end{figure}

Formally, in 2D this means $N-1 = \frac{W}{2\pi}A$, where $A$ is the physical area of the system. Fig.~\ref{HSOparams} shows an example of the relationship between different parameters for the spin-orbit model.

\section{Simulations of nonequilibrium phases and stability analysis}
\label{PDfromSim}
Now we numerically simulate the equations of motion \re{eomHF} and \re{eomHSO} and plot the behavior of $\Delta(t)$ for each of the three phases in Sects.~\ref{subphase1} and \ref{subphase3}. In Sect.~\ref{subphase3}, we also characterize the long time scale of nonintegrable models in Phase~III. In Sect.~\ref{subStAn}, we introduce a stability analysis for the Phases~I and II that gives the conditions under which a nonequilibrium phase transition occurs. 

We will consider several integrability-breaking functions for $f(\eps)$, which appears in the separable BCS equations of motion \eref{eomHF}. All $f(\eps)$ considered here will be even functions, and as we discuss in Sect.~\ref{subStAn}, the particular form of $f(\eps)$ affects which phases occur. With this in mind, we consider the ``Lorentzian'' coupling\cite{barlev3}
\beg
\begin{split}
f_{\textrm{lor}}(\eps,\gamma) = \frac{\gamma}{\sqrt{\gamma^2+\eps^2}},
\end{split}
\label{fLorIntro}
\en
the ``sine'' coupling,
\beg
\begin{split}
f_{\textrm{sin}}(\eps,\gamma) = 1 + \sin^2(\eps/\gamma),
\end{split}
\label{fSinIntro}
\en
and the ``cube root'' coupling,
\beg
\begin{split}
f_{\textrm{cub}}(\eps,\gamma) = \frac{(\gamma^3+|\eps|^3)^{1/3}}{\gamma}.
\end{split}
\label{fCubIntro}
\en
The parameter $\gamma$ is fixed for any particular Hamiltonian, and it characterizes how strongly integrability is broken. For $\gamma \gtrsim W$, we have $f(\eps,\gamma) \sim 1$ in all three cases, which we consider to be ``nearly integrable''. For $\gamma \ll W$, integrability is strongly broken.

We control for finite size effects in our simulations by increasing $N$ until $\Delta(t)$ in the time window of interest no longer changes when $N$ is doubled.  In practice, we find that finite size effects become significant at times $t>t_\mathrm{fs}$, where
\beg
t_\mathrm{fs}\approx \frac{1}{\delta}=\frac{N-1}{W},
\en
is the inverse single-particle level spacing, see also Ref.~\onlinecite{ydgf}. To observe the asymptotic dynamics, $N$ has to be sufficiently large,
so that the relaxation time $\tau< t_\mathrm{fs}$. 

\subsection{Phases I and II}
\label{subphase1}
\begin{figure}
\subfloat[]{
\includegraphics[width=\linewidth]{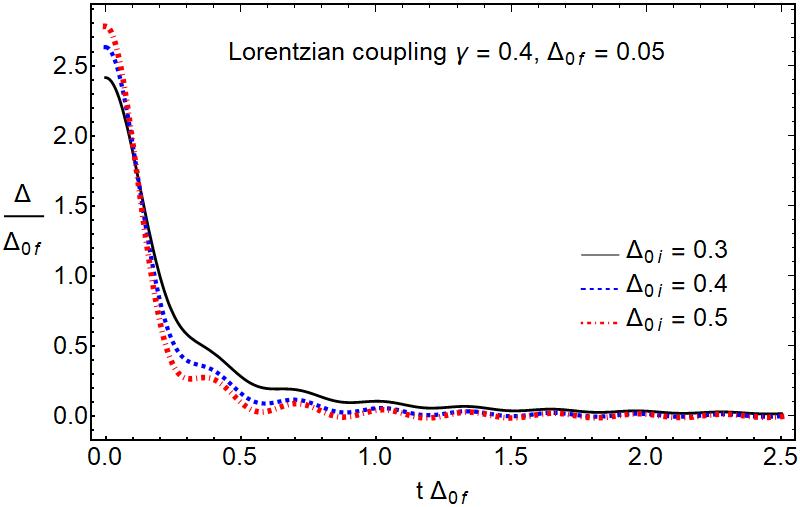}
}

\subfloat[]{
\includegraphics[width=\linewidth]{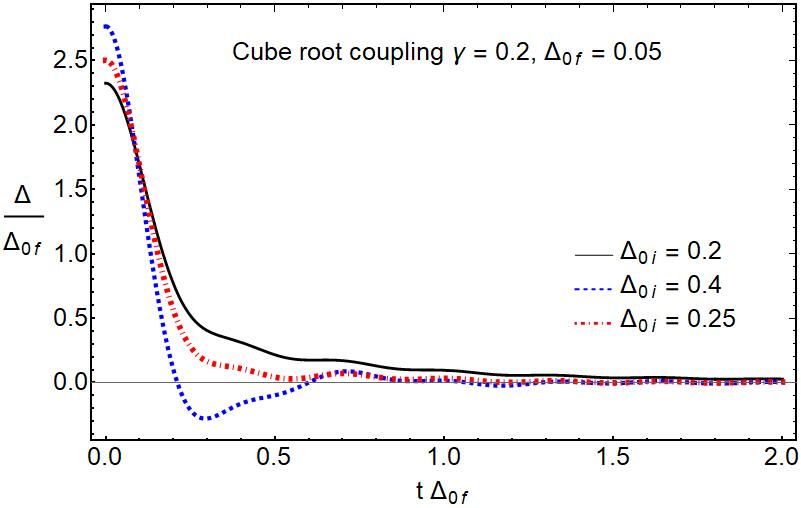}
}
\caption{Examples of Phase~I quenches for separable BCS models. The equilibrium gaps $\Delta_{0i}$, $\Delta_{0f}$ and integrability breaking parameter $\gamma$ are given in units of the bandwidth $W$, and there are $N = 5\times10^4$ (a) and $N=2\times10^5$ (b) spins. The initial rapid decay of $\Delta$ is shown, but out of caution one must simulate to longer times (still smaller than the inverse level spacing) in order to verify that the phase is indeed stable.}
\label{figHFphase1}
\end{figure}

\begin{figure}
\subfloat[]
{
\includegraphics[width=\linewidth]{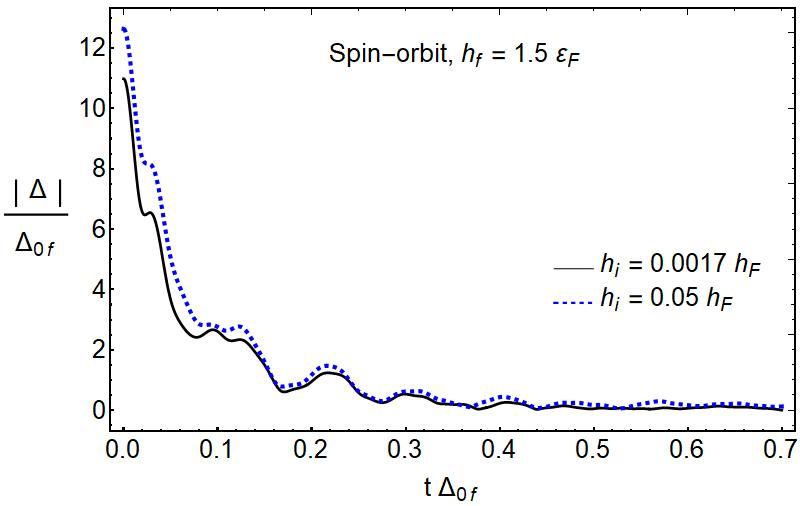}
}

\subfloat[]
{
\includegraphics[width=\linewidth]{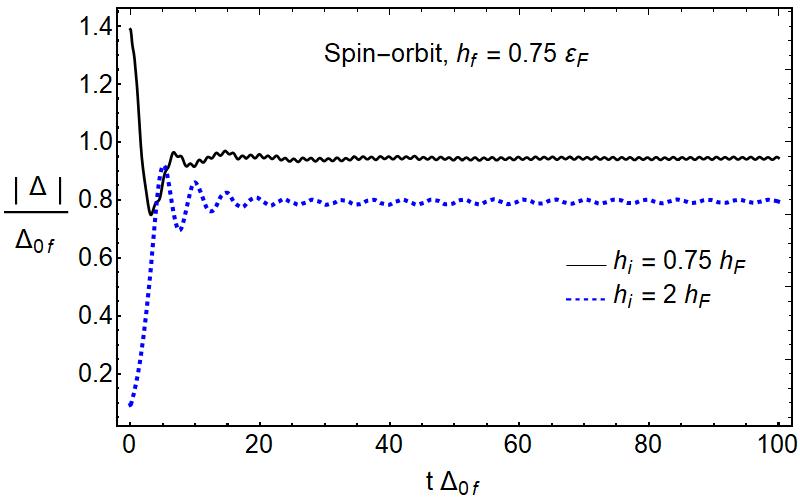}
}
\caption{Quenches in the spin-orbit model that lead to (a) Phase~I and (b) Phase~II. Here the number of single-particle energies is $N = 10^4$, and all other parameters are the same as given in the caption of Fig.~\ref{HSOparams}.}
\label{figHSOphase12}
\end{figure}

\begin{figure}
\subfloat[]
{
\includegraphics[width=\linewidth]{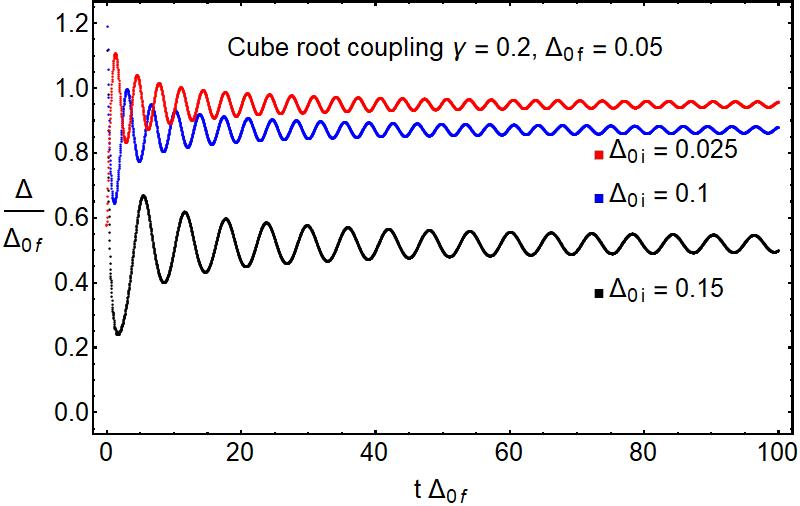}
}

\subfloat[]
{
\includegraphics[width=\linewidth]{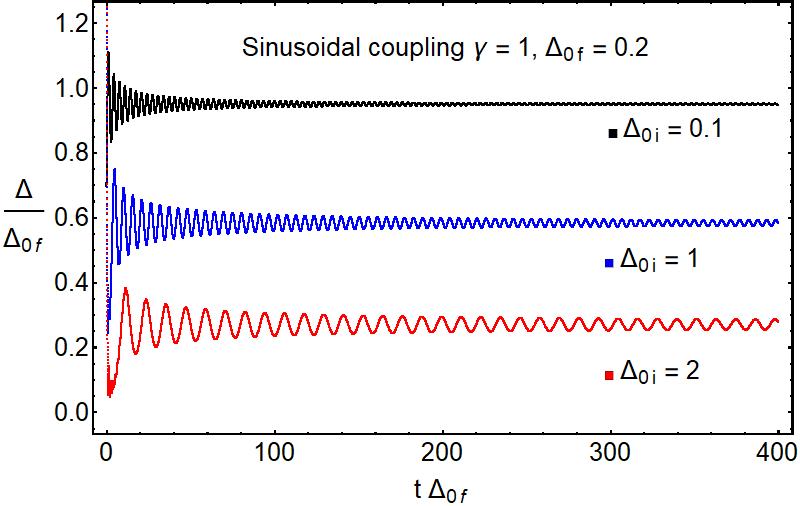}
}
\caption{Examples of Phase~II quenches for separable BCS models. In (a) $N = 2\times10^5$ spins, and the quench from $\Delta_{0i} = 0.15W$ is close to the Phase~I-II boundary. In (b), $N = 5\times10^4$. The oscillatory power law decay to a constant value takes a rather long time, and we have verified out to 
$t\delta = 2$ in (a) and 
$t\delta = 0.5$ in (b) that the amplitude of the oscillations is indeed decreasing to zero with power-law decay. In both plots, $\Delta_0$ and $\gamma$ are expressed in units of the bandwidth.}
\label{figHFphase2}
\end{figure}

Figs.~\ref{figHFphase1}-\ref{figHFphase2} contain examples of Phase~I and Phase~II quenches in both the separable BCS and spin-orbit models.
To heuristically understand the emergence of these two phases, one can insert the prescribed behavior of $\Delta$ into the equations of motion \re{eomHF} and \re{eomHSO}. This examination of the asymptotic solutions to the equations of motion in each phase will be important for the stability analyses of Sect.~\ref{subStAn}.

The following applies to the separable BCS models in the particle-hole symmetric limit, but the analysis is analogous when this symmetry is broken and in the spin-orbit case. In Phase~I, we set $\Delta$ to zero
\beg
\begin{split}
\dot{s}_j^z &= 0,\\
\dot{s}_j^x &= -2\eps_j s_j^y,\\
\dot{s}_j^y &= 2\eps_j s_j^x.
\end{split}
\label{eomHFphase1}
\en
The most general solution that conserves both $s_j^2 = 1/4$ and the time-reversal symmetry \re{treversal} is
\beg
\begin{split}
s_j^z &= z_j,\\
s_j^x &= x_j\cos(2\eps_j t),\\
s_j^y &= x_j \sin(2\eps_j t),\\
z_j^2 &= 1/4 - x_j^2.
\end{split}
\label{eomHFphase1sol}
\en
where $z_j$ is the Phase~I steady state spin-profile. In order for \re{eomHFphase1sol} to make sense as a solution to the actual equations of motion, \eref{self13} must hold, i.e., we must have that $\Delta = g_f\sum_j f_j s_j^-$ equals zero, which is called the self-consistency condition. Strictly speaking, the solution~\re{eomHFphase1sol} violates the self-consistency condition
\beg
\Delta = g_f \sum_j f_j x_j \cos(2\eps_j t) \ne 0,
\label{eomHFphase1SC}
\en
but as the number of single-particle energies $N$ goes to infinity, i.e., in the continuum limit when the sum in \eref{eomHFphase1SC} turns into an integral, $\Delta$ from \eref{eomHFphase1SC} vanishes through dephasing for $1 \ll t \ll 1/\delta = (N-1)/W$. This description is invalid for $t \sim N/W$. In this sense, we refer to the solution \re{eomHFphase1sol} as asymptotically self-consistent, which is a concept we will often use in the remainder of this paper.

Let us now replace $\Delta$ with $\dinf \ne 0$ in \eref{eomHF} to examine the asymptotic solutions corresponding to Phase~II
\beg
\begin{split}
\dot{s}_j^z &= -2f_js_j^y \dinf,\\
\dot{s}_j^x &= -2\eps_j s_j^y,\\
\dot{s}_j^y &= 2\eps_j s_j^x + 2f_j s_j^z \dinf.
\end{split}
\label{eomHFphase2}
\en
The solution which preserves spin length and the  time-reversal symmetry is then
\beg
\begin{split}
s_j^z &= Z_j + \zeta_j \cos(b_j t),\\
s_j^x &= -\frac{f_j \dinf}{\eps_j}Z_j+\frac{\eps_j}{f_j \dinf}\zeta_j \cos(b_j t),\\
s_j^y &= \frac{b_j}{2 f_j \dinf}\zeta_j \sin(b_j t),\\
\label{eomHFphase2sol}
\end{split}
\en
where $Z_j$ is the Phase~II steady state spin profile, which, along with $\dinf$, determines the other constants
\beg
\begin{split}
b_j &= 2\sqrt{\eps_j^2+f_j^2 \dinf^2},\\
\zeta_j^2 &= \frac{f_j^2 \dinf^2}{b_j^2}-\frac{f_j^2 \dinf^2}{\eps_j^2}Z_j^2.
\label{eomHFphase2solParams}
\end{split}
\en
The solution \re{eomHFphase2sol} must be asymptotically self-consistent, i.e., for $N\to\infty$, $\lim_{t\to\infty}\Delta = \dinf$, which implies
\beg
\begin{split}
1 = -g_f\sum_j\frac{f_j^2 Z_j}{\eps_j},
\label{ASCphase2}
\end{split}
\en
which is the nonequilibrium analogue of the ground state self-consistency requirement \re{SelfConHF}.

\subsection{Stability analysis}
\label{subStAn}
Now we consider the stability of Phases~I and II for the separable BCS model by linearizing the equations of motion \re{eomHFphase1} and \re{eomHFphase2} about the asymptotic states given in \re{eomHFphase1sol} and \re{eomHFphase2sol}, respectively. The main result is \eref{freqsPhase2LinNoPH}, which is the equation for frequencies of linearized perturbations to the asymptotic $\Delta(t)$ of either Phase~I or Phase~II. For Phase~I, the appearance of a complex conjugate pair of imaginary frequencies signals an exponential instability. For Phase~II, a solution $\omega_0$ to \eref{freqsPhase2LinNoPH} may enter the band gap, or a complex conjugate pair of frequencies may appear. The former case, which occurs in the integrable $s$-wave and $p+ip$ models, signifies a transition to Phase~III because the linearized gap $\delta\Delta(t)$ oscillates persistently, i.e., it does not dephase. In Appendix~\ref{linkLaxStab}, we show that the nonequilibrium phase transitions predicted by this stability analysis both match and give a physical interpretation to the results obtained in integrable models\cite{ydgf,fdgy}  using tools inextricably linked to exact solvability.

Although the final result \re{freqsPhase2LinNoPH} applies generally, we limit the discussion to the particle-hole symmetric case to simplify the presentation.  Let $\mathbf{s}_j = \mathbf{s}_{j0} + \delta \mathbf{s}_j$, where $\mathbf{s}_{j0}$ is the Phase~I asymptotic solution from \eref{eomHFphase1sol}. Neglecting second and higher order terms, the linearized equations for the spin components are
\beg
\begin{split}
\delta\dot{s}_j^z &= -2f_js_{j0}^y \delta \Delta\\
\delta\dot{s}_j^x &= -2\eps_j \delta s_j^y,\\
\delta\dot{s}_j^y &= 2\eps_j \delta s_j^x + 2f_j z_j \delta \Delta,\\
\delta\Delta  &\equiv g_f \sum_j f_j \delta s_j^x.
\end{split}
\label{eomLinHFphase1}
\en
Expanding $\mathbf{s}_j(t)$ in Fourier components
\beg
\begin{split}
\delta\mathbf{s}_j(t) &= \sum_{\omega}\delta \widetilde{\mathbf{s}}_j(\omega)e^{-i \omega t},\\
\delta\Delta &= \sum_{\omega}\delta \widetilde{\Delta}(\omega)e^{-i \omega t},
\end{split}
\en
and using the Fourier space version of the self-consistency relation  in \eref{eomLinHFphase1}, we find the following equation for the allowable frequencies $\omega$
\beg
\begin{split}
1 &= 4g_f\sum_j\frac{f_j^2 \eps_j z_j}{\omega^2 - 4\eps_j^2}.
\end{split}
\label{freqsPhase1Lin}
\en
The following discussion uses particle-hole symmetry along with the empirical fact that for quenches from the ground state, $z_j \eps_j < 0$ in Phase~I. Upon inspecting \eref{freqsPhase1Lin}, one determines that there are $N/2$ unique $\omega_j^2$, of which all but one lie between consecutive $4\eps_j^2$. The remaining $\omega_0^2$ is less than the smallest $4\eps_j^2$, and can therefore be negative. A negative $\omega_0^2$ corresponds to a pair of conjugate imaginary frequencies, and therefore an exponential instability in $\delta \mathbf{s}_j$. We thus determine the Phase~I boundary in $(\Delta_{0i},\Delta_{0f})$ space to be those values for which $\omega_0^2$ passes through zero.


The stability analysis for Phase~II follows a similar logic. Consider the linearized equations of motion
\beg
\begin{split}
\delta\dot{s}_j^z &= -2f_j s_{j0}^y \delta \Delta -2f_j \dinf\delta s_j^y,\\
\delta\dot{s}_j^x &= -2\eps_j \delta s_j^y,\\
\delta\dot{s}_j^y &= 2\eps_j \delta s_j^x + 2 f_j s_{j0}^z \delta \Delta + 2 f_j \dinf\delta s_j^z,
\label{eomLinHFphase2}
\end{split}
\en
where now $\mathbf{s}_{j0}$ is the Phase~II asymptotic solution from \eref{eomHFphase2sol}. Again changing to the Fourier basis, we solve for $\delta \widetilde{s}_j^x(\omega)$ and apply the self-consistency condition for $\delta \widetilde{\Delta}(\omega)$, which reads
\beg
\begin{split}
&\delta \widetilde{\Delta}(\omega)\bigg(1-4g_f\sum_j\frac{\eps_jf_j^2Z_j}{\omega^2-b_j^2}\bigg) =\\&= \frac{2g_f}{\omega}\sum_j\eps_jf_j^2\zeta_j\bigg(\frac{\delta\widetilde{\Delta}(\omega+b_j)}{\omega+b_j} + \frac{\delta\widetilde{\Delta}(\omega-b_j)}{\omega-b_j}\bigg).
\label{integral_equation}
\end{split}
\en
Although in principle  \eref{integral_equation} can be solved numerically with $Z_j$ and $\dinf$ as input, such an approach is needlessly complex and obscures the mechanism by which Phase~II gives way to Phase~III. The difficulty presented by \eref{integral_equation} stems from the fact that we required exact self-consistency. It turns out that relaxing this requirement to asymptotic self-consistency, defined in Sect.~\ref{subphase1}, suffices to understand the Phase~II-III transition.

We return to \eref{eomLinHFphase2} and solve it in the time domain under the assumption $\delta  {\Delta}(t) = \delta_+e^{-i\omega_0 t}+\delta_-e^{i\omega_0 t}$. We neglect higher order harmonics because the Phase~III oscillations near the II-III boundary are small. Under this ansatz, $\delta s_j^x(t)$ has six frequencies: $\pm\omega_0$ and $\pm\omega_0\pm b_j$. If $\omega_0$ is a real frequency isolated from the continuum of $b_j$ defined in \eref{eomHFphase2solParams}, then the constant $\dinf$ of Phase~II is ``unstable'' in the sense that oscillatory perturbations do not dephase. The self-consistent equation for this harmonic $\delta\Delta(t)$ is
\beg
\begin{split}
&1 = 4g_f\sum_j\frac{f_j^2 \eps_j Z_j}{\omega_0^2 - b_j^2}+\\&+ \frac{2g_f}{\omega_0}\sum_j\bigg(\frac{e^{ib_jt}f_j^2 \eps_j \zeta_j}{\omega_0-b_j}+[b_j\to -b_j]\bigg).
\label{exactSCphaseIIansatz}
\end{split}
\en
This relation cannot hold for arbitrary $t$, but it will in the continuum limit if we require $\omega_0^2<b^2_{\textrm{min}}$ and $t\to\infty$, which allows the harmonic ansatz to be asymptotically self-consistent due to dephasing. Thus the equation for $\omega_0$, the frequency of a harmonic perturbation to $\dinf$ in Phase~II, is
\beg
\begin{split}
1 = 4g_f\sum_j\frac{f_j^2 \eps_j Z_j}{\omega_0^2 - b_j^2}.
\label{freqsPhase2Lin}
\end{split}
\en
\eref{freqsPhase2Lin} generalizes the small  quench linearization method developed in Ref.~\onlinecite{barlev3}, which we recover by replacing $Z_j$ of \eref{freqsPhase2Lin} with the z-component spin profile of the $g_i$ ground state. For the Lorentzian coupling, $\omega_0$
is in the band gap for infinitesimal quenches, so that linearized Phase~III oscillations do not decay\cite{barlev3}.

In order to understand whether the finite quench dynamics admit such an isolated $\omega_0$, consider the implications of \re{freqsPhase2Lin} combined with \re{ASCphase2} for the $\dinf$ of Phase~II. We find
\beg
\begin{split}
\frac{\omega_0^2}{4\dinf^2} &= \frac{I_1(\omega_0^2)}{I_2(\omega_0^2)},\\
I_1(\omega_0^2) &\equiv g_f\sum_j\frac{f_j^4 Z_j}{\eps_j(\omega_0^2 - b_j^2)},\\
I_2(\omega_0^2) &\equiv g_f\sum_j\frac{f_j^2 Z_j}{\eps_j(\omega_0^2 - b_j^2)}.
\label{PhaseIIASC_combined}
\end{split}
\en
It helps to analyze \re{PhaseIIASC_combined} under the simplifying assumption that $Z_j/\eps_j < 0$, which holds exactly for the integrable $s$-wave model, and is therefore applicable in the weak-coupling regime  ($\Delta_{0i},\Delta_{0f}\ll W$) of the general separable case\cite{ydgf}. With this restriction, \eref{freqsPhase2Lin} implies $\omega_0^2$ is real, while \eref{PhaseIIASC_combined} requires $\omega_0^2>0$, i.e., the allowed frequencies $\omega_0$ are purely real. We now examine the effect of the function $f_j$ in determining whether solutions $\omega_0^2$ to \eref{PhaseIIASC_combined} are isolated from the $b_j^2$ continuum.

If $f_j < f(0)$ for all $j$ and $b^2_{\textrm{min}} = 4\dinf^2$, then \eref{PhaseIIASC_combined} has a solution $0 < \omega_0^2 < b^2_{\textrm{min}}$, and oscillations of $\delta\Delta(t)$ do not dephase. In this scenario, Phase~III is the asymptotic state due the presence of persistent periodic oscillations about the Phase~II solution. If $f_j < f(0)$ for all $j$ and $b^2_{\textrm{min}} < 4\dinf^2$, then the relationship between $\omega_0^2$ and $b^2_{\textrm{min}}$ is not immediately obvious from \eref{PhaseIIASC_combined}. The Lorentzian coupling, where $f_j = \gamma(\gamma^2+\eps_j^2)^{-1/2}$, allows for both possibilities: If $\dinf \le \gamma$, then $b^2_{\textrm{min}} = 4\dinf^2$ and Phase~II is not the asymptotic state. If $\dinf > \gamma$, then $b_{\textrm{min}}^2 = 4\gamma(2\dinf-\gamma)$, and we cannot characterize solutions to \eref{PhaseIIASC_combined} without detailed knowledge of $Z_j$ and $\dinf$.

If $f_j\ge f(0)$ for all $j$, then $b^2_{\textrm{min}} = 4\dinf^2$ and we find that solutions $\omega_0^2$ to \eref{PhaseIIASC_combined} are not isolated from the $b_j^2$ continuum. In this case, the harmonic ansatz for $\delta\Delta(t)$ is not asymptotically self-consistent, and there are no persistent small oscillations about Phase~II. The integrable $s$-wave model is defined by $f_j= f(0) = 1$, in which case $\omega_0^2 = 4\dinf^2$ is the only solution to \eref{PhaseIIASC_combined}, which is not isolated. On the other hand, Phase~III exists in the $s$-wave case\cite{ydgf}. Therefore, $f_j\ge f(0)$ does not imply that such models will always reach Phase~II. Indeed, the relaxation to Phase~II is always accompanied by nonperturbative oscillations which persist in the case of Phase~III.

Thus, even under the simplifying assumptions of particle-hole symmetry and $Z_j/\eps_j < 0$, the stability analysis of Phase~II reveals a variety of possible behaviors in the separable BCS models. The nature of $f(\eps)$ near $\eps=0$ (the Fermi surface) is especially crucial to determining whether oscillations fully dephase to Phase~II -- a statement which extends to the non-particle-hole symmetric case in the weak coupling regime.

Upon relaxing the restriction $Z_j/\eps_j < 0$, isolated solutions to \eref{PhaseIIASC_combined} can have nonzero imaginary part, thereby allowing for the possibility of exponential instabilities to Phase~II solutions (see Fig.~\ref{didStudy}). In the non-particle-hole symmetric case, $\Delta(t) = \dinf e^{-2i\mu_{\infty}t}$ in Phase~II, and the equation for the frequencies of harmonic $\delta\Delta(t)$ can be expressed in the form
\beg
\begin{split}
S_2^2(\omega_0) &= \big(S_1(\omega_0)-1\big)^2 + \big(S_1(\omega_0)-1\big)S_3(\omega_0),\\
S_1(\omega_0) &\equiv 4g_f\sum_j\frac{\widetilde{\eps}_jf_j^2Z_j}{\omega_0^2-\widetilde{b}_j^2},\\
S_2(\omega_0) &\equiv 2g_f\omega_0\sum_j\frac{f_j^2Z_j}{\omega_0^2-\widetilde{b}_j^2},\\
S_3(\omega_0) &\equiv 4g_f\Delta_{\infty}^2\sum_j\frac{f_j^4Z_j}{\widetilde{\eps}_j(\omega_0^2-\widetilde{b}_j^2)},
\end{split}
\label{freqsPhase2LinNoPH}
\en
where
\beg
\widetilde{\eps}_j \equiv \eps_j-\mu_{\infty},\quad \widetilde{b}_j \equiv 2\sqrt{\widetilde{\eps}_j^2+f_j^2\dinf^2}.
\en
The self-consistency equation for $\Delta(t)$ in Phase~II has the same form as \eref{ASCphase2}, with the substitution $\eps_j \to \widetilde{\eps}_j$. In the particle-hole symmetric limit, $S_2(\omega_0)=0$ and the correct solution to \eref{freqsPhase2LinNoPH} solves \eref{freqsPhase2Lin}. In the limit $\dinf \to 0$, \re{freqsPhase2LinNoPH} is also the stability equation for Phase~I. In Appendix~\ref{linkLaxStab}, we show that the Phase~I-II and Phase~II-III transitions given by \re{freqsPhase2LinNoPH} are identical to those obtained using exact solvability in the integrable $s$-wave and $p+ip$ models.

\begin{figure}
\includegraphics[width=\linewidth]{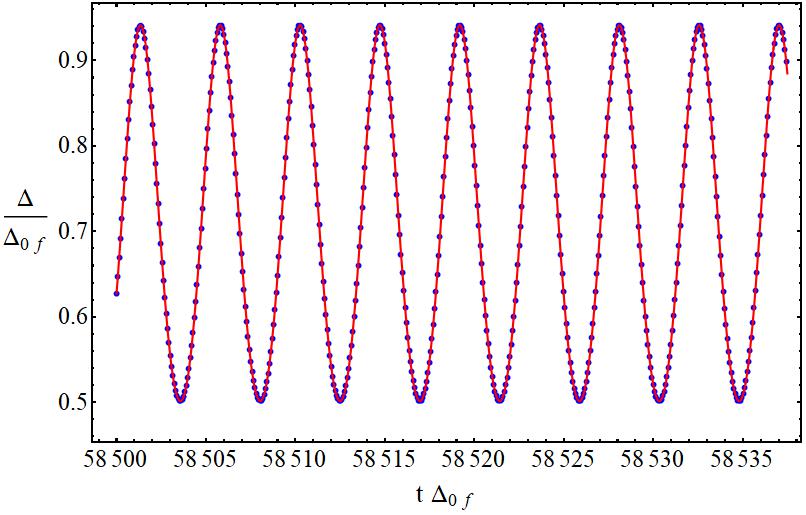}
\caption{The quench in the Lorentzian separable BCS model (blue dots) from Fig.~\ref{time_scales}~(c) and (d) [$\gamma=W$] and the corresponding elliptic function fit (solid red) from \eref{deltaParamPhase3} with $a\approx0.868205$, $\Delta_+\approx0.941415$, $\Delta_- \approx 0.501511$, $\widetilde{\Delta}_+= \widetilde{\Delta}_-^* \approx 0.915740+0.002407i$ and $t_0 = 2.801929$. To obtain these parameters, we fit $\dot{\Delta}$ to \eref{ellipticDefinition} and then shift by the appropriate $t_0$. If a fifth order polynomial is used instead of $P_4[\Delta(t)]$, the coefficient of the $\Delta^5$ term is $-6.08\times 10^{-5}$, providing further evidence that this asymptotic $\Delta(t)$ is indeed an elliptic function. Although only a short time frame is shown, this fit works well for the entire time interval from $t \Delta_{0f}=10^4$, which is the time scale after which the oscillation amplitude stabilizes, to the times shown. In this fitting procedure, $\Delta$ is given in units of $\Delta_{0f}=0.4W$ and time is measured in units of $\Delta_{0f}^{-1}$ as pictured. In terms of the level spacing 
$\delta=5\times 10^{-6}W$, the time domain pictured is $0.73125<t\delta<0.731688$.
}
\label{LorPhaseIII}
\end{figure}

\subsection{Phase III}
\label{subphase3}
\subsubsection{Universality of elliptic oscillations}

\begin{figure}
\includegraphics[width=\linewidth]{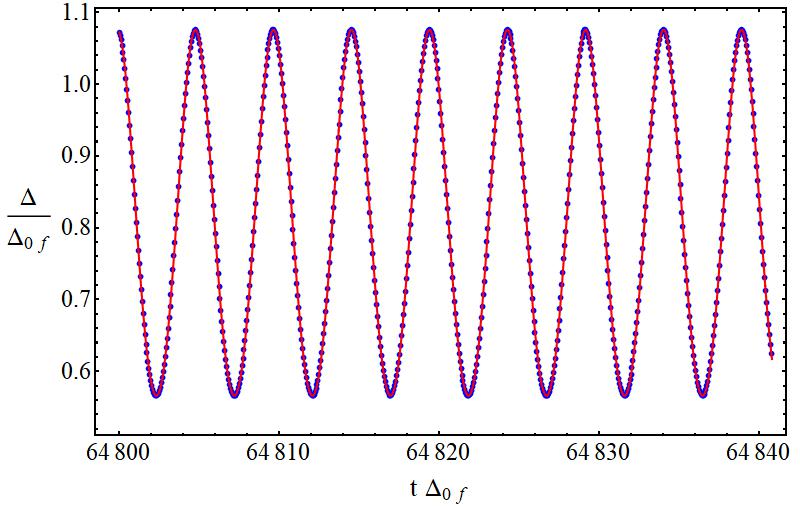}
\caption{A Phase~III quench in a $f(\eps) = \exp[-|\eps|/\gamma]$ separable BCS model (blue dots), where $\gamma = 0.5W$, $N=2\times10^5$, $\Delta_{0i} = 0.04W$, $\Delta_{0f} = 0.8W$. The corresponding elliptic function fit (solid red) from \eref{deltaParamPhase3} has $a\approx0.821896$, $\Delta_+\approx1.075648$, $\Delta_- \approx 0.566069$, $\widetilde{\Delta}_+= \widetilde{\Delta}_-^* \approx 0.010686+1.327633i$ and $t_0 = 2.131916$. To obtain these parameters, we fit $\dot{\Delta}$ to \eref{ellipticDefinition} and then shift by the appropriate $t_0$. If a fifth order polynomial is used instead of $P_4[\Delta]$, the coefficient of the $\Delta^5$ term is $4.22\times 10^{-9}$. In this fitting procedure, $\Delta$ is given in units of $\Delta_{0f}$ and time is measured in units of $\Delta_{0f}^{-1}$ as pictured. In terms of the level spacing $\delta$, the time domain pictured is $0.405<t\delta<0.40525$.
}
\label{ExpPhaseIII}
\end{figure}

\begin{figure}
\includegraphics[width=\linewidth]{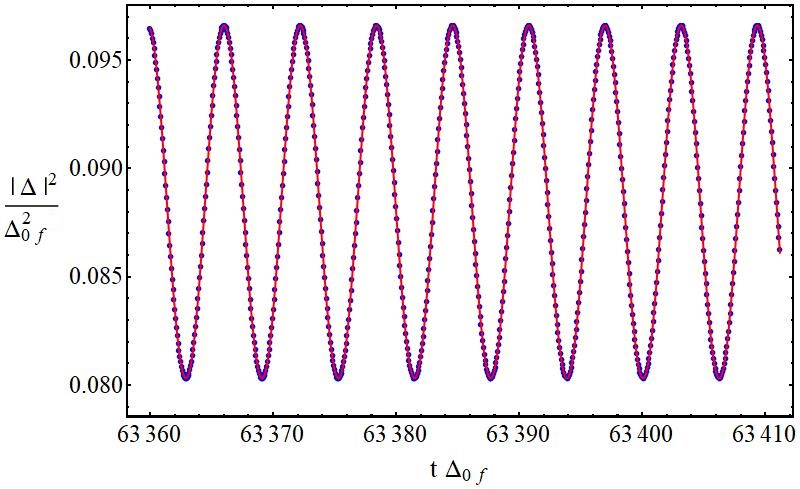}
\caption{A Phase~III quench in the spin-orbit model (blue dots), where in units of the bandwidth $W$: $\eps_F = 0.1$, $\alpha^2 = 0.9$, $g N = 2.315999$, $h_i = 1.998980$, $h_f = 0.801020$. These parameters uniquely determine the initial and final equilibrium gaps and chemical potentials through the use of \eref{HSOzComp} and \eref{SelfConHSO}. The energies $\eps_j$ are uniformly distributed in the interval $[0,W]$, and the number of pseudospins is $N=8\times10^4$. As particle-hole symmetry does not hold, we fit $\Omega\equiv|\Delta|^2$ to the elliptic function definition in \eref{deltaParamPhase3}. The fit is $a\approx0.776633$, $\Delta_+\approx0.096608$, $\Delta_- \approx 0.080316$, $\widetilde{\Delta}_+=\widetilde{\Delta}_-^* \approx 0.873604+0.883872i$ and $t_0 = 3.033272$. The fit (solid red) is good for all $t > \tau$, where $\tau$ is the relaxation time defined in Sect.~\ref{subsubRel}. Here $\tau \Delta_{0f} \approx 3050$. In the fitting procedure, $\Delta$ is given in units of $\Delta_{0f}$ and time is measured in units of $\Delta_{0f}^{-1}$ as pictured. In terms of the level spacing $\delta$, the time domain pictured is $1.472<t\delta<1.473$, shortly after which finite size effects take over.
}
\label{SOPhaseIII}
\end{figure}

The asymptotic Phase~III solution is significantly more complicated than its Phase~I and Phase~II counterparts \re{eomHFphase1sol} and \re{eomHFphase2sol}. We derive this solution in Sect.~\ref{asympPerSol}. Presently we provide evidence that the asymptotic behavior of $\Delta(t)$ can always be described by Jacobi elliptic functions.
Consider first the particle-hole symmetric limit, for which we find
\beg
\dot{\Delta}^2(t)=P_4[\Delta(t)],\quad\textrm{as }t\to\infty,
\label{ellipticDefinition}
\en
where $P_4[\Delta(t)]$ is a generic fourth-order polynomial in $\Delta(t)$. Now parametrize $P_4[\Delta(t)]$ as
\beg
\begin{split}
P_4[\Delta(t)] = -a^2(\Delta(t)-\Delta_+)(\Delta(t)-\Delta_-)\times\\
\times(\Delta(t)+\widetilde{\Delta}_+)(\Delta(t)+\widetilde{\Delta}_-),
\label{p4Param}
\end{split}
\en
where the real coefficients $\Delta_{\pm}$ are the maximum and minimum values of $\Delta(t)$, while $\widetilde{\Delta}_{\pm}$ are either complex conjugate or independent real numbers. This parametrization leads to the following solution for $\Delta(t)$
\beg
\begin{split}
\Delta(t) &= \frac{\widetilde{\Delta}_+(\Delta_++\widetilde{\Delta}_-)\textrm{dn}^2[ab(t-t_0),m]-\widetilde{\Delta}_-(\Delta_++\widetilde{\Delta}_+)}{\Delta_++\widetilde{\Delta}_+-(\Delta_++\widetilde{\Delta}_-)\textrm{dn}^2[ab(t-t_0),m]},\\
m &\equiv \frac{(\Delta_+-\Delta_-)(\widetilde{\Delta}_+-\widetilde{\Delta}_-)}{(\Delta_++\widetilde{\Delta}_-)(\Delta_-+\widetilde{\Delta}_+)},\\
b &\equiv \frac{1}{2}\sqrt{(\Delta_++\widetilde{\Delta}_-)(\Delta_-+\widetilde{\Delta}_+)},
\label{deltaParamPhase3}
\end{split}
\en
where $\textrm{dn}[t,m]$ is the Jacobi-dn function. When particle-hole symmetry does not hold, then one replaces $\Delta(t)$ with $|\Delta(t)|^2$ in \esref{ellipticDefinition}-(\ref{deltaParamPhase3}). In Figs.~\ref{LorPhaseIII} and \ref{ExpPhaseIII} we show that Phase~III oscillations in separable BCS models satisfy \eref{ellipticDefinition} and \eref{deltaParamPhase3}, while Fig.~\ref{SOPhaseIII} shows the same for the spin-orbit model.

As a general rule of thumb, most spin-orbit quenches that superficially appear to relax to Phase~III really have not. Fig.~\ref{SOPhaseIII} is the result of a thorough search of the parameter space in order to find a true Phase~III quench within a computationally achievable time. On the one hand, the final field $h_f$ has to be large enough so as to nonperturbatively break integrability, for small perturbations lead to long relaxation times.  On the other hand, the fields cannot be so large as to suppress the equilibrium gap $\Delta_{0}$ scale, which is the scale of the oscillation frequency. The value of $\alpha$ must also break integrability nonperturbatively, but a larger $\alpha$ also requires a larger number of spins to reach the thermodynamic limit. Finally, it turns out that a smaller Fermi energy relative to the bandwidth promotes a faster relaxation time. We discuss this Phase~III relaxation time further in Sect.~\ref{subsubRel} in the context of the separable BCS models.

For the integrable $s$-wave case it can be shown analytically\cite{ydgf} that $\widetilde{\Delta}_{\pm} = \Delta_{\pm}$ and $a=1$, which greatly simplifies $P_4[\Delta(t)]$ and $\Delta(t)\to\Delta_+\textrm{dn}[\Delta_+(t-t_0),1-\frac{\Delta^2_-}{\Delta^2_+}]$. The mechanism behind the emergence of the three phases in the $s$-wave Hamiltonian is a dynamical reduction in the number of degrees of freedom. The Phase~III asymptotic solution for $\Delta(t)$ is identical with that of a 2-spin $s$-wave Hamiltonian, while Phases~II and I correspond to 1-spin and 0-spin solutions, respectively. In Phase~III, this technique does not work for the separable BCS models. In Appendix~\ref{AppReductionMech}, we show that the 2-spin solution for these nonintegrable models is identical to that of the integrable case, up to a rescaling of time, while the general asymptotic solution that we observe is \eref{deltaParamPhase3}. Thus, if a reduction mechanism exists in the nonintegrable cases, the form of the $m$-spin Hamiltonian must also change.
\label{subsubUniv}

\subsubsection{Relaxation time}
\label{subsubRel}
In Sect.~\ref{subStAn} we saw that there are examples of nonintegrable separable BCS models where the constant $\Delta_{\infty}$ of Phase~II is unstable to harmonic perturbations, and in Sect.~\ref{subsubUniv} we gave evidence that the Phase~III oscillations of these models are elliptic functions. This behavior is typical of integrable models as well, although the form of the elliptic functions changes once integrability is broken. A more important difference, however, is that a long relaxation time scale $\tau$ emerges before the system truly reaches Phase~III.

\begin{figure*}
\subfloat[]
{
\includegraphics[width=.48\linewidth]{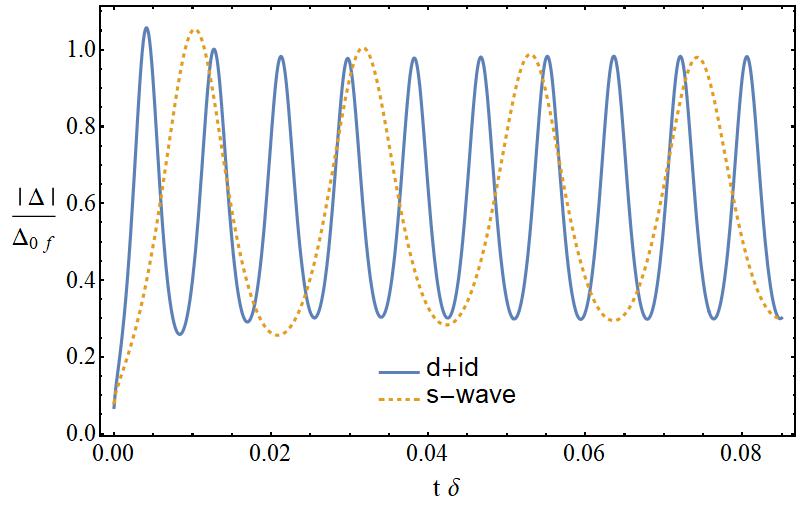}
}
\hfill
\subfloat[]
{
\includegraphics[width=.48\linewidth]{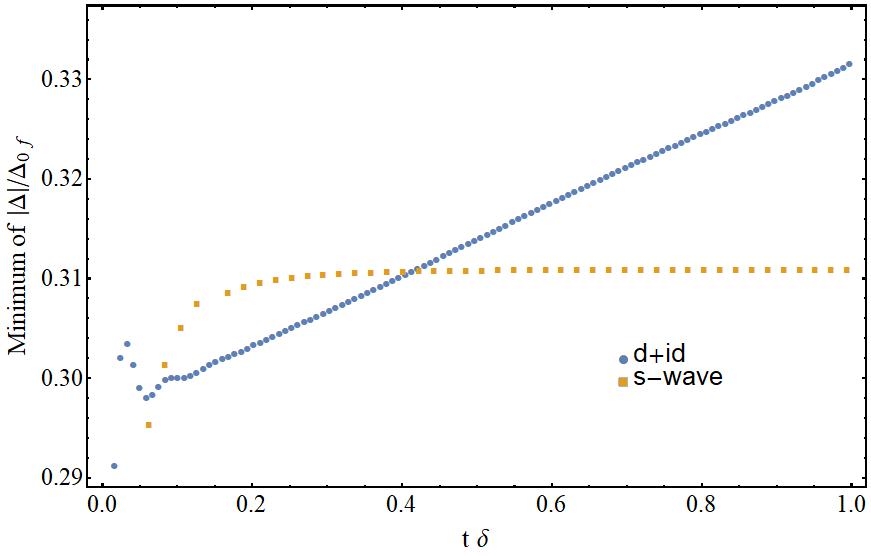}
}
\caption{Example of a deceitful quench in the $f(\eps) = \eps$ ($d+id$) separable BCS model, which at short times seems to enter Phase~III on a similar time scale as the corresponding integrable $s$-wave quench with the same parameters. Part (b) shows that minimum of the $d+id$ $|\Delta(t)|$ is actually evolving over the entire time scale considered, and it is not clear what the asymptotic phase is. For both models, we used $4\times10^4$ single-particle energies $\eps_j$ uniformly distributed on the interval $[0,W]$, $\Delta_{0f} = 0.00625W$, $\Delta_{0i}=0.05\Delta_{0f}$, $\eps_F = 0.25W$ \cite{note11}. In Fig.~\ref{didStudy}, we explore similar quenches in the $d+id$ model at larger energy scales, where the dynamics are faster.}
\label{did_short}
\end{figure*}

\begin{figure*}
\subfloat[]{
\includegraphics[width=.48\linewidth]{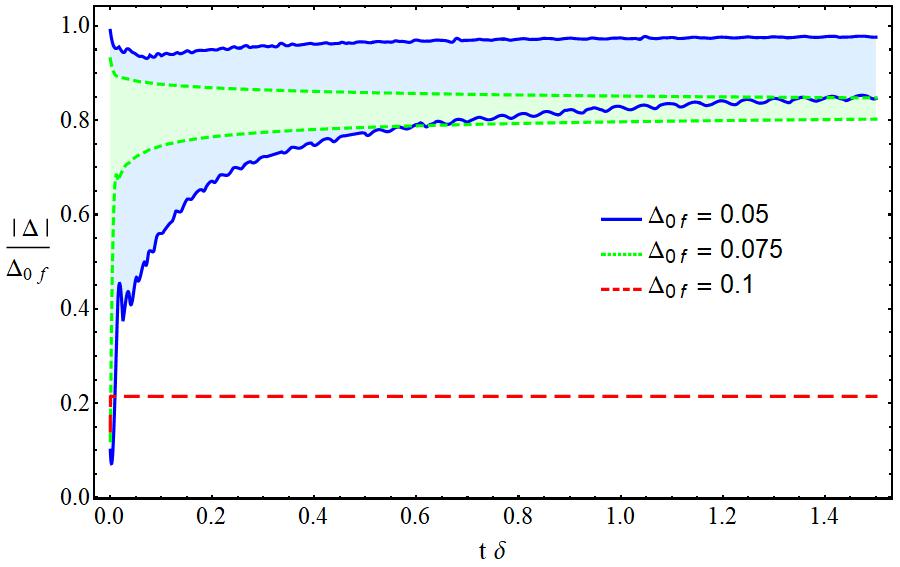}
}
\hfill
\subfloat[]{
\includegraphics[width=.48\linewidth]{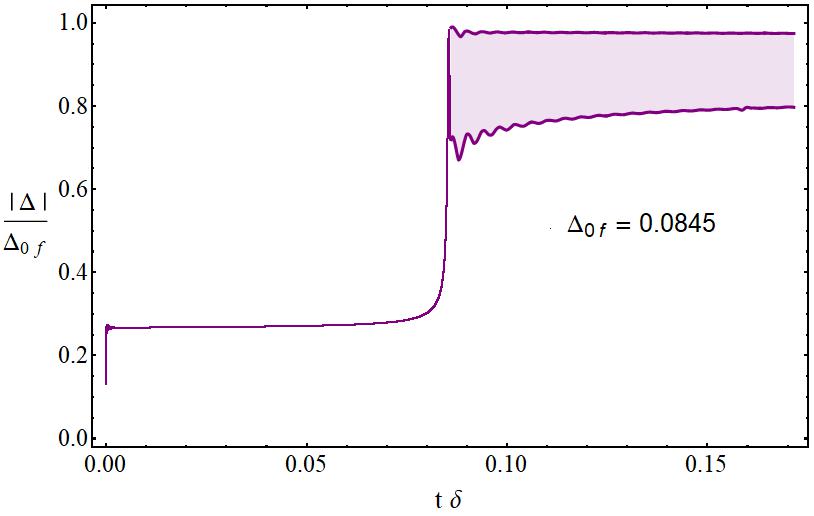}
}
\caption{Study of the long time dynamics of $d+id$ model quenches, continued from Fig.~\ref{did_short}. We keep the same parameters and the same ratio $\Delta_{0i}/\Delta_{0f} = 0.05$ while varying $\Delta_{0f}$. Pictured are the maxima and minima of oscillations of $|\Delta|$. Part (a) shows that below a certain critical $\Delta_{0f} \sim 0.0845W$, the amplitude of $|\Delta|$ oscillations evolves over an extremely long time scale. When $\Delta_{0f} = 0.05W$, there are also multiple incommensurate frequencies, and it is unclear whether the asymptotic state is Phase~II, III, or something else entirely. When $\Delta_{0f} = 0.075W$, the decay in amplitude of $|\Delta|$ resembles typical decays to Phase~II seen in other models (see Fig.~\ref{figHFphase2}). At $\Delta_{0f} = 0.1W$, the system rapidly enters Phase~II at a smaller $\Delta_{\infty}$ than would be inferred from the other two cases, indicating that we have crossed a transition point. Part (b) shows a quench at this transition point, where the Phase~II state seen for $\Delta_{0f} = 0.1W$ exhibits an exponential instability and moves to an oscillatory state with unknown asymptotic behavior.  The integrable $s$-wave BCS model, $f(\eps)=1$, is deep
in Phase~III for all these values of $\Delta_{0f}$ and $\Delta_{0i}$.}
\label{didStudy}
\end{figure*}

Fig.~\ref{did_short} gives an example of the long relaxation time in the $d+id$ model, which is the separable BCS model with $f(\eps) = \eps$. The initial dynamics at weak coupling seem to indicate\cite{didShortTime} that $|\Delta(t)|$ oscillates with a single frequency reminiscent of Phase~III. Upon closer inspection, however, the amplitude of the oscillations slowly changes with no indication of stabilizing. In Fig.~\ref{didStudy}, quenches at higher energies provide further evidence that the long-time asymptotic state is difficult to determine based on the short-time dynamics.

\begin{figure*}
\subfloat[]
{
\includegraphics[width=.48\linewidth]{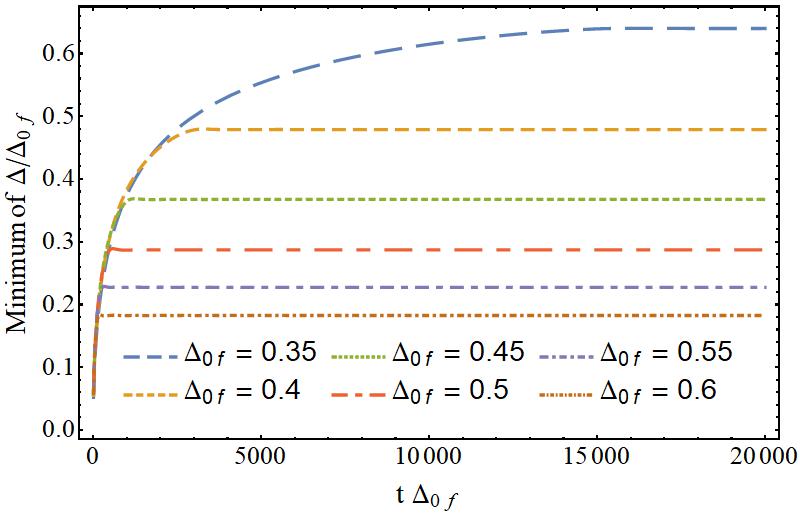}
}
\hfill
\subfloat[]
{
\includegraphics[width=.48\linewidth]{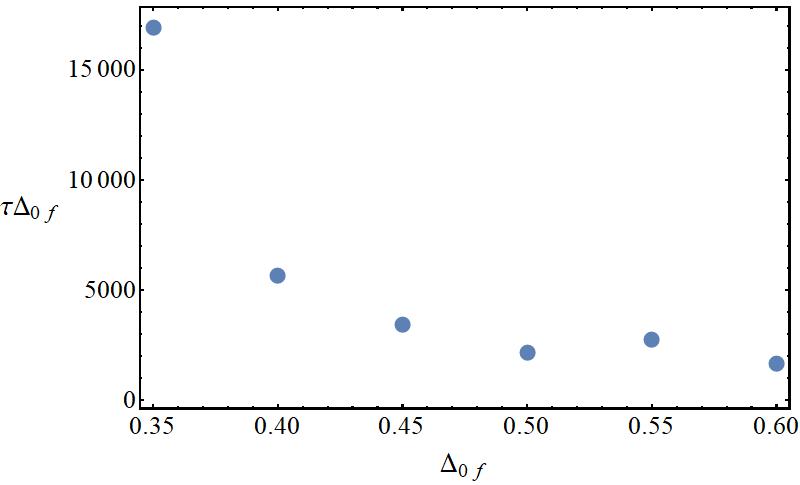}
}

\subfloat[]
{
\includegraphics[width=.48\linewidth]{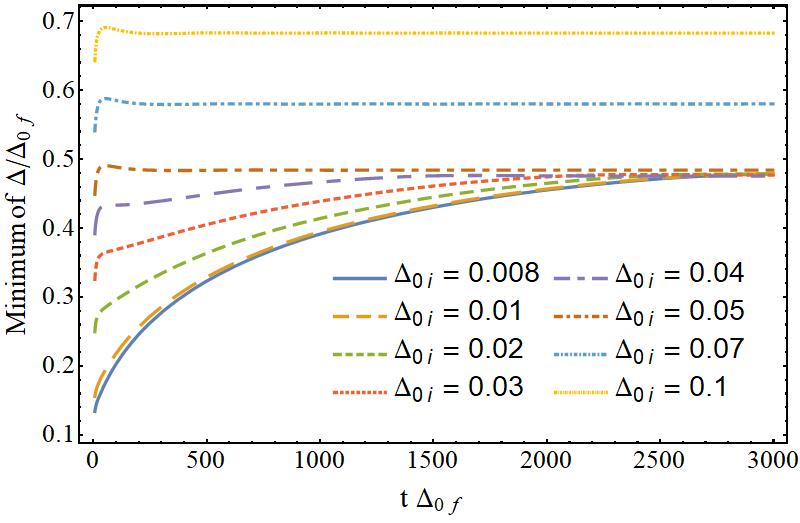}
}
\hfill
\subfloat[]
{
\includegraphics[width=.48\linewidth]{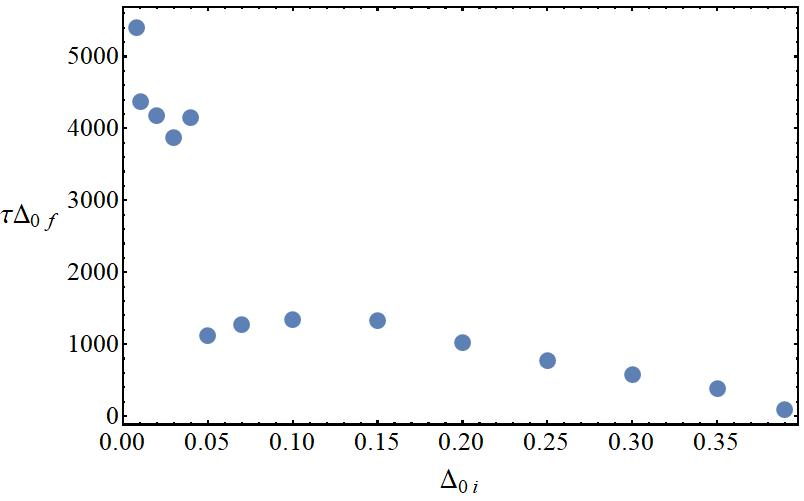}
}
\caption{Nonintegrable pairing models exhibit an extremely long relaxation time $\tau$ when the asymptotic state is Phase~III, which is most prominent in the evolution of the minima of the oscillations of $\Delta(t)$. Pictured is a study of $\tau$ as a function of $\Delta_{0f}$, at fixed $\Delta_{0i}=10^{-3}W$ (a,b), and $\tau$ as a function of $\Delta_{0i}$ at fixed $\Delta_{0f}=0.4W$ (c,d) in the Lorentzian model at $\gamma=0.8W$ in the particle-hole symmetric case. The time $\tau$ is not monotonic in either case, but it is generally a decreasing function of the initial and final coupling strengths $g_i$ and $g_f$. In all plots, $\Delta_{0i}$ and $\Delta_{0f}$ are given in units of the bandwidth $W.$ In (a,b) $2.4\times10^4 > N > 1.2\times10^4$ and in (c,d) $N = 8400$.}
\label{lor_tau_df}
\end{figure*}

Let us now explore the dependence of the relaxation time $\tau$ on $\Delta_{0i}$, $\Delta_{0f}$ and $\gamma$ in the Lorentzian separable BCS model defined in \eref{fLorIntro}. We define $\tau$ as the minimum time after which the minimum of $|\Delta(t)|$ oscillations stays within $\eta=10^{-4}$ of its asymptotic value. This definition of $\tau$ and the precise value of $\eta$ are somewhat arbitrary, but empirically we find that the minima of $|\Delta(t)|$ take longer to relax to the stationary value than the maxima. Typically, the minimum will increase for a time until it begins to oscillate with decreasing amplitude about a final value. Most importantly, this definition of $\tau$ delineates clearly the difference between integrable and nonintegrable behavior. Fig.~\ref{lor_tau_df} shows the dependence of $\tau$ on the values of $\Delta_{0i}$ and $\Delta_{0f}$, with one or the other fixed. Generally speaking, we find that quenches at lower energy scales increase $\tau$.

\begin{figure*}
\subfloat[]
{
\includegraphics[width=.48\linewidth]{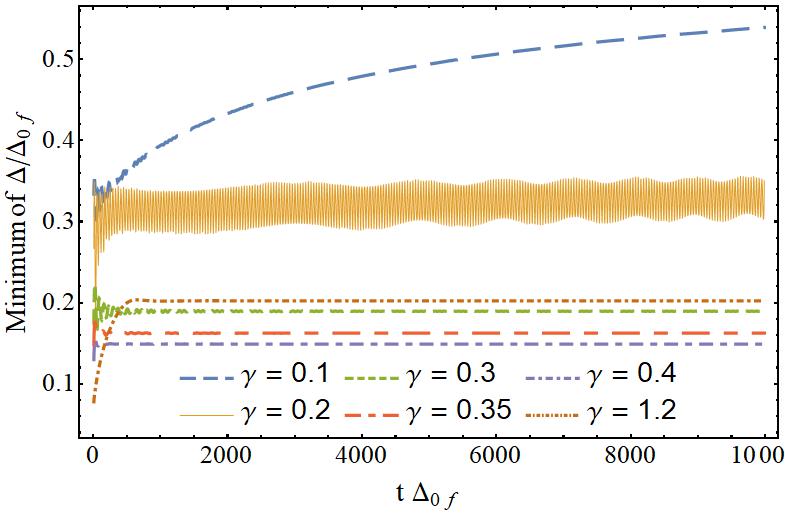}
}
\hfill
\subfloat[]
{
\includegraphics[width=.48\linewidth]{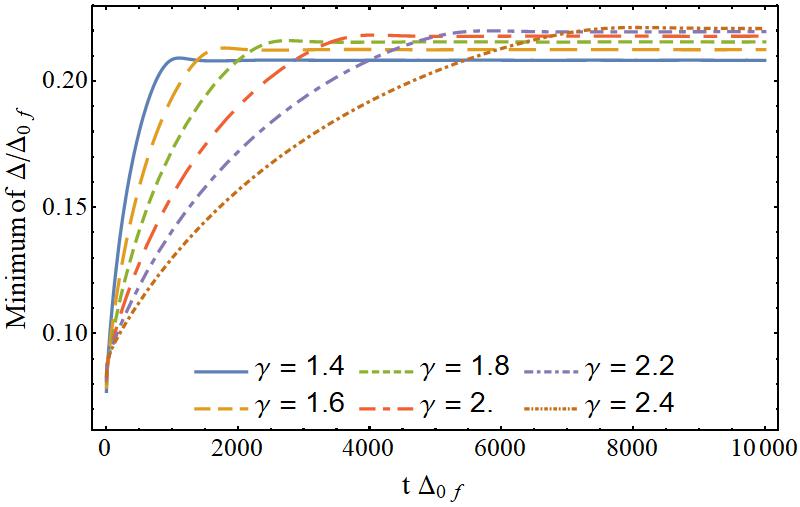}
}

\subfloat[]
{
\includegraphics[width=.48\linewidth]{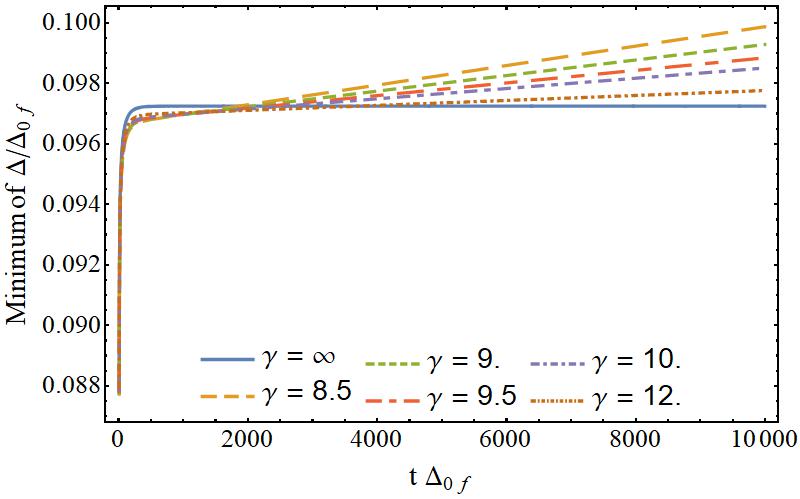}
}
\hfill
\subfloat[]
{
\includegraphics[width=.48\linewidth]{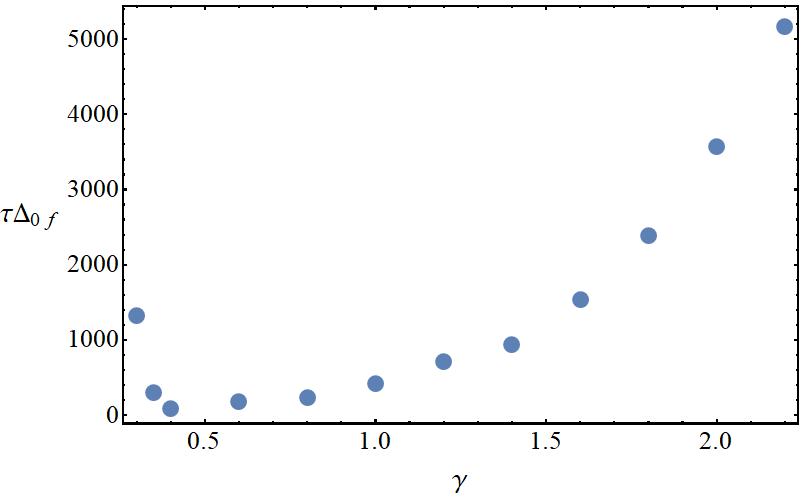}
}
\caption{Study of the relaxation time $\tau$, see Fig.~\ref{lor_tau_df}, in the Lorentzian model as a function of the integrability breaking parameter $\gamma$ at fixed $\Delta_{0i} = .005W$ and $\Delta_{0f} = 0.6W$, where $\gamma=\infty$ is the integrable $s$-wave model. For these quench parameters, both the Lorentzian and $s$-wave models enter Phase~III. Parts (a)-(c) show how the minimum of $\Delta(t)$ slowly evolves and reaches an asymptote, while part (d) gives $\tau$ near $\gamma=0.4W$, where the minimum satisfies $\tau_{\min}\Delta_{0f} \approx 89$. This minimum is still greater than the relaxation time of the $s$-wave case, where $\tau\Delta_{0f}\approx65$. The relaxation time increases sharply away from $\gamma=0.4W$, especially in the direction of decreasing gamma, where $\tau\Delta_{0f} \approx 64500$ at $\gamma = 0.11W$. In all plots, $\gamma$ is given in units of the bandwidth $W$ and $N=5500$.}
\label{lor_tau_gam}
\end{figure*}

More interesting is the dependence of $\tau$ on $\gamma$, the integrability-breaking parameter, at fixed $(\Delta_{0i},\Delta_{0f})$. First, let us examine quenches that lead to Phase~III in both the Lorentzian and integrable $s$-wave models. Fig.~\ref{lor_tau_gam} shows that $\tau$ has single minimum for $\gamma \sim 0.4W$ and increases away from this point both as $\gamma\to0$ and as $\gamma\to\infty$. In all cases, the relaxation time of quenches in the integrable $s$-wave model, which is the $\gamma\to\infty$ limit of our separable BCS Hamiltonians,  is far smaller.  We believe that the increase of $\tau$ as $\gamma\to\infty$ is indicative of nonperturbative behavior of the dynamics in the vicinity of the integrable limit, see Sect.~\ref{theConclusion}.

The behavior of $f(\eps)$ as $\gamma\to 0$ is model dependent; in the case of the Lorentzian model, the stability analysis of Sect.~\ref{subStAn} indicates that Phase~II is unstable to harmonic perturbations if $\gamma > \Delta_{\infty}$; otherwise, Phase~II could be stable. We observe in Fig.~\ref{lor_tau_gam} large oscillations in the evolution of the minimum of $\Delta(t)$ at $\gamma=0.2W$, behavior which occurs in the range $0.13W\lesssim\gamma\lesssim0.26W$ For $\gamma \lesssim 0.13W$, the minima oscillations disappear and $\tau$ begins to dramatically increase. 
Despite this qualitative change in the evolution of $|\Delta(t)|$, down to at least $\gamma = 0.11W$ we still find that the system eventually enters Phase~III with a reduced amplitude of oscillation.
\begin{figure}
\subfloat[]
{
\includegraphics[width=\linewidth]{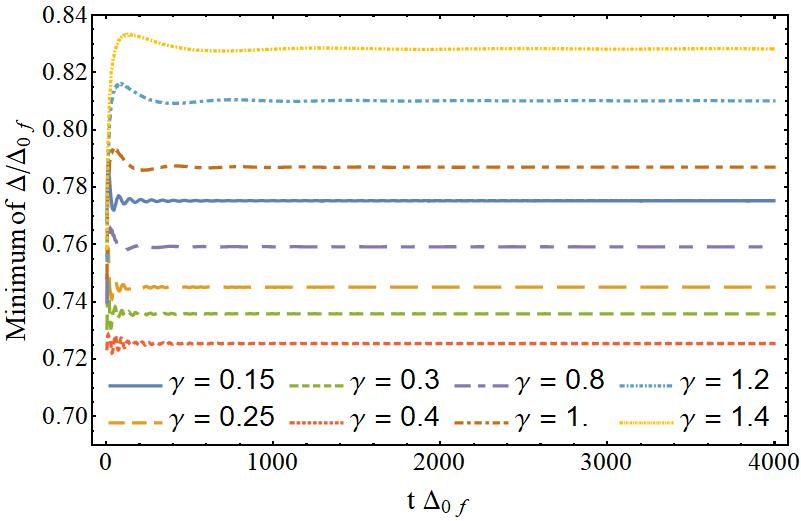}
}

\subfloat[]
{
\includegraphics[width=\linewidth]{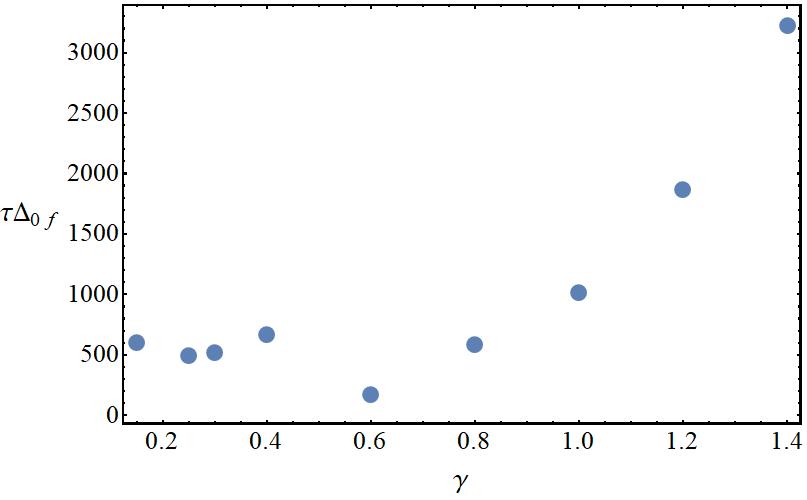}
}

\caption{Study of the relaxation time $\tau$ in the Lorentzian model as a function of the integrability breaking parameter $\gamma$ at fixed $\Delta_{0i} = 0.2W$ and $\Delta_{0f} = 0.6W$. For these quench parameters, the $s$-wave model enters Phase~II, while the Lorentzian enters Phase~III. Part (a) shows how the minimum of $\Delta(t)$ slowly evolves and reaches an asymptote, while part (b) gives $\tau$  near $\gamma=0.6W$, where $\tau_{\min}\Delta_{0f} \approx 175$. The relaxation time increases away from $\gamma=0.6W$ in both directions. In all plots, $\gamma$ is given in units of the bandwidth $W$ and $N=2800$.}
\label{lor_tau_gam5}
\end{figure}

Fig.~\ref{lor_tau_gam5} is similar to Fig.~\ref{lor_tau_gam}, except we now choose $\Delta_{0i}$ and $\Delta_{0f}$ such that the (integrable) $s$-wave model enters Phase~II. The behavior of $\tau$ with respect to $\gamma$ is qualitatively similar, except there is no regime where the minimum of $\Delta(t)$ undergoes large oscillations.

The spin-orbit model also has a very long relaxation time to Phase~III. In order to observe this asymptotic state, as is shown in Fig.~\ref{SOPhaseIII}, one must carefully choose model and quench parameters, otherwise $\tau$ is simply too large for our present numerical study.

\section{Phase III asymptotic solution}
\label{asympPerSol}
We now explore the structure of the Phase~III asymptotic  state. First, we treat $\Delta(t)$ as a periodic external driving and show that there is always a periodic solution for the classical pseudospins (and auxiliary functions in the spin-orbit model), and then we provide evidence that the class of periodic $\Delta(t)$  that are also self-consistent are elliptic functions.
\subsection{External driving}
\label{subExtDr}
In the separable BCS model, the mean-field dynamics can be described alternatively by a Gaussian wave function with complex Bogoliubov amplitudes $u_j(t)$ and~$v_j(t)$
\beg
\begin{split}
| \psi \rangle = \prod_j [u_j^*(t) + v_j^*(t)\hat{c}_{j\uparrow}^{\dag}\hat{c}_{j\downarrow}^{\dag}]| 0 \rangle,
\label{GaussianWF}
\end{split}
\en
where normalization requires $|v_j|^2+|u_j|^2 = 1$. The equations of motion for $u(t)$ and $v(t)$ follow from the time-dependent Schr{\"o}dinger equation $i\frac{\partial}{\partial t}| \psi \rangle = \hat{H} | \psi \rangle$ applied to \re{GaussianWF} with the mean-field Hamiltonian from \re{hamMF},
\begin{equation}
\begin{split}
i\frac{d}{dt}\left(\begin{array}{c}
u_j(t)\\
v_j(t)
\end{array} \right) = \left(\begin{array}{cc}
\eps_j & f_j \Delta \\
f_j \Delta^* & -\eps_j
\end{array} \right) \left(\begin{array}{c}
u_j(t)\\
v_j(t)
\end{array} \right),
\end{split}
\label{HFbogEOM}
\end{equation}
where we shifted the Hamiltonian by a constant $\hat{H} = \hat{H}_f - \sum_j \eps_j$ in order to make  it   traceless. The mapping to the classical pseudospins is
\beg
\begin{split}
s_j^- = u_jv_j^*,\quad s_j^z = \frac{|v_j|^2 - |u_j|^2}{2}.
\label{Bog2Spins}
\end{split}
\en
We shall discuss the nature of the asymptotic Phase~III $\Delta(t)$ in terms of $v(t)$ and $u(t)$. To do so, consider first \eref{HFbogEOM} with a periodic $\Delta(t)=\Delta(t+T)$ that is not necessarily self-consistent, which decouples each pair of $(u_j,v_j)$ from all the others. The abstract form of \eref{HFbogEOM} is
\begin{equation}
i\frac{d}{dt}\left(\begin{array}{c}
\mathbf{u}\\
\mathbf{v}
\end{array} \right) = \mathbf{h}(t) \left(\begin{array}{c}
\mathbf{u}\\
\mathbf{v}
\end{array} \right)
\label{HFbogEOMgen2}
\end{equation}
with
\begin{equation}
\mathbf{h}(t) = \left(\begin{array}{cc}
\mathbb{A} & \mathbb{B}(t)\\
\mathbb{B}^{\dag}(t) & -\mathbb{A}
\end{array} \right),
\label{HFbogEOMgen3}
\end{equation}
where $\mathbf{u}$ and $\mathbf{v}$ are $m$-dimensional vectors, $\mathbb{A}$ is a constant real symmetric $m\times  m$ matrix, $\mathbb{B}(t)$ is a complex $m\times m$ matrix periodic in $t$ with period $T$, and we dropped the index $j$ for simplicity. The forthcoming discussion is valid for all systems of this form, see also Ref.~\onlinecite{ydgf}. For example, the spin-orbit dynamics admit such a representation with $m=4$, while $m=1$ in the separable BCS model.

As $\mathbf{h}(t)$ is periodic by assumption, the Floquet theorem applies. There are thus $2m$ independent solutions $\boldsymbol{\psi}_i(t)$ to \eref{HFbogEOMgen2} of the form
\begin{equation}
\boldsymbol{\psi}_i(t) = e^{\delta_it}\left(\begin{array}{c}
\mathbf{U}_i(t)\\
\mathbf{V}_i(t)
\end{array} \right),\quad i=1\dots 2m,
\label{floquet1}
\end{equation}
where the $\mathbf{U}_i(t)$ and $\mathbf{V}_i(t)$ are periodic with the same period as $\mathbf{h}(t)$ and the $\delta_i$ are complex numbers known as Floquet exponents. The solutions $\boldsymbol{\psi}_i(t)$ therefore have the property
\begin{equation}
\boldsymbol{\psi}_i(t+T) = \rho_i\boldsymbol{\psi}_i(t), \quad\rho_i \equiv e^{\delta_iT},
\label{floquet2}
\end{equation}
where the $\rho_i$ are known as Floquet multipliers. Because $\mathbf{h}(t)$ is Hermitian, \eref{HFbogEOMgen2} conserves the norm of the solutions $\boldsymbol{\psi}_i(t)$, which implies $|\rho_i|=1$ and $\delta_i = i \nu_i$ for $\nu_i$ real. Furthermore, the particular form of $\mathbf{h}(t)$ implies that if $\boldsymbol{\psi}=(\mathbf{u},\mathbf{v})^T$ is a solution then so is $\widetilde{\boldsymbol{\psi}} = (\mathbf{v}^*,-\mathbf{u}^*)^T$. This pairing of solutions implies that if $\delta_i$ is a Floquet exponent, then so is $-\delta_i$. In Sect.~\ref{subSpinSol}, we will use this latter fact to prove that there is always a periodic spin solution to \eref{eomHF} for a given periodic $\Delta(t)$.

Before continuing, we note that the Phase~III asymptotic $\Delta(t)$ is only periodic in the particle-hole limit of the separable BCS model. In the general case, $\Delta(t)=F(t)e^{-2i\mu_{\infty}t}$, where $F(t)$ is periodic. Nonetheless, we can still reduce this problem, where $\mathbf{h}(t)$ is not periodic, to the periodic case by absorbing the phase $2\mu_{\infty}t$ in the following manner:
\begin{equation}
\begin{split}
\mathbf{v}' &= \mathbf{v}\,e^{-i\mu_{\infty}t},\\
\mathbf{u}' &= \mathbf{u}\,e^{i\mu_{\infty}t},\\
\mathbb{A}' &= \mathbb{A} - \mu_{\infty}\mathbb{1},
\label{FloquetAbsorbPhase}
\end{split}
\end{equation}
so that the time evolution of $(\mathbf{u}',\mathbf{v}')^T$ is described by \eref{HFbogEOMgen2} with periodic $\mathbf{h}(t)$ of the form given in \eref{HFbogEOMgen3} where $\mathbb{A}$ is replaced by $\mathbb{A}'$. In terms of the pseudospin representation of the dynamics, this transformation amounts to an overall time-dependent rotation about the z-axis with frequency $2\mu_{\infty}$.

\subsection{Phase~III spin solution in the separable BCS model}
\label{subSpinSol}
Now we draw our attention to the behavior of the spin solutions to the separable BCS model for the periodic external $\Delta(t)$ considered in the previous section. The dimension of the matrix $\mathbf{h}(t)$ is now $2m=2$ and there are two independent solutions to the Floquet problem
$$
\boldsymbol{\psi}_{1j}(t) =  e^{i\nu_j t}\left(\begin{array}{c}
U_j(t)\\ V_j(t) \end{array} \right),\quad \boldsymbol{\psi}_{2j}(t) =  e^{-i\nu_j t}\left(\begin{array}{c}
V_j^*(t)\\ -U_j^*(t) \end{array} \right)\!,
\label{FloquetHFspins1}
$$
where $U_j(t)$ and $V_j(t)$ are periodic and we restored the index $j$. Using $\boldsymbol{\psi}_{1j}(t)$ and \eref{Bog2Spins}, we can construct a periodic spin solution $\boldsymbol{\sigma}_j(t)$ [i.e., a periodic solution of  \eref{eomHF} for the given external $\Delta(t)$ that does not necessarily satisfy \eref{self13}],
\begin{equation}
\begin{split}
\sigma_j^-(t) &= U_j(t)V_j^*(t),\\
\sigma_j^z(t) &= \frac{|V_j(t)|^2-|U_j(t)|^2}{2}.
\end{split}
\label{FloquetHFspins2}
\end{equation}
We will now show that the most general spin solution $\mathbf{s}_j(t)$ precesses about the periodic solution $\boldsymbol{\sigma}_j(t)$ with a variable angular velocity. First we write the general solution $\boldsymbol{\Psi}_j(t)$ as a linear combination of $\boldsymbol{\psi}_{1j}(t)$ and $\boldsymbol{\psi}_{2j}(t)$
\begin{equation}
\boldsymbol{\Psi}_j(t) = \cos\frac{\theta_j}{2}\boldsymbol{\psi}_{1j}(t) + \sin\frac{\theta_j}{2}\boldsymbol{\psi}_{2j}(t).
\label{FloquetHFspins3}
\end{equation}
Although the coefficients of linear combination are in principle complex, we can drop the constant overall phase of $\boldsymbol{\Psi}_j(t)$ as well as absorb $\frac{1}{2}\times$the remaining constant relative phase into the definitions of $U_j(t)$ and $V_j(t)$. Once again using \re{Bog2Spins}, we now write $\boldsymbol{\Psi}_j(t)$ in terms of spin variables. It is helpful to first parametrize $U_j(t)$ and $V_j(t)$ as
\beg
\begin{split}
U_j(t) &= |U_j(t)|e^{\frac{i}{2}[\alpha_j(t)-2\nu_j t-\beta_j(t)]},\\
V_j(t) &= |V_j(t)|e^{\frac{i}{2}[\alpha_j(t)-2\nu_j t+\beta_j(t)]},
\end{split}
\label{FloquetHFspins4}
\en
whence
\beg
\begin{split}
s_j^- &= \cos\theta_j\,\sigma_j^-+\sin\theta_j\frac{\sigma_j^-}{|\sigma_j^-|}\bigg(\sigma_j^z\cos\alpha_j-\frac{i}{2}\sin\alpha_j\bigg),\\
s_j^z &= \cos\theta_j\,\sigma_j^z-\sin\theta_j\,|\sigma_j^-|\cos\alpha_j.
\end{split}
\label{FloquetHFspins5}
\en
Note that $\theta_j$ is the only time-independent quantity in \eref{FloquetHFspins5}. A geometric interpretation of the motion of the general solution $\mathbf{s}_j(t)$ with respect to the periodic solution $\boldsymbol{\sigma}_j(t)$ becomes clear once we use \eref{FloquetHFspins5} to express $\mathbf{s}_j(t)$ in the body coordinate system of $\boldsymbol{\sigma}_j(t)$. Let $\hat{\mathbf{z}}'_j = \hat{\boldsymbol{\sigma}}_j$, while $\hat{\mathbf{x}}'_j$ lies along the line defined by the intersection of the plane spanned by \{$\hat{\mathbf{z}}'_j,\hat{\mathbf{z}}_j$\} and that perpendicular to $\hat{\mathbf{z}}'_j$. Finally, $\hat{\mathbf{y}}'_j$ satisfies $\hat{\mathbf{y}}_j'\cdot\hat{\mathbf{x}}_j'=\hat{\mathbf{y}}'_j\cdot\hat{\mathbf{z}}'_j=0$ and $\hat{\mathbf{x}}_j'\times\hat{\mathbf{y}}_j'=\hat{\mathbf{z}}_j'$. These definitions lead to
\beg
\begin{split}
\hat{\mathbf{x}}'_j &= \frac{2}{|\sigma_j^-|}\bigg( \sigma_j^z\sigma_j^x \hat{\mathbf{x}}_j + \sigma_j^z\sigma_j^y \hat{\mathbf{y}}_j - |\sigma_j^-|^2 \hat{\mathbf{z}}_j \bigg),\\
\hat{\mathbf{y}}'_j &= -\frac{\sigma_j^y}{|\sigma_j^-|} \hat{\mathbf{x}}_j + \frac{\sigma_j^x}{|\sigma_j^-|} \hat{\mathbf{y}}_j.
\end{split}
\label{BodyCoors2}
\en
The general spin solution $\mathbf{s}_j(t)$ in this new coordinate system is then
\beg
\begin{split}
\mathbf{s}_j(t) &= \cos\theta_j \boldsymbol{\sigma}_j(t) + \sin\theta_j \boldsymbol{\sigma}_{j\perp}(t),\\
\boldsymbol{\sigma}_{j\perp}(t) &\equiv \frac{\cos\alpha_j(t)}{2}\hat{\mathbf{x}}'_j + \frac{\sin\alpha_j(t)}{2}\hat{\mathbf{y}}'_j,
\end{split}
\label{BodyCoors3}
\en
where $\boldsymbol{\sigma}_j\cdot\boldsymbol{\sigma}_{j\perp}=0$ and $\boldsymbol{\sigma}_{j\perp}$ is not periodic. We see from \eref{BodyCoors3} that $\mathbf{s}_j(t)$ makes a constant angle $\theta_j$ with the periodic solution and rotates about it with a variable angular frequency $\dot\alpha_j(t)$. From \eref{FloquetHFspins4} and the periodicity of $U_j(t)$ and $V_j(t)$, we conclude that $\alpha_j(t)-2\nu_j t$ is also periodic with the same period as the external $\Delta(t)$ driving the system.

\subsection{Asymptotic self-consistency}
\label{subAsympSC}
Thus far, we have considered $\Delta(t)$ to be an external periodic driving that is not necessarily self-consistent. We showed for any such external driving, there is a corresponding periodic spin solution $\boldsymbol{\sigma}_j(t)$ with the same period as $\Delta(t)$. Furthermore, we derived in \eref{BodyCoors3} that the general spin solution $\mathbf{s}_j(t)$ precesses in a simple manner about $\boldsymbol{\sigma}_j(t)$. In the true quench dynamics, however, $\Delta(t)$ must be self-consistent, and we now show that this requirement implies that there always exists a set of constants $\theta_j$, such that the following integral equation holds for the asymptotic periodic $\Delta(t)$:
\beg
\begin{split}
\Delta(t) = g_f\sum_jf_j\sigma_j^-[\Delta(t)]\cos\theta_j,
\end{split}
\label{selfConsistencyPeriodicDelta}
\en
The notation $\boldsymbol{\sigma}_j = \boldsymbol{\sigma}_j[\Delta]$ emphasizes that the periodic spin solution is some complicated nonlocal function of $\Delta(t)$. An analogous expression to \eref{selfConsistencyPeriodicDelta} exists for the spin-orbit model.

\eref{selfConsistencyPeriodicDelta} is simply asymptotic self-consistency, as introduced in Sect.~\ref{PDfromSim}, applied to the Floquet problem studied in Sects.~\ref{subExtDr} and \ref{subSpinSol}. To see this, suppose that we observe some Phase~III asymptotic periodic $\Delta(t)$ after a quench from the ground state of the separable BCS model, as discussed in Sect.~\ref{subphase3}. This $\Delta(t)$ is self-consistent by definition, i.e.,
\beg
\begin{split}
\Delta(t) = g_f\sum_jf_js_j^-(t),
\end{split}
\label{selfConsistencyPeriodicDelta1}
\en
which we write in terms of the underlying periodic spin solution $\boldsymbol{\sigma}_j$ by using \eref{FloquetHFspins5}
\beg
\begin{split}
\Delta &= g_f\sum_jf_j\bigg(\sigma_j^-[\Delta]\cos\theta_j + \sigma_{j\perp}^-[\Delta]\sin\theta_j\bigg),\\
\sigma_{j\perp} &\equiv \frac{\sigma_j^-}{|\sigma_j^-|}\bigg(\sigma_j^z\cos\alpha_j-\frac{i}{2}\sin\alpha_j\bigg),\\
\alpha_j(t) &= A_j(t) + 2\nu_j t,\quad A_j(t+T) = A_j(t),
\end{split}
\label{selfConsistencyPeriodicDelta2}
\en
where $\nu_j$ is the imaginary part of the Floquet exponent as introduced in \eref{floquet1}. As in our analysis of self-consistency in Phases~I and II, \eref{selfConsistencyPeriodicDelta2} cannot hold exactly, this time because the sum over $\sigma_{j\perp}^-[\Delta]$ is the only non-periodic term. Nonetheless, under the reasonable assumption that $\nu_{j+1}-\nu_j\sim \delta$, where $\delta$ is the level spacing, the sum over $ \sigma_{j\perp}^-[\Delta]$ dephases in $N\to\infty$ limit as $t\to\infty$ (the $N\to\infty$ limit comes first), leading to \eref{selfConsistencyPeriodicDelta}.

\subsection{Self-consistent solutions in the separable BCS model}
\label{subSCHF}
\begin{figure}[h]
\subfloat[]
{
\includegraphics[width=\linewidth]{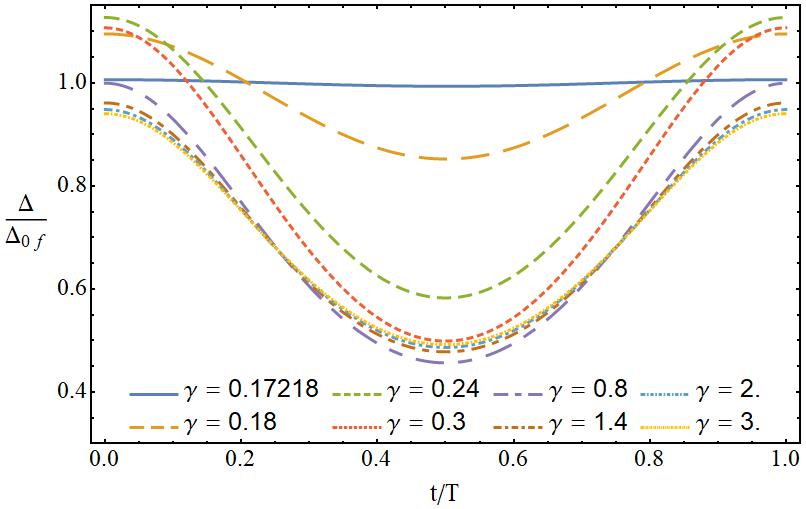}
}

\subfloat[]
{
\includegraphics[width=\linewidth]{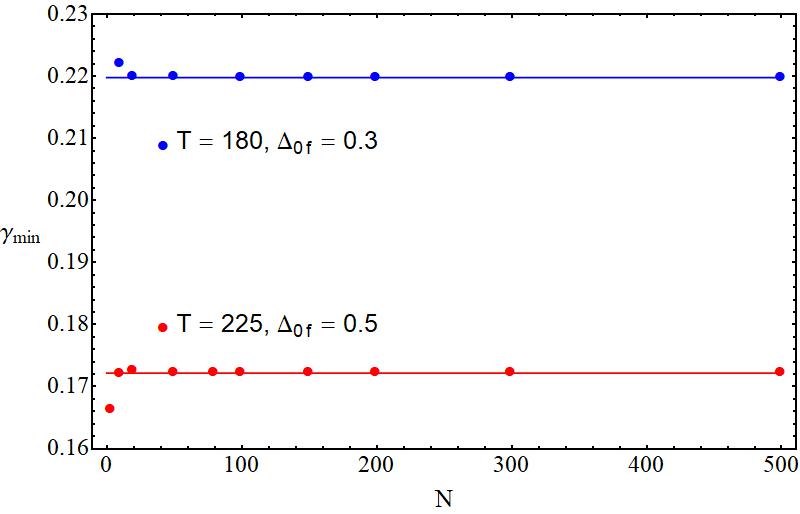}
}
\caption{(a) Examples of exactly self-consistent, periodic $\Delta(t)$'s for the Lorentzian separable BCS equations of motion for different values of $\gamma$ at fixed $\Delta_{0f} = 0.5W$, period $T = 225/W$, and $N = 500$. For these fixed parameters, below $\gamma_{\textrm{min}} \sim 0.172W$ the only exactly self-consistent, periodic $\Delta(t)$ is a constant in time equal to the equilibrium value. (b) Convergence of $\gamma_{\textrm{min}}$ as a function $N$. In both plots, $\Delta_{0f}$ and $\gamma$ are given in units of $W$ and $T$ in units of $W^{-1}$.}
\label{Fig_selfcon}
\end{figure}

\begin{figure}[h!]
\includegraphics[width=\linewidth]{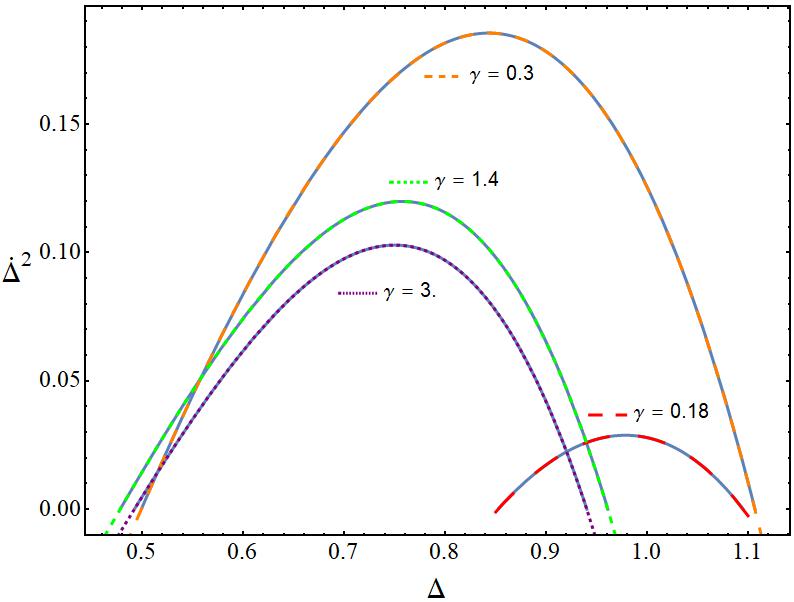}
\caption{Evidence that the self-consistent periodic $\Delta(t)$ from Fig.~\ref{Fig_selfcon} are elliptic functions. Squared time derivatives $\dot{\Delta}(t)$ as a function of $\Delta(t)$ are given by solid blue lines. These lines overlap strongly with the dashed lines, which are the fits to the defining differential equation for elliptic functions \eref{ellipticDefinition}. If a $\Delta^5$ coefficient is included in the fits, it is several orders of magnitude smaller than those for the 4th order fit shown here, providing strong evidence that $\dot{\Delta}^2(t)$ is indeed a 4th order polynomial in $\Delta(t)$. In this plot, $\gamma$ is given in units of $W$ and $\Delta$ in units of $\Delta_{0f}$}
\label{FigdelDotSq}
\end{figure}

We have seen that an asymptotically self-consistent periodic $\Delta(t)$ satisfies the functional equation \re{selfConsistencyPeriodicDelta} in the separable BCS model. We now will give evidence that solutions to \eref{selfConsistencyPeriodicDelta} are elliptic functions. In order to generate such solutions, fix a period $T$ and write $\Delta(t)$ as a Fourier series
\beg
\Delta(t) = \sum_{n=-\infty}^{\infty}c_ne^{2\pi i n\frac{t}{T}},
\label{fourierDeltaSC}
\en
which we truncate to some $n_{\textrm{max}}$, such that $c_{n}=0$ if $|n|>n_{\textrm{max}}$. In the particle-hole symmetric limit, $\Delta(t)$ is a real quantity  that satisfies $\Delta(t) = \Delta(-t)$ [see \eref{treversal}], so that $c_n$ is real and equals $c_{-n}$.

For a fixed set of coefficients $c_n$, we determine $\sigma^x_j[\Delta(t)]$ by solving the equations of motion \re{eomHF} from $t=0$ to $t=T$ with $\Delta(t)$ given by \re{fourierDeltaSC}. If the choice of $c_n$ produces a self-consistent $\Delta(t)$, then it will be equal to the quantity $\Delta_{\textrm{comp}}(t)$ defined as
\beg
\begin{split}
\Delta_{\textrm{comp}}(t) = g_f\sum_jf_j\sigma_j^x[\Delta(t)]\cos\theta_j,
\end{split}
\label{PeriodicDeltaComp}
\en
for some set of $\theta_j$. For most choices of $c_n$, however, \eref{PeriodicDeltaComp} will not hold. As both $\Delta(t)$ and $\Delta_{\textrm{comp}}(t)$ are periodic functions of  time with the same period, we define a distance $r(\{c_n\})$  as
\beg
\begin{split}
r^2(\{c_n\}) = \int_0^T \bigg(\Delta_{\textrm{comp}}(t) - \Delta(t)\bigg)^2dt.
\end{split}
\label{distanceSC}
\en
A given $\Delta(t)$ is asymptotically self-consistent if and only if $r(\{c_n\}) = 0$.

We now explore the results of this procedure for the Lorentzian coupling of the separable BCS model for various values of the integrability breaking parameter $\gamma$. It turns out that this procedure works when we fix $\cos\theta_j = 1$, i.e., we find exactly (and not just asymptotically) self-consistent solutions. In order to find such solutions, we start from the known values of the Fourier coefficients of the $s$-wave ($\gamma = \infty$) solution, which are close to the Fourier coefficients of the $\gamma \gg 1$ solutions. We then progressively lower $\gamma$ while finding Fourier coefficients that minimize $r(\{c_n\})$. Typically we obtain values of $r \sim 10^{-12}-10^{-11}$ before declaring the solution self-consistent.

Fig.~\ref{Fig_selfcon} gives of examples of such solutions at fixed $\Delta_{0f}$ and period $T$. Notably, there is a minimum $\gamma = \gamma_{\textrm{min}}$ below which the amplitude of oscillation vanishes. As $\gamma$ is increased from this minimum, the amplitude of oscillations increases to a maximum and then decreases to a nonzero limiting value as $\gamma\to\infty$. Fig.~\ref{Fig_selfcon} also shows the fast convergence of $\gamma_{\textrm{min}}$ as a function of $N$ for two examples of this procedure.

In Sect.~\ref{subphase3}, we argued through example quenches that the $\Delta(t)$ of Phase~III are always elliptic functions, i.e., they satisfy \eref{ellipticDefinition}. We show in Fig.~\ref{FigdelDotSq} that the exactly self-consistent $\Delta(t)$ from Fig.~\ref{Fig_selfcon} also satisfy \eref{ellipticDefinition} to a high degree of accuracy. The Floquet analysis of the equations of motion from Sect.~\ref{subExtDr} applies to any periodic $\Delta(t)$. From Fig.~\ref{FigdelDotSq}, we conclude that the self-consistency requirement \re{selfConsistencyPeriodicDelta} is essential to selecting elliptic functions amongst all possible periodic functions.

\begin{figure}[h!]
\subfloat[]
{
\includegraphics[width=\linewidth]{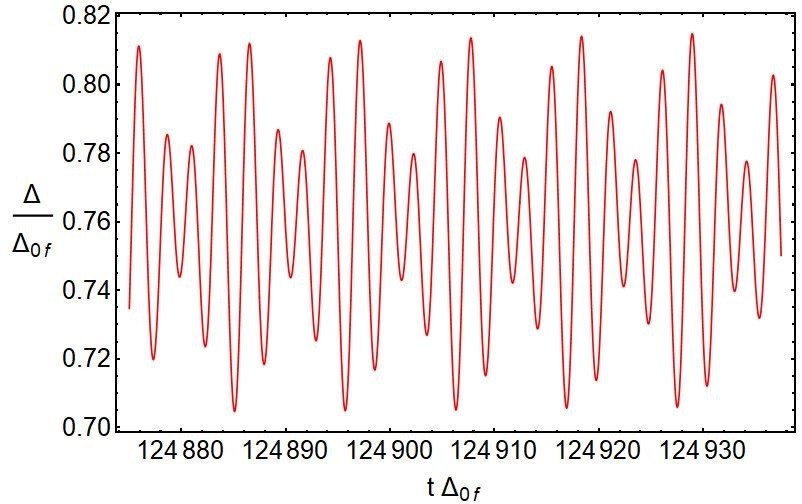}
\label{quasi1}
}

\subfloat[]
{
\includegraphics[width=\linewidth]{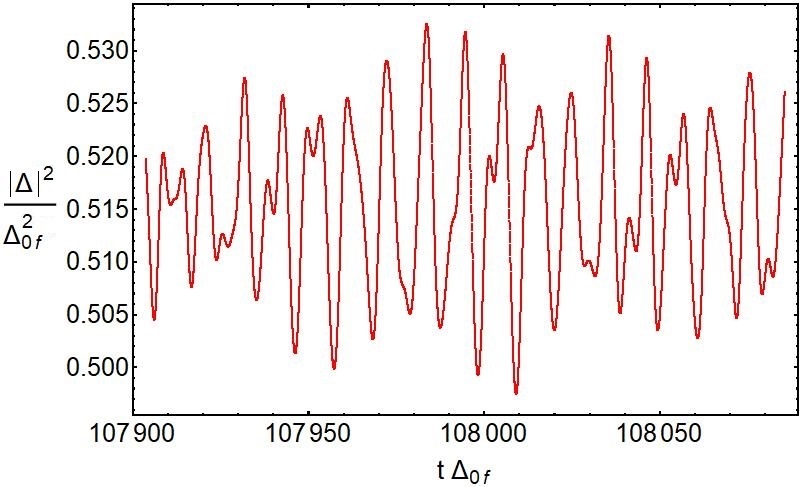}
\label{quasi2}
}
\caption{Quenches of nonintegrable separable BCS and spin-orbit models that do not conform to Phases I, II or III.  This quasiperiodic   dynamics of the order parameter
emerge early and persist for the entire time of the simulation, see also Fig.~\ref{quasilong}.
 Plot (a) is the particle-hole symmetric  separable BCS model with sine coupling from \eref{fSinIntro} and $N = 4\times10^5$ spins. In units of the bandwidth, the integrability breaking parameter is $\gamma = 0.075$, while $\Delta_{0i} = 0.05$ and $\Delta_{0f} = 0.5$. Part (b) is the spin-orbit model with $N=2\times10^5$ spins. In units of the bandwidth: $\eps_F = 0.4$, $\alpha^2=0.4$, $g N = 2$, $h_i = 2$, and $h_f = 0.514256$. }
\label{quasi}
\end{figure}

\begin{figure}[t!]
\subfloat[]
{
\includegraphics[width=\linewidth]{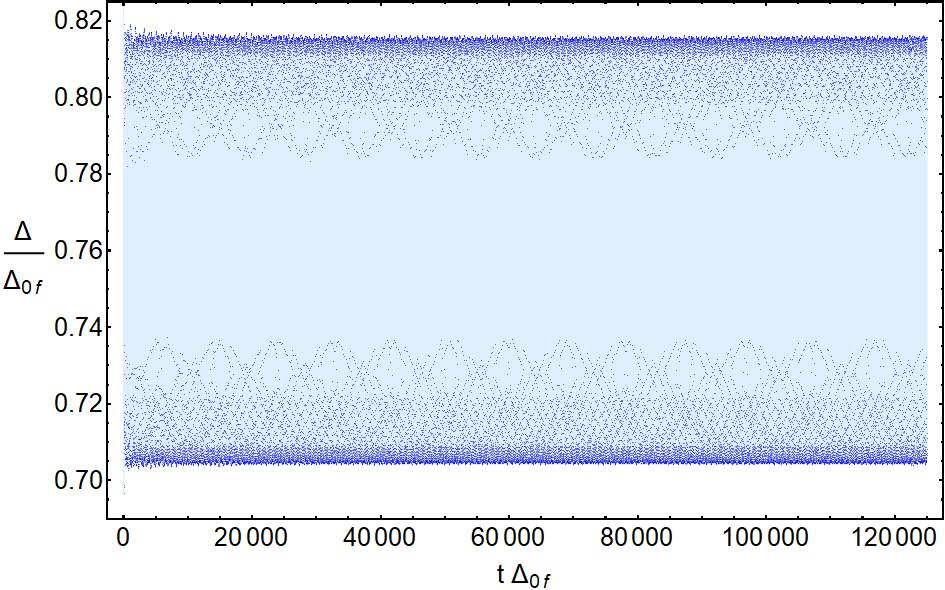}
\label{quasilong1}
}

\subfloat[]
{
\includegraphics[width=\linewidth]{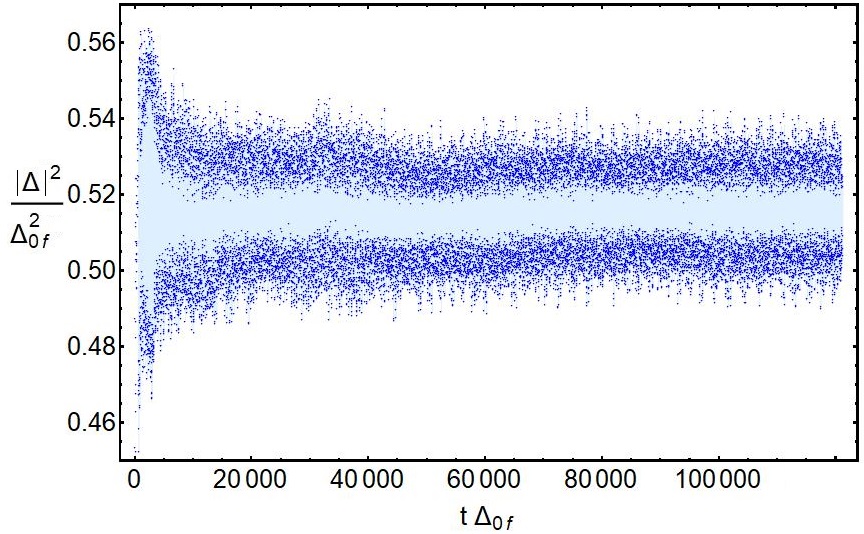}
\label{quasilong2}
}
\caption{Darker blue points are local minima and maxima of the oscillations for the quenches from  Fig.~\ref{quasi} for the entire time of
the simulations.  These plots suggest that there are regions of quasiperiodicity (Phase IV) in the quantum quench phase diagrams of nonintegrable pairing models. Part (a) is the same quench as in Fig.~\ref{quasi1}, part (b) corresponds to Fig.~\ref{quasi2}. In terms of the inverse level spacing, the time evolution goes out to $t_{\max}=0.625 \delta^{-1}$ in plot (a) and to $t_{\max}= \delta^{-1}$ in plot (b).}
\label{quasilong}
\end{figure}

\section{Quasiperiodic \protect Phase IV} 
\label{quasisect}

 Quenches that do not conform to Phases~I-III are another intriguing consequence of integrability breaking. We present two such examples in Figs.~\ref{quasi} and \ref{quasilong}. Figs.~\ref{quasi1} and \ref{quasilong1} show a particle-hole symmetric quench of the separable BCS Hamiltonian with sine coupling from
 \eref{fSinIntro}. Figs.~\ref{quasi2} and \ref{quasilong2}  depict a  quench of the Zeeman  field in the spin-orbit model~\re{HSOSpin}.  The   quasiperiodic behavior 
 of $\Delta(t)$ in
 Fig.~\ref{quasi1} sets in  very early on, as corroborated by Fig.~\ref{quasilong1},  and persists with no appreciable changes at least until the times shown in the figure.  Similarly, Fig.~\ref{quasi2} is representative of the long-time  spin-orbit $|\Delta(t)|^2$ as evidenced by Fig.~\ref{quasilong2}.  Based on our preliminary analysis of the Fourier spectrum of $|\Delta(t)|^2$ for this quench and of the maximal Lyapunov exponent with the method of local divergence rates\cite{HR}, we believe that it too is quasiperiodic. However, a more careful study    is needed to unambigously distinguish between quasiperiodicity and chaos in this case.  Such a study is beyond the scope of the present paper, where we mainly focus on the properties of Phases I-III.

 Note that the simulation times in Figs.~\ref{quasi} and \ref{quasilong} are
 enormous compared to the characteristic   time of a single oscillation and even to  typical Phase III relaxation times $\tau\Delta_{0f}\sim 10^3 $ we observed in Sect.~\ref{subsubRel}, cf. Fig.~\ref{lor_tau_gam} and the caption to Fig.~\ref{SOPhaseIII}. 
Thus,  both of these examples do not belong to Phases I, II, or III. We therefore conclude that there are regions of quasiperiodicity in the
 quantum quench phase diagrams of nonintegrable pairing models, which we call Phase IV.

\section{Conclusion}
\label{theConclusion}

The far-from-equilibrium steady states reached by nonintegrable pairing models after a quantum quench admit a similar taxonomy as do the integrable cases. We have shown that some or all of Phases I-III may occur in the separable BCS models and spin-orbit model defined in \eref{ham1}. The persistent periodic oscillations characterizing Phase~III are always elliptic functions, regardless of whether the model is integrable. Moreover, we have developed a stability analysis of the three phases, summarized in \eref{freqsPhase2LinNoPH}, which generalizes known results in the integrable cases and elucidates the mechanism of nonequilibrium phase transitions using the language of linear analysis.

Despite these striking similarities, important consequences accompany integrability breaking. As argued in Sect.~\ref{subStAn}, some nonintegrable models may not exhibit all three phases. At the same time, an entirely new quasiperiodic Phase IV emerges in certain models.
 Another key byproduct of integrability breaking is the emergence of a new, extremely long relaxation time scale $\tau$ when the asymptotic state either is or appears to be Phase~III. For $t < \tau$, $\Delta$ can oscillate with more than one fundamental frequency and a slowly varying amplitude. This time scale is a generic feature of nonintegrable models, and its existence renders short-time analyses inadequate for determining the long-time dynamics.  Moreover, $\tau$ diverges
 as we approach integrable points (e.g., as $\gamma^{-1}\to0$ in the separable pairing models of Sect.~\ref{PDfromSim}), and it is often too large for the practical determination of the true asymptotic state.

While the squared modulus of $\Delta(t)$ [and $\Delta(t)$ itself in the particle-hole symmetric case]  is always an elliptic function in Phase~III, its parametrization is more complicated in nonintegrable models. As a result, the reduction mechanism discussed in Appendix~\ref{AppReductionMech}, which explains how Phase~III manifests itself in the integrable models, does not apply to nonintegrable models. Nonetheless, we demonstrated in Sect.~\ref{asympPerSol} that the common structure of the nonintegrable models implies the existence of a  periodic solution to the classical pseudospin equations of motion if $\Delta(t)$ is taken to be a generic periodic external driving. Using numerical examples, we argued that further requiring $\Delta(t)$ to be self-consistent selects elliptic functions amongst all possible periodic functions.

It is instructive to discuss the BCS quench dynamics in terms of bifurcation theory\cite{Intro_bifurc,Kuznetsov,Hilborn}. For example, consider the particle-hole symmetric separable BCS models with real $\Delta$. For fixed initial conditions \re{GSHF} and any function   $\Delta(t)$ with fixed $\Delta(0)$, the equations of motion  \re{eomHF} have a unique solution ${\bf s}_j[\Delta(t)]\equiv {\bf s}[\eps_j, \Delta(t)]$. 
\eref{self13} then provides a closed nonlinear integral equation for $\Delta(t)$ [cf. \eref{selfConsistencyPeriodicDelta}],
\beg
\Delta(t)=g_f\!\int\, d\eps  \,s_x[\eps, \Delta(t)].
\label{intSep19}
\en
Phase~I is a fixed point, $\Delta=0$, of this equation\cite{noteSep20_2},  while Phase~II corresponds to two fixed points $\Delta_\infty$ and $e^{i\pi}\Delta_\infty=-\Delta_\infty$. In Phase~III   we end up on one of  two limit cycles related to each other by a rotation by $\pi$ around the z-axis [change of sign of $\Delta(t)$]. The Phase~I to II and  II to III transitions  correspond to supercritical pitchfork and Hopf bifurcations, respectively, in this language\cite{noteSep20_1}. The same results apply to the spin-orbit model \re{HSOSpin}.  We also note that this quantum quench phase diagram is surprisingly  similar to the nonequilibrium phase diagram of two atomic condensates coupled to a heavily damped   cavity mode\cite{aniket1,aniket2}. The mean-field dynamics of the latter system are described by
the driven-dissipative variant of the Bloch equations \re{eomHF} for two classical spins representing individual condensates. Moreover,  
there are   islands of quasiperiodicity  in the phase diagram of the two coupled condensates, where the dynamics are very similar to that shown in Figs.~\ref{quasi} and \ref{quasilong}.

Bifurcation theory also offers a plausible explanation for the divergence of the relaxation time $\tau$ near integrable points. Consider Phase~III for an integrable pairing Hamiltonian, such as the particle-hole symmetric $s$-wave BCS model. Suppose the corresponding limit cycle loses stability as soon as integrability is broken and another limit cycle emerges as an attractor.  An example of such behavior is the transcritical bifurcation\cite{Intro_bifurc,Kuznetsov,Hilborn}. Because the
 instability is weak for weak integrability breaking and because the evolution   starts near the unstable limit cycle, the system takes a very long time $\tau$ to reach the attractor. The weaker the integrability breaking, the closer we are to the bifurcation and the longer the time~$\tau$.
 
 An interesting open problem is to explore the newly discovered quasiperiodic Phase IV. In particular, one needs to investigate the possibility that asymptotic oscillations of $|\Delta(t)|$ for certain quenches may be  chaotic, rather than quasiperiodic, i.e., the potential existence of a chaotic phase in addition to the quasiperiodic one. Let us also mention that quasiperiodic
   $|\Delta(t)|$ also occurs in integrable models, but only when the initial (pre-quench) state is a highly excited state instead of the ground state\cite{multi}.
 
 In this paper, we employed reduced BCS Hamiltonians~\re{ham1} to model pairing dynamics. This description is valid
only at times $t\ll \Gamma^{-1}$, where $\Gamma$ is the highest among the rates of processes such Hamiltonians neglect. These processes include pair-breaking collisions\cite{volkov,galperin,barlev0,ydgf}, three-body losses in ultracold gases\cite{matt}, thermal fluctuations\cite{sasha}, etc. Thus, to reach the asymptotic state before these effects influence the dynamics, we need $\Gamma^{-1}\gg \tau$. In Phases~II and III, this requirement is much more stringent than $\Gamma^{-1}\gg T_\Delta$ typically quoted in the literature on  collisionless pairing dynamics. 
Here $T_\Delta$ is the characteristic period of $\Delta(t)$ oscillations ($T_\Delta$ is of the order of the inverse equilibrium gap
$\Delta_{0f}$ in
our separable BCS models).   Another limitation is the parametric instability of Phase III with respect to spontaneous eruptions of spatial inhomogeneities\cite{turbulence,mattSep20,chern,comment}. To avoid this instability, the system size has to be smaller than the superconducting coherence length.

 \begin{acknowledgments}

We thank M. Dzero, A. J. Millis and A. Patra for helpful discussions. This work was supported by the National Science Foundation Grant DMR-1609829. J.A.S. was supported by a Rutgers University Louis Bevier Dissertation Completion Fellowship.

\end{acknowledgments}

\appendix
 
 \section{Mean-field equations of motion}
\label{appA}
The pseudospin equations of motion for the separable BCS model \re{eomHF} obtain simply from the Heisenberg equations of motion $\frac{d}{dt}\hat{A} = i\,[\hat{H},\hat{A}]$ applied to the mean-field $\hat{H}_f$ in \eref{hamMF} and the pseudospin operators $\hat{\mathbf{s}}$ defined in \eref{FtoPs}. The classical spin variables $\mathbf{s}$ are the expectation values of the pseudospin operators $\mathbf{s}=\langle\hat{\mathbf{s}}\rangle$, and the time-dependent order parameter $\Delta$ is determined self-consistently according to \eref{self13}.

The generalized pseudospin representation of the spin-orbit Hamiltonian $\hat{H}_{so}$ from \eref{ham1} requires more work\cite{dky}. First, we diagonalize the kinetic part of $\hat{H}_{so}$ through the following unitary transformation to new fermionic operators $\hat{a}_{\mathbf{k}\pm}$
\begin{equation}
\begin{split}
&U_{\mathbf{k}}\left(\begin{array}{c}
\hat{c}_{\mathbf{k}\up}\\
\hat{c}_{\mathbf{k}\downarrow}
\end{array} \right) = \left(\begin{array}{c}
\hat{a}_{\mathbf{k}+}\\
\hat{a}_{\mathbf{k}-}
\end{array} \right)\\
&U_{\mathbf{k}} = \left(\begin{array}{cc}
\cos\frac{\phi_k}{2} & -i\,e^{-i \theta_{\mathbf{k}}}\sin\frac{\phi_k}{2}\\
\sin\frac{\phi_k}{2} & i\,e^{-i \theta_{\mathbf{k}}}\cos\frac{\phi_k}{2}
\end{array} \right),
\end{split}
\label{HSOdiagonalization}
\end{equation}
where $\mathbf{k} = k e^{i \theta_{\mathbf{k}}}$ and $\phi_k$ is defined in terms of the model parameters in \eref{SOparams}. One can check that the new elementary excitation energies are $\eps_{\mathbf{k}\pm}\equiv\eps_{\mathbf{k}}\mp R_{\mathbf{k}}$. \eref{HSOdiagonalization} implies
\begin{equation}
\begin{split}
\hat{c}_{-\mathbf{k}\downarrow}\hat{c}_{\mathbf{k}\up} = &\frac{-i\,e^{i \theta_{\mathbf{k}}}}{2}\bigg(\sin\phi_{\mathbf{k}}(\hat{a}_{-\mathbf{k}+}\hat{a}_{\mathbf{k}+}-\hat{a}_{-\mathbf{k}-}\hat{a}_{\mathbf{k}-})+\\
&+\cos\phi_{\mathbf{k}}(\hat{a}_{\mathbf{k}-}\hat{a}_{-\mathbf{k}+}+\hat{a}_{\mathbf{k}+}\hat{a}_{-\mathbf{k}-})+\\
&+\hat{a}_{\mathbf{k}+}\hat{a}_{-\mathbf{k}-}+\hat{a}_{-\mathbf{k}+}\hat{a}_{\mathbf{k}-}\bigg).
\end{split}
\label{HSOdiagonalization3}
\end{equation}
Upon summing over $\mathbf{k}$, the last two terms in parentheses cancel with momenta of opposite sign. Therefore, the interaction term of \re{ham1} in this new basis becomes
\begin{equation}
\begin{split}
&g\sum_{\mathbf{k}\mathbf{k}^{\prime}}\hat{c}^{\dag}_{\mathbf{k}\up}\hat{c}^{\dag}_{-\mathbf{k}\downarrow}\hat{c}_{-\mathbf{k}^{\prime}\downarrow}\hat{c}_{\mathbf{k}^{\prime}\up} = \frac{1}{g}\hat{\Delta}^{\dag}\hat{\Delta},\\
&\hat{\Delta} \equiv \frac{g}{2}\sum_{\mathbf{k}\lambda}e^{i \theta_{\mathbf{k}}}\bigg(\lambda\sin\phi_{\mathbf{k}}\hat{a}_{-\mathbf{k}\lambda}\hat{a}_{\mathbf{k}\lambda}+\cos\phi_{\mathbf{k}}\hat{a}_{\mathbf{k}\lambda}\hat{a}_{-\mathbf{k}\bar{\lambda}}\bigg),
\end{split}
\label{HSOdiagonalization4}
\end{equation}
and upon taking the mean-field approximation $\hat{c}^{\dag}\hat{c}^{\dag}\hat{c}\hat{c}\approx \langle\hat{c}^{\dag}\hat{c}^{\dag}\rangle\hat{c}\hat{c} + \hat{c}^{\dag}\hat{c}^{\dag}\langle\hat{c}\hat{c}\rangle - \langle\hat{c}^{\dag}\hat{c}^{\dag}\rangle\langle\hat{c}\hat{c}\rangle$, the interaction term becomes
\begin{equation}
\begin{split}
\hat{\Delta}^{\dag}\hat{\Delta} &\approx \Delta^*\hat{\Delta}+\Delta\hat{\Delta}^{\dag}-\Delta^*\Delta,\\
\Delta &\equiv \langle\hat{\Delta}\rangle.
\end{split}
\label{HSOdiagonalization5}
\end{equation}
Neglecting the constant term $\Delta^*\Delta/g$, we arrive at the mean-field spin-orbit Hamiltonian $\hat{H}_{so}$ in the $\hat{a}$ basis found in \re{hamMF}.
Similar to the separable BCS model, we now search for a set of quadratic fermionic operators whose equations of motion are closed. Define the following operators
\begin{equation}
\begin{split}
\hat{S}_{\mathbf{k}\lambda}^z & = \frac{1}{2}\big( \hat{a}_{\mathbf{k}\lambda}^{\dag}\hat{a}_{\mathbf{k}\lambda} + \hat{a}_{-\mathbf{k}\lambda}^{\dag}\hat{a}_{-\mathbf{k}\lambda} - 1 \big),\\
\hat{S}_{\mathbf{k}\lambda}^- &= \lambda \eta_{\mathbf{k}} \hat{a}_{-\mathbf{k}\lambda} \hat{a}_{\mathbf{k}\lambda},\\
\hat{L}_{\mathbf{k}\lambda}^z &= -\frac{\lambda}{4}\big(  \hat{a}_{\mathbf{k}+}^{\dag}\hat{a}_{\mathbf{k}-} + \hat{a}_{-\mathbf{k}+}^{\dag}\hat{a}_{-\mathbf{k}-} +\\&+ \hat{a}_{\mathbf{k}-}^{\dag}\hat{a}_{\mathbf{k}+} + \hat{a}_{-\mathbf{k}-}^{\dag}\hat{a}_{-\mathbf{k}+} \big),\\
\hat{L}_{\mathbf{k}\lambda}^- &= \frac{\eta_{\mathbf{k}}}{2}\big(  \hat{a}_{\mathbf{k}+}\hat{a}_{-\mathbf{k}-} + \hat{a}_{\mathbf{k}-}\hat{a}_{-\mathbf{k}+}  \big),\\
\hat{T}_{\mathbf{k}} &= \frac{i}{4} \big(  -\hat{a}_{\mathbf{k}+}^{\dag}\hat{a}_{\mathbf{k}-} - \hat{a}_{-\mathbf{k}+}^{\dag}\hat{a}_{-\mathbf{k}-} +\\&+ \hat{a}_{\mathbf{k}-}^{\dag}\hat{a}_{\mathbf{k}+} + \hat{a}_{-\mathbf{k}-}^{\dag}\hat{a}_{-\mathbf{k}+} \big),
\end{split}
\label{HSOoperators}
\end{equation}
where $\eta_{\mathbf{k}} = e^{i\,\theta_{\mathbf{k}}} = -\eta_{-\mathbf{k}}$ and, as usual, $\hat{S}^- = \hat{S}^x -i\,\hat{S}^y$ and $\hat{L}^- = \hat{L}^x -i\,\hat{L}^y$.

One can check that $\hat{\mathbf{S}}_{\mathbf{k}\lambda}$, $\hat{\mathbf{L}}_{\mathbf{k}\lambda}$ and $\hat{T}_{\mathbf{k}}$ are Hermitian operators. There is reflection symmetry in $\mathbf{k}$-space: $\hat{A}_{-\mathbf{k}\lambda} = \hat{A}_{\mathbf{k}\lambda}$ for all operators $\hat{A}_{\mathbf{k}\lambda}$ in \re{HSOoperators}, as well as the following band symmetry for $\hat{\mathbf{L}}_{\mathbf{k}\lambda}$: $\hat{L}_{\mathbf{k}+}^- = \hat{L}_{\mathbf{k}-}^-$ and $\hat{L}_{\mathbf{k}+}^z = -\hat{L}_{\mathbf{k}-}^z$.

We apply the Heisenberg equations of motion to \re{HSOoperators} and $\hat{H}_{so}$ from \re{hamMF} and then take expectation values to arrive at the generalized pseudospin equations of motion~\re{eomHSO}. The time-dependent order parameter $\Delta = \langle\hat{\Delta}\rangle$ as a function of the new variables can be found in \eref{fieldsHSO}. The factor $\eta_{\mathbf{k}}$ does not appear in \eref{eomHSO}, implying that the dynamics preserve any radial symmetry found in the initial state. As all initial states considered in this work are radially symmetric, one can opt to label the generalized pseudospin variables by their single-particle energies rather than their momentum vector.

\section{Integrable limit of spin-orbit quenches}
\label{appB}
The authors of Ref.~\onlinecite{ddgp} created a full nonequilibrium phase diagram of the spin-orbit model for quenches of the magnetic field $h_i\to h_f$ as a function of $h_i$ and $h_f$. However,  this phase diagram needs to be revised by running simulations to  much longer times 
$t>\tau$, which, in particular, may modify the Phase II-III boundary\cite{note22}. The phase diagram of Ref.~\onlinecite{ddgp} is also missing the quasiperiodic   Phase IV
discovered in the present work.

In Ref.~\onlinecite{dky}, an attempt was made to analyze interaction and external field quenches  to the integrable limit $h_f=0$, but mistakes led to an incorrect phase diagram for the interaction quenches. Here we correct those mistakes and generate a correct phase diagram.

When the external field $h$ is set to zero, $H_{so}$ from \re{HSOSpin} becomes equivalent to the integrable $s$-wave model with a dispersion relation $\eps_{\mathbf{k}\lambda} = \frac{k^2}{2} - \lambda \alpha k$. This becomes clear in the equations of motion \re{eomHSO} with $\cos\phi_k = 0$ and $\sin\phi_k = 1$, where the spin degrees of freedom $\mathbf{S}_{\mathbf{k}\lambda}$ decouple from the others and $\Delta$ depends only on $\mathbf{S}_{\mathbf{k}\lambda}$. In what follows, the initial state of the system will be the ground state for some $h_i\ge0$ given by \re{GSHSO}, and the Hamiltonian for $t\ge0$ is
\begin{equation}
\begin{split}
H &= \sum_{\mathbf{k}\lambda}2\eps_{k\lambda}S^z_{\mathbf{k}\lambda}-2|\Delta|^2/g_f,\\
\Delta &= \frac{g_f}{2}\sum_{\mathbf{k}\lambda}S^-_{\mathbf{k}\lambda},\quad \eps_{\mathbf{k}\lambda} = \frac{k^2}{2} - \lambda \alpha k.
\end{split}
\label{HSOint}
\end{equation}
We use the integrability of $H$ to construct the exact phase diagram using a technique imported from Refs.~\onlinecite{ydgf} which we now summarize briefly. The analysis centers around a quantity $\mathbf{L}(u)$ called the Lax vector (not to be confused with the variables $\mathbf{L}_{\mathbf{k}\lambda}$)
\begin{equation}
\begin{split}
\mathbf{L}(u) = -\frac{2}{g_f}\hat{\mathbf{z}} + \sum_{\mathbf{k}\lambda}\frac{\mathbf{S}_{\mathbf{k}\lambda}}{u - \eps_{k\lambda}}.
\end{split}
\label{Lax1}
\end{equation}
The integrability of $H$ follows from the fact that $L^2(u)$ is conserved by the time evolution for arbitrary $u$, which implies conservation of the $2N$ roots of $L^2(u)$, which we call $u_j$. As demonstrated in Ref.~\onlinecite{ydgf}, each of the asymptotic nonequilibrium phases corresponds a unique number of isolated complex pairs of $u_j$ in the continuum limit. Phase~I corresponds to zero isolated $u_j$, Phase~II corresponds to one pair, and Phase~III corresponds to two pairs. As the $u_j$ are constants of the motion, we can evaluate $L^2(u)$ at $t=0$ to determine the number of isolated pairs of $u_j$ and thus generate the phase diagram for a given $h_i$.

Let us first start with the case when $h_i=0$ and we quench the interaction $g_i\to g_f$. In this case the ground state self-consistency relationship is
\beg
\begin{split}
\frac{2}{g_{i,f}} = \sum_{\mathbf{k}\lambda}\frac{1}{2E_{\mathbf{k}\lambda}},\quad E_{\mathbf{k}\lambda}=\sqrt{(\eps_{\mathbf{k}\lambda}-\mu_{i,f})^2+\Delta_{0i,f}^2}.
\end{split}
\label{SelfConHSOint}
\en
Using \eref{SelfConHSOint} along with the initial state given by \eref{GSHF}, we find that the initial Lax vector has the form
\begin{equation}
\begin{split}
\mathbf{L}(u) &= \bigg( \Delta_{0i}L_x(u),\,0,\,(\mu_i-u)L_x(u) - \widetilde{\beta}\bigg),\\
 L_x(u) &= \sum_{\mathbf{k}\lambda}\frac{1}{2(u-\eps_{\mathbf{k}\lambda})E_{\mathbf{k}\lambda}},\quad\widetilde{\beta} \equiv \frac{2}{g_f}-\frac{2}{g_i}.
\end{split}
\label{Lax2}
\end{equation}
If $g_f = g_i$, i.e., the zero quench, then $\widetilde{\beta}=0$ and the only complex pair of roots is $u_{\pm} = \pm i\,\Delta_{0i} + \mu$. This is the degenerate Phase~II case, where $\Delta(t) = \Delta_{0i}$ identically. When $g_f\ne g_i$, $L^2(u)=0$ implies
\begin{equation}
\begin{split}
\sum_{\mathbf{k}\lambda}\frac{1}{(u-\eps_{\mathbf{k}\lambda})\sqrt{(\eps_{\mathbf{k}\lambda}-\mu_i)^2+\Delta_{0i}^2}} = -\frac{2\widetilde{\beta}}{u-\mu_i \pm i\,\Delta_{0i}}.
\end{split}
\label{Lax3}
\end{equation}

\begin{figure}[h!]
\includegraphics[width=\linewidth]{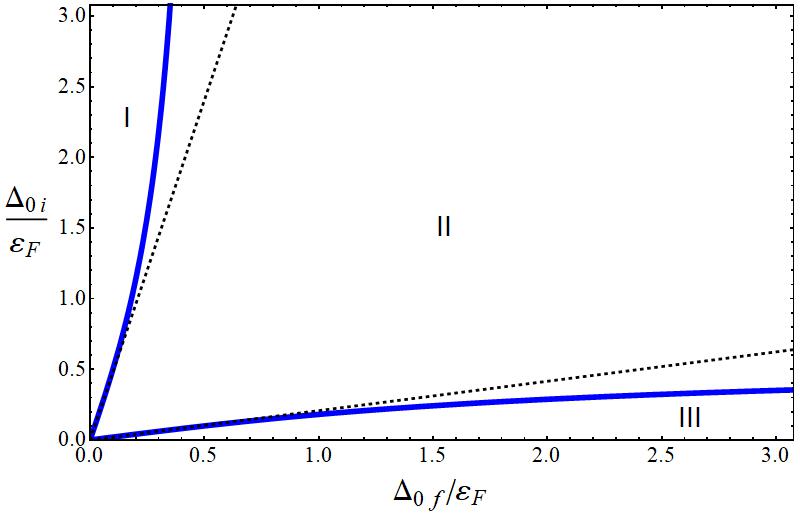}
\caption{Phase diagram for interaction quenches $g_i\to g_f$ in the integrable limit $h_f = h_i = 0$ of the spin-orbit model. Apart from the varying coupling constant $g$, the model parameters are the same as found in Fig.~\ref{HSOparams}. The black dotted lines $\Delta_{0i} = e^{\pm\pi/2}\Delta_{0f}$ indicate the weak coupling limit ($\Delta \ll W$) phase boundaries\cite{ydgf}. The thick blue lines mark the true phase boundaries, which are characterized by the appearance of a new pair of complex roots of \eref{hZeroHSOintEqs2} when passing from Phase~I to Phase~II or Phase~II to Phase~III.}
\label{spinOrbitPDhZero}
\end{figure}

We now construct the phase diagram shown in Fig.~\ref{spinOrbitPDhZero} for the $h_i = h_f = 0$, $g_i\to g_f$ quenches in the spin-orbit model. As we will not utilize particle-hole symmetry, the chemical potential $\mu$ must be calculated from the fermion number \eref{HSOzComp}, which in the present case reads
\beg
\begin{split}
N_f = \sum_{\mathbf{k}\lambda}\bigg(-\frac{\eps_{\mathbf{k}\lambda}-\mu}{2\sqrt{(\eps_{\mathbf{k}\lambda}-\mu)^2+\Delta_{0i}^2}}+\frac{1}{2}\bigg).
\end{split}
\label{nfAppendix}
\en
In the continuum limit, we have the following translation from sums over $\mathbf{k}\lambda$ to integrals over the continuum for arbitrary functions $F(\eps_{\mathbf{k}\lambda})$
\beg
\begin{split}
\sum_{\mathbf{k}\lambda}F(\eps_{\mathbf{k}\lambda})&=\frac{N}{W}\int_{-\eps_b}^{W_-}F(x) \nu_{\alpha}(x)dx,\\
\nu_{\alpha}(x) &=\begin{cases}
\frac{2}{\sqrt{1 + x/\eps_b}}, & -\eps_b\le x\le 0\\
2, & 0\le x\le W_+\\
1-\frac{1}{\sqrt{1 + x/\eps_b}}, & W_+\le x\le W_-
\end{cases},\\
\eps_b &\equiv \alpha^2/2,\quad W_{\lambda} \equiv W - 2\lambda\sqrt{\eps_b W}.
\end{split}
\label{sumTranslation}
\en
Thus, the spin-orbit coupling $\alpha$ at $h=0$ has the simple effect of introducing a peculiar density of states $\nu_{\alpha}(x)$ to the $s$-wave problem. Let $\widetilde{B} = \lim_{N\to\infty}\widetilde{\beta}/N$ and $n = \lim_{N\to\infty}N_f/N$, the latter of which is fixed for the entire phase diagram. For a given pair $(\Delta_{0f},\Delta_{0i})$, we first  solve for $(\mu_f,\mu_i)$ and then for $\widetilde{B}$ through the following integral equations:
\beg
\begin{split}
2n &= \int_X\bigg(1-\frac{x-\mu_{i,f}}{\sqrt{(x-\mu_{i,f})^2+\Delta_{0i,f}^2}}\bigg),\\
2\widetilde{B} &= \int_X\bigg(\frac{1}{\sqrt{(x-\mu_f)^2+\Delta_{0f}^2}}-\frac{1}{\sqrt{(x-\mu_i)^2+\Delta_{0i}^2}}\bigg),\\
\int_X(\cdot) &\equiv \frac{1}{W}\int_{-\eps_b}^{W_-}(\cdot)\nu(x)dx.
\end{split}
\label{hZeroHSOintEqs}
\en
We then use $\widetilde{B}$ and $\mu_i$ as input for the following integral equation:
\beg
\begin{split}
\int_X\frac{1}{(u-x)\sqrt{(x-\mu_i)^2+\Delta_{0i}^2}} = -\frac{2\widetilde{B}}{u-\mu_i \pm i\,\Delta_{0i}},
\end{split}
\label{hZeroHSOintEqs2}
\en
which we solve for $u$. The number of complex pairs of roots to \eref{hZeroHSOintEqs2} determines which nonequilibrium phase the system enters.

\begin{figure}
\includegraphics[width=\linewidth]{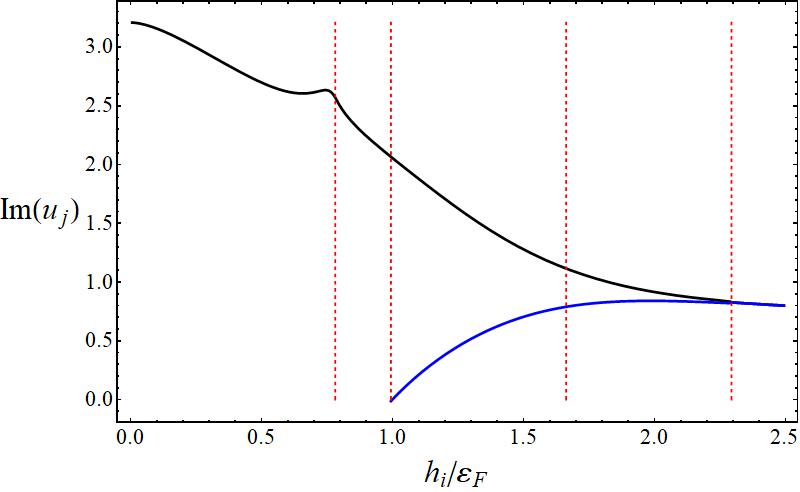}
\caption{Behavior of the roots of $L^2(u)$ for quenches from the ground state of $h_i\ne0$ to $h_f=0$ in the continuum limit with spin-orbit parameters as given in Fig.~\ref{HSOparams}. Each solid line is the absolute value of the imaginary part of a pair of complex conjugate roots. Regions of $h_i$ with one such line indicate that the asymptotic state is Phase~II, while the region where there are two separate lines indicate Phase~III. The vertical dashed lines indicate various critical values of $h_i$ where the system undergoes a phase transition or crossover. From left to right, $h_1 = 0.7813\eps_F$ is the topological transition of the ground state, $h_2 = 0.9938\eps_F$ is a Phase~II-III transition, $h_3 = 1.6625\eps_F$ is the BCS-BEC crossover, and $h_4 = 2.2938\eps_F$ is a Phase~III-II transition which also appears to correspond to $\Delta_{0i} = 0$ being the only self-consistent initial equilibrium gap. These critical values of $h_i$ depend in general on the various spin-orbit model parameters.}
\label{spinOrbitZerohf}
\end{figure}

Quenches from $h_i\ne0$ to $h_f=0$ still undergo integrable dynamics, except now the initial state is no longer the $s$-wave ground state. We consider the behavior of the zeros of $L^2(u)$ with respect to $h_i$ in the continuum limit with the spin-orbit parameters given in Fig.~\ref{HSOparams}. The Lax vector is still as defined in \eref{Lax1}, but we now enter the spin-orbit ground state \re{GSHSO} into the equation $L^2(u_j) = 0$, which implies $L^x(u_j) = \pm i\,L^z(u_j)$. The spin components of the $h_i\ne0$ ground state are functions of the form $F_{\lambda}(\eps_{\mathbf{k}})$ instead of $F(\eps_{\mathbf{k}\lambda})$; we therefore do not use \re{sumTranslation} for the continuum limit, but rather
\beg
\sum_{\mathbf{k}\lambda}F_{\lambda}(\eps_{\mathbf{k}})=\frac{N}{W}\int_0^W \bigg( F_+(x) + F_-(x)\bigg)dx.
\label{sumTranslation2}
\en
The result of the root calculation is given in Fig.~\ref{spinOrbitZerohf}, where we plot the absolute value of the imaginary part of each root pair. For small $h_i$,   there is only one pair of complex roots, i.e., the asymptotic phase is Phase~II. At a certain critical $h_i$, a second pair of complex roots appears, and the system enters Phase~III. For larger $h_i$, the two pairs of roots merge into one and the system reenters Phase~II. Phase~I does not occur in $h_f=0$ quenches for the parameters we used.

\section{Integrability breaking forbids asymptotic reduction}
\label{AppReductionMech}

An important property of the quench dynamics of integrable $s$ and $(p+ip)$-wave Hamiltonians   is the dynamical reduction in the number of   degrees of freedom at $t\to+\infty$ in the thermodynamic limit\cite{ydgf,fdgy}. In particular, Phase ~III in these models corresponds to the motion of two   collective classical spins ${\bf S}_1$ and  ${\bf S}_2$ governed by a Hamiltonian of the same form. The   asymptotic  order parameter $\Delta(t)$ in  Phase~III coincides with that of the  $2$-spin problem. Further, there are special \textit{reduced solutions} of equations of motion with the same $\Delta(t)$   that are of the  form
\beg
{\bf s}_j= \alpha_j {\bf S}_1+\beta_j {\bf S}_2+\eta_j \hat {\bf z},
\label{eayAppred}
\en
where $\alpha_j$, $\beta_j$ and $\eta_j$ are time-independent and $\hat {\bf z}$ is a unit vector along the z-axis.
These observations lead to an analytical expression for $\Delta(t)$ and, moreover, help to construct the full asymptotic spin configuration in Phase~III. We note also that, as we will see below, for the $s$-wave BCS model in the particle-hole symmetric case, \re{eayAppred} is equivalent to the ansatz of Ref.~\onlinecite{barlev0}.

We will now show that the above reduction mechanism  relies on integrability and breaks down for nonintegrable separable BCS models.
We will prove two independent statements: (i) reduced solutions exist only when $f^2(x)=C_1+C_2 x$, i.e. only when the Hamiltonian is integrable\cite{richardson,ortiz2005} and (ii) $\Delta(t)$ for a 2-spin separable BCS Hamiltonian with an arbitrary choice of new $\eps_{1,2}$,
$f_{1,2}$ and $g$ does not match  the asymptotic $\Delta(t)$ we obtained in Sect.~\ref{subphase3}.

\subsection{Existence of reduced solutions implies integrability and vice versa}

We will follow the same steps as in the derivation of the 2-spin solutions in  Ref.~\onlinecite{ydgf} and show that it only works for  special choices of $f(x)$.   First, we treat the general non-particle-hole symmetric case. 

Let
\beg 
\Delta=\Omega e^{-i\Phi}.
\en
The 2-spin (reduced) Hamiltonian is
\beg
\begin{split}
H_\mathrm{red} =\sum_{j=1}^2 2\widetilde{\eps}_j S^z_j- \widetilde{g} \sum_{j,k} \widetilde{f}_j \widetilde{f}_k S^-_jS^+_k= \\
=\sum_{j=1}^2 2\widetilde{\eps}_jS^z_j-\frac{|\Delta|^2}{\widetilde{g}},  
\end{split}
\label{eayAppH2spin}
\en
where $\Delta=\widetilde{g}(\widetilde{f}_1 S^-_1+\widetilde{f}_2 S^-_2)$. We take both $\widetilde{f}_k$ to be nonzero, because otherwise the two spins simply decouple and rotate uniformly around the z-axis.

Energy and $S^z_1+S^z_2$ are conserved. Since there are two conservation laws and two degrees of freedom, $H_\mathrm{red}$  is integrable. For more than two spins, integrability persists only for special choices of $\widetilde{f}_k$. This fact alone already distinguishes the 2-spin problem from that of a generic $N$-spin separable BCS Hamiltonian.

Conservation of energy and $S^z_1+S^z_2$  read
\beg
\begin{split}
2\widetilde{\eps}_1 S^z_1+2\widetilde{\eps}_2 S^z_2=\widetilde{E}+\frac{\Omega^2}{\widetilde{g}},\\
S^z_1+S^z_2=\mathrm{const},\\
\end{split}
\label{eayAppconslaws}
\en
We need $\widetilde{\eps}_1\ne \widetilde{\eps}_2$ or $|\Delta|$ will be constant.
We use \eref{eayAppconslaws} to express $S^z_k$ in terms of $\Omega^2$,
\beg
S^z_k=  \widetilde{a}_k\Omega^2+  \widetilde{b}_k,\quad k=1,2;
\label{eayAppsz2spin}
\en
where $\widetilde{a}_k$ and $\widetilde{b}_k$ are time-independent and $\widetilde{a}_1=-\widetilde{a}_2\ne0$. Furthermore, \eref{eayAppred} implies
a similar expression for $s_j^z$ in terms of the order parameter amplitude, 
\beg
s^z_j=  a_j\Omega^2+   b_j.
\label{eayAppszred}
\en
Conservation of the energy
\beg
E=\sum_j 2 \eps_js^z_j-\frac{|\Delta|^2}{ g},
\en
and of $J_z=\sum_j s_j^z$ require
\beg
\sum_j a_j=0,\quad \sum_j 2 \eps_j a_j=\frac{1}{g}.
\label{eayAppconstr}
\en

We write the Bloch equations for the separable BCS Hamiltonian   as
\begin{align}
\label{eayAppeom1} \dot s_j^z=-if_j(s_j^- \Delta^*-s_j^+\Delta),\\  
 \label{eayAppeom2} \dot s_j^-=-2i f_j s_j^z\Delta-2i\eps_j s_j^-.
 \end{align}
Since the equations of motion and \esref{eayAppsz2spin} and \re{eayAppszred}   for the reduced solution and the 2-spin problem have the same form, we can treat
both of them simultaneously.

  Substituting  \eref{eayAppszred}     into \eref{eayAppeom1}, we find
\beg
s_j^-e^{i\Phi}-s_j^+e^{-i\Phi}=2i\frac{a_j}{f_j}\dot\Omega.
\label{eayAppminus}
\en
Next, we multiply   \eref{eayAppeom2} by $e^{i\Phi}$ and add the resulting equation to its complex conjugate, 
\beg
\frac{d\phantom{t}}{dt}\left(s_j^-e^{i\Phi}+s_j^+e^{-i\Phi}\right)=\frac{4a_j\eps_j}{f_j}\dot\Omega-2\frac{a_j}{f_j}\dot\Phi\dot\Omega,
\label{eayAppinter}
\en
where we made use of \eref{eayAppminus}.  Integrating  and adding the resulting equation and \eref{eayAppminus}, we obtain
\beg
s_j^-e^{i\Phi}=\frac{2a_j\eps_j}{f_j}\Omega-\frac{a_j}{f_j} A+i\frac{a_j}{f_j}\dot\Omega+\frac{a_jc_j}{f_j},
\label{eayApps-}
\en
where $\frac{a_j c_j}{f_j}$ is the integration constant and  $A=\int dt \dot\Phi\dot\Omega$.  The self-consistency condition $\Delta=g\sum_j f_j s_j^-$,
combined with \eref{eayAppconstr}, implies $\sum_j a_j c_j=0$. 

The analogous expressions for the 2-spin problem are
\beg
S_k^-e^{i\Phi}=\frac{2\widetilde{a}_k \widetilde{\eps}_k}{\widetilde{f}_k}\Omega-\frac{\widetilde{a}_k}{\widetilde{f}_k} A+i
\frac{\widetilde{a}_k}{\widetilde{f}_k}\dot\Omega+\frac{\widetilde{a}_k \widetilde{c}_k}{\widetilde{f}_k},
\label{eayAppS-}
\en
and $\widetilde{a}_1 \widetilde{c}_1+\widetilde{a}_2 \widetilde{c}_2=\widetilde{a}_1 (\widetilde{c}_1-\widetilde{c}_2)=0$. Therefore, 
$\widetilde{c}_1=\widetilde{c}_2$ and the  last term in \eref{eayAppS-} can be absorbed into $A$, which is defined up to a constant anyway, i.e.,
\beg
S_k^-e^{i\Phi}=\frac{2\widetilde{a}_k \widetilde{\eps}_k}{\widetilde{f}_k}\Omega-\frac{\widetilde{a}_k}{\widetilde{f}_k} A+i
\frac{\widetilde{a}_k}{\widetilde{f}_k}\dot\Omega.
\label{eayAppS-1}
\en
Since $s_j^-$ is related to $S_1^-$ and $S_2^-$ via \eref{eayAppred}, this also eliminates the last term in \eref{eayApps-}, i.e.,
\beg
s_j^-e^{i\Phi}=\frac{2a_j\eps_j}{f_j}\Omega-\frac{a_j}{f_j} A+i\frac{a_j}{f_j}\dot\Omega.
\label{eayApps-1}
\en
Combining the conservation of the spin norm, $s_j^2=(s_j^z)^2+|s_j^-|^2$, with \esref{eayAppszred} and \re{eayApps-1}, we derive  the following differential equation for $\Omega$:
\beg
(a_j\Omega^2+b_j)^2+\frac{(2a_j\eps_j\Omega-a_j A)^2+a_j^2\dot\Omega^2}{f_j^2}=s_j^2,
\en
or, equivalently,
\beg
\begin{split}
\dot\Omega^2+f_j^2\Omega^4+\Omega^2\left(2\frac{f_j b_j}{a_j}+4\eps_j^2\right) -4\eps_j A\Omega\\
 +A^2+\frac{f_j^2(b_j^2-s_j^2)}{a_j^2}=0.\\
 \end{split}
\label{eayAppmain}
\en
This equation implies, among other things, that $A$ is a function of $\Omega$. Indeed, consider a set of numbers $x_j$, such that $\sum_j x_j=0$. Multiplying \eref{eayAppmain} by $x_j$ and summing over $j$, we find
\beg
A\Omega=\lambda \Omega^4+2\mu\Omega^2+\kappa,
\label{eayAppA}
\en
where $\lambda, \mu$ and $\kappa$ are real constants. Substituting this back into \eref{eayAppmain}, we obtain
\beg
\begin{split}
\frac{\dot w^2}{4}+\lambda^2 w^4 + (f_j^2-4\lambda\xi_j)w^3
 +\left(\frac{2f_j b_j}{a_j} +
2\lambda\kappa+4\xi_j^2\right) w^2+\\
\left(\frac{f_j^2(b_j^2-s_j^2)}{a_j^2}-4\kappa\xi_j\right)w+\kappa^2=0,
\end{split}
\label{eayAppw}
\en
where $w=\Omega^2$ and $\xi_j=\eps_j-\mu$. These equations are consistent only when the coefficients of powers of $w$ are $j$-independent.
In particular, we must have $f_j^2=4\lambda\xi_j+\mathrm{const.}$, i.e.,
\beg
f_j^2=C_1+C_2\eps_j,
\label{eayAppfj}
\en
where $C_1$ and $C_2$ are real constants.
This is the most general form of $f_j$ for which the separable BCS Hamiltonian~\re{HFSpin} is known to be integrable \cite{richardson,ortiz2005}. In particular, $C_2$=0 corresponds to the $s$-wave and $C_1=0$ to the $(p+ip)$-wave models. Conversely, when \eref{eayAppfj} holds and the separable Hamiltonian is therefore integrable, the $j$-independence of coefficients at $w^2$ and $w$ determines $a_j$ and $b_j$, and  \eref{eayAppw} means that
$w=|\Delta|^2$ is a certain elliptic function of time.

\subsection{Asymptotic $\Delta(t)$ does not match the 2-spin solution in nonintegrable cases}

In Sect.~\ref{subphase3} we numerically determined $\Delta(t)$ in two nonintegrable separable BCS Hamiltonians, see \eref{deltaParamPhase3}. Here we show that $\Delta(t)$ for the most general separable 2-spin Hamiltonian \re{eayAppH2spin} cannot match \eref{deltaParamPhase3}.

Since $\Delta(t)$ in \eref{deltaParamPhase3} is real, we take $\Delta$ in the 2-spin problem to be real as well, though we do not  a priori assume particle-hole symmetry in the 2-spin problem. All we need is to specialize the derivation of the previous subsection to the case of real $\Delta$. Then, the Bloch equations become
\beg
\begin{split}
\dot{S}_j^z &= -2\widetilde{f}_j S_j^y \Delta,\\
\dot{S}_j^x &= -2\widetilde{\eps}_j S_j^y,\\
\dot{S}_j^y &= 2\widetilde{\eps}_j S_j^x + 2\widetilde{f}_j S_j^z \Delta.
\end{split}
\label{eayAppeom2spin}
\en
Substituting \eref{eayAppsz2spin} into the first two equations of motion, we obtain
\beg
 S_k^y=-\frac{\widetilde{a}_k}{\widetilde{f}_k} \dot\Delta,
 \label{eayAppsy2spin}
 \en
 and
 \beg
 S_k^x=\frac{2\widetilde{\eps}_k \widetilde{a}_k}{\widetilde{f}_k}  \Delta+ \frac{\widetilde{a}_k \widetilde{c}_k}{\widetilde{f}_k},
  \label{eayAppsx2spin}
 \en
 where $\frac{\widetilde{a}_k \widetilde{c}_k}{\widetilde{f}_k}$ is the integration constant.
 As before,  the  self-consistency condition $\widetilde{g}(\widetilde{f}_1S_1^-+\widetilde{f}_2S_2^-)=\Delta$ together with  $\widetilde{a}_1 =-\widetilde{a}_2$ imply $\widetilde{c}_1 =\widetilde{c}_2 \equiv \widetilde{c}$, and the conservation of spin length $(S_k^x)^2+(S_k^y)^2+(S_k^z)^2=S_k^2$ yields
 \beg
\dot \Delta^2+\left( 2\widetilde{\eps}_k\Delta+ \widetilde{c} \right)^2+\left(\widetilde{f}_k \Delta^2+\frac{\widetilde{b}_k \widetilde{f}_k}{ \widetilde{a}_k}\right)^2= \frac{S_k^2 \widetilde{f}_k^2}{\widetilde{a}_k^2}.
\label{eayAppdiff1}
\en
Equating the coefficients at different powers of $\Delta$ for $k=1$ and 2, we find $\widetilde{c}(\widetilde{\eps}_1-\widetilde{\eps}_2)=0 \Rightarrow \widetilde{c}=0$, 
\beg
\widetilde{f}_1=\widetilde{f}_2 \equiv \widetilde{f},
\label{eayAppfcons}
\en 
and two more relationships that constrain $\widetilde{a}_k$ and $\widetilde{b}_k$. The constraint \re{eayAppfcons} is a consequence of the requirement
that $\Delta$ be real. Now \eref{eayAppdiff1}   is of the form
\beg
\dot\Delta^2=-\widetilde{f}^2(\Delta^2-\Delta_+^2)(\Delta^2-\Delta_-^2).
\label{eayAppdn}
\en
This is the same as the equation for the asymptotic $\Delta(t)$ for the integrable  $s$-wave BCS Hamiltonian in the particle-hole symmetric case up to rescaling $\Delta_\mathrm{new}=  \widetilde{f} \Delta $. This is not surprising because  $\widetilde{f}_1=\widetilde{f}_2=\widetilde{f}$ and the factor of $\widetilde{f}^2$ in \eref{eayAppH2spin} can be absorbed into the coupling constant, $\widetilde{g}_\mathrm{new}=\widetilde{f}^2\widetilde{g} $ resulting in an integrable $s$-wave BCS Hamiltonian for two spins with $\Delta_\mathrm{new}=\widetilde{g}_\mathrm{new}(S_1^-+S_2^-)=\widetilde{f}\Delta$. The solution of \eref{eayAppdn} is $\Delta(t)=\Delta_+\mathrm{dn} [\widetilde{f}\Delta_+(t-t_0), 1-\frac{\Delta_+^2}{\Delta_-^2}]$. As we saw in Sect.~\ref{subphase3}, in the nonintegrable  case we find instead a more general differential equation \eref{ellipticDefinition} with the solution given by \eref{deltaParamPhase3}. 

\section{The link between Lax constructions and the stability analysis}
\label{linkLaxStab}
As mentioned above, the separable BCS model is integrable when $f_j^2 = C_1 \eps_j + C_2$. Two important cases are the $s$-wave model where $f_j = 1$ and the $p+ip$ model where $f_j = \sqrt{\eps_j}$. In past work\cite{ydgf,fdgy}, integrability has been exploited to determine the nonequilibrium asymptotic phases through the use of Lax constructions. These techniques are useful for constructing phase diagrams, but the physical interpretation of the phase transitions is obscured by the use of exact solvability. We demonstrate here that the stability equation \eref{freqsPhase2LinNoPH}, which applies to the nonintegrable cases as well, both predicts the same transition points and clarifies the physical meaning of the  Lax construction.

In the following, we will assume the quantities $Z_j$, $\Delta_{\infty}$ and $\mu_{\infty}$ are given. They are functions of the quench parameters $\Delta_{0i}$, $\Delta_{0f}$, the particle number $N_f$, and the Fermi energy $\eps_F$.

 \subsection{Lax norms}
In the $s$-wave model, the Lax vector is\cite{ydgf}
\begin{equation}
\begin{split}
\mathbf{L}_s(u) = -\frac{\hat{\mathbf{z}}}{g_f} + \sum_j\frac{\mathbf{s}_j}{u - \eps_j},
\end{split}
\label{StabilityLaxSwave}
\end{equation}
while in the $p+ip$ model its components are\cite{fdgy}
\beg
\begin{split}
L_p^+(u) &= \sum_j\frac{\sqrt{\eps_j}s_j^+}{u - \eps_j},\\
L_p^-(u) &= \sum_j\frac{\sqrt{\eps_j}s_j^-}{u - \eps_j},\\
L_p^z(u) &= \sum_j\frac{\eps_js_j^-}{u - \eps_j} - \frac{1}{g_f},
\end{split}
\en
where $u$ is a  complex (spectral) parameter.

We focus on the norms of these quantities, defined as $L^2(u) = L_x^2(u) + L_y^2(u) + L_y^2(u)$ in the $s$-wave case and $L_2(u) = u L^+(u)L^-(u) +[L^z(u)]^2$ for $p+ip$. Integrability follows from the fact that the $L^2(u)$ and $L_2(u)$ are conserved by the time evolution for arbitrary $u$, which implies conservation of their roots $u_j$. As demonstrated in Refs.~\onlinecite{ydgf} and \onlinecite{fdgy} and discussed in Appendix~\ref{appB}, each of the asymptotic nonequilibrium phases corresponds a unique number of isolated complex pairs of $u_j$ in the continuum limit. Phase~I corresponds to zero isolated $u_j$, Phase~II corresponds to one pair, and Phase~III corresponds to two pairs.

The main result of this Appendix is that  the roots of the Lax norm $u$ and the frequencies $\omega$ of $\delta\Delta(t)$ are related by $u - u_r = \pm\frac{1}{2}\sqrt{\omega^2 - b_{\textrm{min}}^2}$, where $u_r$ is the real part of the root (cf. 
Refs.~\onlinecite{yta,ydgf}), and $b_{\textrm{min}}$ is the band edge in the frequency spectrum ($b_{\textrm{min}} = 0$ in Phase~I). Thus, the new pair of complex conjugate Lax roots appears at the same time that $\omega$ emerges into the band gap 
(i.e., $\omega^2<b_{\textrm{min}}^2$ in Phase II and $\omega^2<0$ in Phase~I). Here and below in this Appendix, we use the same notation $u$ for the roots and for generic values of the spectral parameter.

One may plug into the Lax norms the asymptotic spin solution \re{eomHFphase2sol} for Phase~II, but we shall use solutions that do not impose particle-hole symmetry. Letting $\widetilde{\eps}_j = \eps_j - \mu_{\infty}$, and noting that sums over the time-dependent terms dephase in the $t\to\infty$ limit, we find
\begin{equation}
\begin{split}
L^2(u) &= \bigg(-\frac{1}{g_f}+\sigma_1\bigg)^2 + \Delta_{\infty}^2\sigma_2^2,\\
\sigma_1 &\equiv \sum_j\frac{Z_j}{u-\eps_j}, \quad \sigma_2 \equiv \sum_j\frac{Z_j}{\widetilde{\eps}_j(u-\eps_j)},\\
L_2(u) &= \bigg(-\frac{1}{g_f}+ p_1\bigg)^2 + u \Delta_{\infty}^2p_2^2\\
p_1 &\equiv \sum_j\frac{\eps_jZ_j}{u - \eps_j}, \quad p_2 \equiv\sum_j\frac{\eps_jZ_j}{\widetilde{\eps}_j(u - \eps_j)}.\\
\end{split}
\label{StabilityLaxSwave2b}
\end{equation}
\eref{StabilityLaxSwave2b} reduces to the Phase~I Lax norms when $\Delta_{\infty} = 0$ and by convention $Z_j \to z_j$. In Phase~II, \eref{StabilityLaxSwave2b} is supplemented by the self-consistency relationship
\beg
1  = -g_f\sum_j\frac{f_j^2Z_j}{\widetilde{\eps}_j}.
\label{selfConAppNoPH}
\en

\subsection{Phase I-II transition}

In the $s$-wave case, and in Phase~I, we compare the stability equation \eref{freqsPhase2LinNoPH} to the vanishing of the Lax norm $L^2(u) = 0$. After some algebra, \esref{freqsPhase2LinNoPH} and $L^2(u) = 0$ become
\begin{subequations}
\begin{align}
\frac{1}{g_f} & = \sum_j\frac{z_j}{\pm\frac{1}{2}\omega_0+\mu_{\infty}-\eps_j},
\label{stabilitySwavePhase1}\\
\frac{1}{g_f} &= \sum_j\frac{z_j}{u-\eps_j},
\label{LaxSwavePhase1}
\end{align}
\end{subequations}
respectively. We argued in Sect.~\ref{subStAn} that the Phase~I-II transition occurs when a purely imaginary pair of complex conjugate $\omega_0$ emerges as solutions to \eref{stabilitySwavePhase1}, implying an exponential instability to Phase~I. The Lax construction stipulates that the same transition occurs when an isolated pair of complex conjugate $u$ solve \eref{LaxSwavePhase1}. In order for these two methods to match, we must make the identification $u -\mu_{\infty} = \pm\frac{1}{2}\omega_0$, i.e., the real part of the emergent Lax norm pair of roots must be $\mu_{\infty}$. We prove this is the case in Sect.~\ref{LaxRealProofs}.

The corresponding equations for Phase~I in the $p+ip$ model are
\begin{subequations}
\begin{align}
\frac{1}{g_f} & = \sum_j\frac{\eps_jz_j}{\pm\frac{1}{2}\omega_0+\mu_{\infty}-\eps_j},
\label{stabilitySwavePhase1b}\\
\frac{1}{g_f} &= \sum_j\frac{\eps_jz_j}{u-\eps_j},
\label{LaxSwavePhase1b}
\end{align}
\end{subequations}
and the same identification reconciles the two approaches.

\subsection{Phase II-III transition}

In Phase~II, one applies the self-consistency relationship \re{selfConAppNoPH} to the Lax norms \re{StabilityLaxSwave2b}. In the $s$-wave case, $L^2(u) = 0$ becomes
\beg
0 = \big[(u-\mu_{\infty})^2+\Delta_{\infty}^2\big]\bigg(\sum_j\frac{Z_j}{\widetilde{\eps}_j(u-\eps_j)}\bigg)^2,
\label{LaxSwavePhase2}
\en
and we see the single pair of isolated conjugate roots are $u_{\pm} = \mu_{\infty} \pm i\Delta_{\infty}$. The equation for the second pair of isolated roots that would signal a transition to Phase~III is therefore
\beg
0 = \sum_j\frac{Z_j}{\widetilde{\eps}_j(u-\eps_j)}.
\label{LaxSwavePhase3}
\en
After applying \eref{selfConAppNoPH} to the quantities $S_j(\omega_0)$ in the stability equation \re{freqsPhase2LinNoPH}, we find for the $s$-wave model
\beg
\begin{split}
S_1(\omega) -1 = \bigg(\frac{\omega^2}{4\Delta_{\infty}^2}-1\bigg)S_3(\omega).
\end{split}
\label{sol2Repeat}
\en
This simplifies \eref{freqsPhase2LinNoPH} to
\begin{equation}
\begin{split}
0 = \sum_j\frac{Z_j}{\widetilde{\eps}_j(\pm y + \mu_{\infty} -\eps_j)},\quad y = \frac{1}{2}\sqrt{\omega^2 - 4\Delta_{\infty}^2}.
\end{split}
\label{StabilityLaxSwave3c}
\end{equation}
Matching \re{StabilityLaxSwave3c} to \re{LaxSwavePhase3}, we make the correspondence $u - \mu_{\infty} = \pm\frac{1}{2}\sqrt{\omega_0^2 - 4\Delta_{\infty}^2}$. As we discussed in Sect.~\ref{subStAn}, an $\omega_0$ emerging out of the continuum and into the band gap signals the transition to Phase~III. The band edge in the $s$-wave model is precisely $2\Delta_{\infty}$. We show in Sect.~\ref{LaxRealProofs} that the new pair of conjugate Lax roots has real part $\mu_{\infty}$. Therefore, the two approaches predict the same phase transition.

In the $p+ip$ case, $L_2(u) = 0$ couples with \re{selfConAppNoPH} to give
\beg
0 = \big[u \Delta_{\infty}^2 + (u - \mu_{\infty})^2\big]\bigg(\sum_j\frac{\eps_jZ_j}{\widetilde{\eps}_j(u - \eps_j)}\bigg)^2.
\en
The single pair of isolated roots of Phase~II is then
\beg
u_{\pm} =  u_c  \pm i\Delta_{\infty}\sqrt{ \mu_{\infty} - \frac{\Delta_{\infty}^2}{4}};\quad u_c\equiv \mu_{\infty} - \frac{\Delta_{\infty}^2}{2},
\label{StabilityLaxPip6}
\en
and the emergent pair of conjugate roots solves
\beg
\begin{split}
0=\sum_j\frac{\eps_jZ_j}{\widetilde{\eps}_j(u - \eps_j)}.
\end{split}
\label{StabilityLaxPipy}
\en

To show that the stability analysis reproduces \eref{StabilityLaxPipy}, we will need two relations. The first holds in general by applying the self-consistency relation \re{selfConAppNoPH} to the sums in \re{freqsPhase2LinNoPH}
\beg
\begin{split}
S_1(\omega) - 1&= \omega^2 S_4(\omega) - S_3(\omega),\\
 S_4(\omega)&\equiv g_f\sum_j\frac{f_j^2 Z_j}{\widetilde{\eps}_j(\omega^2-\widetilde{b}_j^2)},
\label{S1m1Transformed}
\end{split}
\en
while the second is specific to the $p+ip$ model
\beg
S_2(\omega) = -2\omega \mu_{\infty} S_4(\omega) + \frac{\omega}{2\Delta_{\infty}^2}S_3(\omega).
\label{S2TransformedPip}
\en
We substitute \esref{S1m1Transformed}-\re{S2TransformedPip} into \eref{freqsPhase2LinNoPH}, which becomes a quadratic function of $S_3$ and $S_4$. The solution is
\begin{equation}
\begin{split}
0 = \sum_j\frac{Z_j}{\widetilde{\eps}_j(\pm y + u_c - \eps_j)},\quad y = \frac{1}{2}\sqrt{\omega^2 - B_1^2},
\end{split}
\label{StabilityLaxPipFinal}
\end{equation}
where $B_1 = \sqrt{4\mu_{\infty}\Delta_{\infty}^2-\Delta_{\infty}^4}$ is the band edge when $u_c \ge 0$. In this parameter range, we identity $u - u_c = \pm\frac{1}{2}\sqrt{\omega^2 - B_1^2}$. We show in Sect.~\ref{LaxRealProofs} that the real part of the emergent Lax roots is $u_c$, and therefore the stability analysis and Lax constructions give the same Phase~II-III transition. When $u_c < 0$, the band edge is no longer $B_1$, and we believe there to be no Phase~II-III transition in that case.

\subsection{Real parts of Lax roots at the transitions}
\label{LaxRealProofs}

The equivalence between the Lax construction and the stability analysis relies on the fact that the real parts of the emerging Lax roots are equal to $\mu_{\infty}$ at the Phase~I-II transition in both integrable models, $\mu_{\infty}$ at the Phase~II-III transition in the $s$-wave model, and $\mu_{\infty} - \frac{\dinf^2}{2}$ at the Phase~II-III transition in the $p+ip$ model. In other words, the emergent second pair of isolated roots has the same real part as the first pair of isolated roots.

The Phase~I-II transition real parts can be understood by a continuity argument. In the $s$-wave model, \eref{LaxSwavePhase2} implies that the single pair of roots can be written as $u_{\pm} = \mu_{\infty}\pm i\dinf$. As we approach the I-II boundary, $\dinf$ decreases continuously to zero, which implies the real part of both roots at the boundary is $\mu_{\infty}$. In the $p+ip$ case, a similar argument follows from \eref{StabilityLaxPip6}.

\subsubsection{$s$-wave, II-III}

We use results from the spin reduction mechanism, discussed in Appendix~\ref{AppReductionMech}, of the $s$-wave model to obtain the real parts of the Lax roots at the Phase~II-III transition. This discussion quotes several results directly from Sect.~II B 3 of Ref.~\onlinecite{ydgf}. The isolated roots in Phase~III of $L^2(u)$ are given by the roots of the 2-spin spectral polynomial\cite{ydgf} $Q_4(u)$
\beg
Q_4(u) = \big[(u-\mu)^2-\rho\big]^2-\kappa(u-\mu)-\chi.
\label{spectral2SpinSwave}
\en
We determine the real parameters $\mu$, $\rho$, $\kappa$ and $\chi$ at the transition, which will then give the roots of $Q_4(u)$. To do so, we use the differential equation and solution for the 2-spin $\Delta$, which is identical to that of the Phase~III asymptotic $\Delta$ of the many-body problem, which we write as $\Delta = |\Delta|e^{-i \Phi}$. Let $w = |\Delta|^2 = \Lambda^2 +h_1$, where $h_1$ is a constant. The differential equation for $w$ is
\beg
\begin{split}
0 = \dot{w}^2+4w^3+16\rho w^2+16\chi w+4\kappa^2,
\label{spectral2SpinSwaveB}
\end{split}
\en
while the equation for the phase $\Phi$ is
\beg
\begin{split}
\dot{\Phi} = 2\mu - \frac{\kappa}{\Lambda^2+h_1}.
\label{spectral2SpinSwavePhase}
\end{split}
\en
Upon rewriting \re{spectral2SpinSwaveB} as an equation for $\Lambda$, we find
\beg
\begin{split}
\dot{\Lambda}^2 &=-(\Lambda_+^2-\Lambda^2)(\Lambda_-^2-\Lambda^2),
\end{split}
\label{LambdaDiffEq}
\en
where the constants $\Lambda_{\pm}$ are the maximum and minimum of the $\Lambda$ oscillations which are functions of the constants $\rho$, $\chi$ and $\kappa$. The solution of interest to \eref{LambdaDiffEq} is
\beg
\Lambda = \Lambda_+\textrm{dn}\bigg[\Lambda_+(t-t_0),1-\frac{\Lambda^2_-}{\Lambda^2_+}\bigg].
\label{LambdaDiffEqSol}
\en
Near the II-III transition, the oscillations of $\Lambda$ are small and it sufficient to keep only the first harmonic of \eref{LambdaDiffEqSol}
\beg
\begin{split}
\Lambda &\approx \Lambda_0 + \delta \cos\big[{\omega_0 (t-t_0)}\big],\\
\delta &\ll \Lambda_0, \quad \omega_0 \approx 2 \Lambda_0.
\label{LambdaDiffEqSol2}
\end{split}
\en

As we approach the II-III transition, $\Delta \to \dinf e^{-2i\mu_{\infty}t}$. Because $|\Delta|^2 = \Lambda^2 + h_1$ has the same frequency as $\Lambda^2$, and the frequency of small oscillations of $|\Delta|^2$ at the II-III transition is $2\dinf$, we conclude $\Lambda_0 = \dinf$ and $h_1 = 0$. Using \eref{spectral2SpinSwavePhase}, we also find $\kappa = 0$ and $\mu = \mu_{\infty}$.

It remains to determine the constants $\rho$ and $\chi$, which we do by plugging \re{LambdaDiffEqSol2} into \re{spectral2SpinSwaveB} and considering the $\mathcal{O}(\delta^0)$ and $\mathcal{O}(\delta)$ terms separately. The result is $\rho = -\frac{\dinf^2}{2}$ and $\chi = \frac{\dinf^4}{4}$. The roots of the spectral polynomial $Q_4(u)$ from \eref{spectral2SpinSwave} at the Phase~II-III transition therefore solve
\beg
0 = \bigg[(u-\mu_{\infty})^2+\frac{\dinf^2}{2}\bigg]^2-\frac{\dinf^4}{4}.
\label{spectral2SpinSwaveD}
\en
One solution to \re{spectral2SpinSwaveD} is $u_{\pm} = \mu_{\infty} \pm i \dinf$, which is the single isolated pair characteristic of Phase~II. The other solution is a double root at $u = \mu_{\infty}$, i.e., the new pair of roots that emerges in Phase~III has real part $\mu_{\infty}$.

\subsubsection{$p+ip$, II-III}

In order to prove that the Lax construction and stability analysis predict the same $p+ip$ Phase~II-III transition, we needed to assume that the real part of the emerging second pair of roots equals that of the first pair of roots $u_{\pm}$ from \re{StabilityLaxPip6}. Using results from Ref.~\onlinecite{fdgy}, we now show that this is indeed the case.

For brevity, our derivation will use the conventions of Ref.~\onlinecite{fdgy}, where the definitions of some quantities differ by numerical factors. One redefines $\eps\to 2\eps$, $2G\to g$, $\sqrt{2}\Delta\to \Delta$ and $u\to2u$ in order to translate quantities from Ref.~\onlinecite{fdgy} to those in this work. While some details of the derivation depend on such conventions, the conclusion does not. We also assume $u_c \equiv \textrm{Re}[u_{\pm}] \ge 0$, which is the parameter regime where we show the equivalence of the Lax construction and stability analysis for the $p+ip$ model. 

Eq.~(4.3) of Ref.~\onlinecite{fdgy} gives the isolated pair of roots in Phase~II to be $u_{\pm} =u_c \pm 2i\Emin$, where $\Emin$ is the minimum of the asymptotic dispersion relation [see text below Eq.~(5.29) in Ref.~\onlinecite{fdgy}]. According to Eq.~(4.39) in Ref.~\onlinecite{fdgy} the frequency of small oscillations in Phase III close to the Phase II-III boundary is
\beg
\Omega_c=\sqrt{(u_{\rr}-u_c)^2+4\Emin^2},
\label{omegac}
\en
where $u_{\rr}$ is the real part of the pair of roots absent in Phase~II. The frequency $\Omega_c$ should match the frequency of dephasing oscillations in Phase~II close to the boundary. The text below Eq.~(3.53) in Ref.~\onlinecite{fdgy} says that the latter frequency is
\beg
\Omega= 2\Emin.
\en
Setting $\Omega_c = \Omega$, implies that on the Phase~II-III boundary
\beg
u_{\rr}=u_c.
\en


\begin{thebibliography}{99}
\bibitem{note} Not all three phases are necessarily present in a given model, e.g., there are models where Phase II\cite{barlev3} or Phase III\cite{psc} are absent, see also the discussion in Sect.~\ref{subStAn}
\bibitem{note11} One can choose the parameters differently in the two models, e.g., choose a different $\eps_F$ in the $s$-wave model, so that
the first few oscillations almost coincide\cite{didShortTime}. However, such fine-tuning seems unwarranted  and, in any case, the longer-time dynamics will disagree.
\bibitem{noteSep20_2} In equilibrium, i.e., when $g_f=g_i$, \eref{intSep19} is simply the BCS gap equation with two fixed points $\Delta=0$ 
and $\Delta=\pm\Delta_{0i}$ corresponding to the normal and the BCS ground states, respectively.
\bibitem{noteSep20_1} In the absence of particle-hole symmetry, Phase~II is a simple limit cycle $\Delta_\infty e^{-2i\mu_\infty t}$ (circle in the plane of complex $\Delta$). Phase~III is a 2D  (two fundamental frequencies) torus (annulus in the plane of complex $\Delta$).
Phase~I-II and II-III transitions are then analogous to supercritical Hopf and Neimark-Sacker (torus) bifurcations.
\bibitem{note22} Note that Ref.~\onlinecite{ddgp}  has a different convention of naming Phases~I and III. The only example of
Phase~III $|\Delta(t)|$ dynamics shown in Ref.~\onlinecite{ddgp}   is for the integrable case $h_f=0$, when the long relaxation time $\tau$  is absent.
\end{thebibliography}
\end{document}